\documentclass[11pt,a4paper]{article}

\pdfoutput=1

\usepackage{jheppub}

\usepackage[T1]{fontenc}
\usepackage{epsfig}
\usepackage{amscd}
\usepackage{verbatim}
\usepackage{latexsym}
\usepackage{amsfonts,amsthm,amsmath,mathtools}
\usepackage{color}
\usepackage{xcolor}
\usepackage{booktabs}
\usepackage{float}
\usepackage{graphicx}
\usepackage{lscape}
\usepackage{slashed}
\usepackage{subfigure}
\usepackage{multirow,makecell}

\usepackage{tikz}
\usepackage{etoolbox}

\usepackage{ulem}

\newcommand{\circled}[2][]{%
  \tikz[baseline=(char.base)]{%
    \node[shape = circle, color = blue, draw, inner sep = 1pt]
    (char) {\phantom{\ifblank{#1}{#2}{#1}}};%
    \node at (char.center) {\makebox[0pt][c]{#2}};}}
\newcommand{\dcircled}[1]{\circled[-----.]{#1}}

\numberwithin{equation}{section}

\def \be {\begin{equation}}
\def \ee {\end{equation}}
\def \ba {\begin{array}}
\def \ea {\end{array}}
\def \bea{\begin{eqnarray}}
\def \eea{\end{eqnarray}}
\def \nn {\nonumber}

\def \a {\alpha}
\def \b {\beta}
\def \g {\gamma}

\def \G {\Gamma}
\def \d {\delta}

\def \e {\epsilon}
\def \ve {\varepsilon}
\def \m {\mu}
\def \n {\nu}
\def \k {\kappa}
\def \l {\lambda}
\def \L {\Lambda}
\def \s {\sigma}

\def \r {\rho}

\def \th {\theta}
\def \vth {\vartheta}
\def \Th {\Theta}
\def \t {\tau}
\def \z {\zeta}

\def \mA {{\mathcal A}}

\def \mL {{\mathcal L}}

\def \mN {{\mathcal N}}
\def \mO {{\mathcal O}}
\def \mP {{\mathcal P}}
\def \mQ {{\mathcal Q}}

\def \mV {{\mathcal V}}

\def \mZ {{\mathcal Z}}

\def \cA {{\mathcal A}}
\def \cB {{\mathcal B}}

\def \cI {{\mathcal I}}
\def \cJ {{\mathcal J}}
\def \cK {{\mathcal K}}
\def \cL {{\mathcal L}}

\def \cN {{\mathcal N}}

\def \cP {{\mathcal P}}

\def \rC {{\mathrm C}}

\def \rO {{\mathrm O}}

\def \rR {{\mathrm R}}
\def \rS {{\mathrm S}}

\def \rZ {{\mathrm Z}}

\def \td {{\tilde d}}

\def \p {\partial}
\def \f {\frac}

\def \lt {\left}
\def \rt {\right}

\def \lra {\leftrightarrow}
\def \sr {\sqrt}
\def \td {\tilde}

\def \inf {\infty}

\def \lag {\langle}
\def \rag {\rangle}
\def \lagg {\langle\!\langle}
\def \ragg {\rangle\!\rangle}

\def \hi  {{\hat\imath}}
\def \hj  {{\hat\jmath}}

\def \ho  {{\hat 1}}
\def \hw  {{\hat 2}}

\def \ep {\mathrm{e}}
\def \ii {\mathrm{i}}

\def \bo {\boldsymbol{1}}

\def \Tr {{\textrm{Tr}}}

\def \diag {{\textrm{diag}}}

\def \AdS {{\textrm{AdS}}}

\def \gabjm {{\textrm{AdS}_4\times\textrm{S}^7/\textrm{Z}_k}}
\def \goabjm {{\textrm{AdS}_4\times\textrm{S}^7/(\textrm{Z}_{rk}\times\textrm{Z}_{r})}}

\def \Z {{\textrm{Z}}}

\def \AdS {{\textrm{AdS}}}

\def \bos {{\textrm{bos}}}
\def \fer {{\textrm{fer}}}

\def \CS {{\textrm{CS}}}

\begin{document}

\title{BPS Wilson loops in $\mathcal N \geq 2$ superconformal Chern--Simons--matter theories}
\author[a]{Andrea Mauri}
\author[bc]{\!\!,~Hao Ouyang}
\author[a]{\!\!,~Silvia Penati}
\author[de]{\!\!,~Jun-Bao Wu}
\author[a]{\!\!,~Jiaju Zhang}
\affiliation[a]{Dipartimento di Fisica, Universit\`a degli Studi di Milano-Bicocca, \& INFN, Sezione di Milano-Bicocca, Piazza della Scienza 3, I-20126 Milano, Italy}
\affiliation[b]{Institute of High Energy Physics, \& Theoretical Physics Center for Science Facilities,\\Chinese Academy of Sciences, 19B Yuquan Road, Beijing 100049, China}
\affiliation[c]{School of Physical Sciences, University of Chinese Academy of Sciences,\\19A Yuquan Road, Beijing 100049, China}
\affiliation[d]{Center for Joint Quantum Studies, School of Science, Tianjin University,\\135 Yaguan Road, Tianjin 300350, China}
\affiliation[e]{Center for High Energy Physics, Peking University, 5 Yiheyuan Road, Beijing 100871, China}
\emailAdd{andrea.mauri@mi.infn.it, ouyangh@ihep.ac.cn, silvia.penati@mib.infn.it, junbao.wu@tju.edu.cn, jiaju.zhang@unimib.it}

\abstract{
In $\mathcal N \geq 2$ superconformal Chern--Simons--matter theories we construct the infinite family of  Bogomol'nyi--Prasad--Sommerfield (BPS) Wilson loops featured by constant parametric couplings to scalar and fermion matter, including both line Wilson loops in Minkowski spacetime and circle Wilson loops in Euclidean space. We find that the connection of the most general BPS Wilson loop cannot be decomposed in terms of double--node connections. Moreover, if the quiver contains triangles, it cannot be interpreted as a supermatrix inside a superalgebra. However, for particular choices of the parameters it reduces to the well--known connections of 1/6 BPS Wilson loops in  Aharony--Bergman--Jafferis--Maldacena (ABJM) theory and 1/4 BPS Wilson loops in $\mathcal N = 4$ orbifold ABJM theory. In the particular case of $\mathcal N = 2$ orbifold ABJM theory we identify the gravity duals of a subset of operators. We investigate the cohomological equivalence of fermionic and bosonic BPS Wilson loops at quantum level by studying their expectation values, and find strong evidence that the cohomological equivalence holds quantum mechanically, at framing one. Finally, we discuss a stronger formulation of the cohomological equivalence, which implies non--trivial identities for correlation functions of composite operators in the defect CFT defined on the Wilson contour and allows to make novel predictions on the corresponding unknown integrals that call for a confirmation.}


\maketitle


\section{Introduction}

The study of Bogomol'nyi--Prasad--Sommerfield Wilson loops (BPS WLs) have yielded tremendous insights into supersymmetric gauge theories. In particular,
vacuum expectation values of BPS WLs can be computed exacly using localization techniques \cite{Pestun:2007rz,Kapustin:2009kz}, so providing
functions of the coupling constants that interpolate between weak and strong coupling regimes.
Therefore, for theories admitting holographic dual descriptions BPS WLs represent one of the most important tests of the AdS/CFT  correspondence \cite{Maldacena:1997re,Gubser:1998bc,Witten:1998qj,Aharony:2008ug,Aharony:2008gk}.

In three-dimensional superconformal Chern--Simons-matter (SCSM) theories many interesting results on BPS WLs have been obtained in recent years.
One of the most important aspects is that one can construct BPS operators either generalizing the gauge connection to include couplings solely to matter bosons (bosonic WLs) \cite{Gaiotto:2007qi,Berenstein:2008dc,Drukker:2008zx,Chen:2008bp,Rey:2008bh}   or including couplings both to bosonic and fermionic fields (fermionic WLs) \cite{Drukker:2009hy}.
While the construction of BPS WLs in $\mathcal N \geq 2$ SCSM quiver theories has been extensively investigated
\cite{Gaiotto:2007qi,Berenstein:2008dc,Drukker:2008zx,Chen:2008bp,Rey:2008bh,Drukker:2009hy,%
Ouyang:2015qma,Cooke:2015ila,Ouyang:2015iza,Ouyang:2015bmy}, most of the results on fermionic BPS WLs have been limited to operators with connections that can be written as $2\times2$ block diagonal matrices. When taking the trace, these operators correspond to linear combinations of WLs connecting adjacent nodes.
Recently, new BPS WLs in $\mathcal N = 4$ circular quiver SCSM theories with alternating levels have been constructed in \cite{Mauri:2017whf}, which are described by more general connections that cannot be decomposed as linear combinations of double--node connections.
This result suggests that the general form of BPS fermionic WLs may have a richer structure waiting to be explored.

The main goal of this paper is to investigate the most
general BPS WL in $\mathcal N \geq 2$ SCSM theories featured by parametric couplings to scalar and fermion matter\footnote{In this paper we consider only line and circle WLs with constant couplings to scalars, although more general WLs with contour dependent couplings have been also studied \cite{Cardinali:2012ru,Bianchi:2014laa,Bianchi:2018bke}.}.
For a generic quiver $\mathcal N = 2$ SCSM theory with (anti)bifundamental and/or (anti)fundamental matter fields, we write the most general expression for a WL containing arbitrary couplings to bosons and fermions and study under which conditions the operator preserves half of the supersymmetries.  It turns out that {with fixed preserved supercharges} there is only one bosonic $1/2$  BPS WL, while there is an infinite family of parametric fermionic $1/2$ BPS WLs whose connection is in general a non--block--diagonal matrix. In addition, in the $\mathcal N = 2$ case the connection does not have necessarily the structure of a superconnection of a given supergroup. This is the main novelty of our classification.

As a check of our construction we reproduce the already known operators for the Aharony--Bergman--Jafferis(--Maldacena) (ABJ(M)) and $\mN=4$ orbifold ABJM theories \cite{Ouyang:2015iza,Ouyang:2015bmy,Mauri:2017whf}. As a new result we provide the classification of fermionic 1/2 BPS WLs for the $\mN=2$ orbifold ABJM theory \cite{Benna:2008zy}. For the subset of operators that can be obtained by an orbifold quotient of the 1/2 BPS WL in ABJ(M) theory, we identify the corresponding gravity duals following the orbifold decomposition strategy in \cite{Mauri:2017whf}.

For the new infinite family of fermionic BPS WLs that we construct, it is mandatory to investigate how they behave at quantum level and understand how the parametric dependence enters their expectation values.  At classical level, generalizing what happens in ABJ(M) theory and $\mN=4$ SCSM theory \cite{Drukker:2009hy,Ouyang:2015qma,Cooke:2015ila,Ouyang:2015bmy}, we prove that

 all the fermionic WLs, independently of their couplings, are cohomologically equivalent to the bosonic one, i.e. $W_\fer= W_\bos + {\cal Q} ({\rm something})$ where ${\cal Q}$ is a supercharge preserved by all the operators. Whether and how this relation gets promoted at quantum level is a crucial question to be answered when comparing fermionic and bosonic expectation values. In fact, if the cohomological equivalence survives quantum corrections, it implies that the expectation values of all the fermionic BPS WLs are equal to the expectation value of the bosonic operator, and therefore they can all be computed by the same matrix model \cite{Kapustin:2009kz,Marino:2009jd,Drukker:2010nc}.
However, one important subtlety that we have to take into account when comparing expectation values is framing \cite{Kapustin:2009kz}. In fact, since the localization procedure always leads to framing--one results, we expect this to be the correct regularization scheme where the classical  cohomological equivalence translates into
$\langle W_\fer \rangle_1 = \langle W_\bos \rangle_1$. This problem has been already investigated at first few perturbative orders for particular kinds of WLs both in ABJ(M) and $\mN=4$ models \cite{Kapustin:2009kz,Marino:2009jd,Drukker:2010nc,Bianchi:2013zda,Bianchi:2013rma,Griguolo:2013sma,Griguolo:2015swa,
Bianchi:2016yzj,Bianchi:2016gpg,Bianchi:2016vvm}. In this paper we extend this anaysis to general $\mN=2$ SCSM theories. Up to two loops, {using the arguments and speculations in \cite{Bianchi:2016gpg} we find that the bosonic and fermionic BPS WLs have the same framing--one expectation values not only in ABJ(M) theory but also in a generic $\mN=2$ SCSM theory.}

This is already a strong indication that the cohomological equivalence might be valid at quantum level, at framing one, in any $\mN=2$ SCSM theory, although a truly non--trivial check would come at higher orders where the particular choice of the superpotential characterizing the model would enter.

At classical level the expansion of $W_\fer$ in powers of its bosonic and fermionic couplings leads to a stronger cohomological equivalence that translates into an infinite number of non--trivial ${\cal Q}$--identities \cite{Ouyang:2015qma}.
 For any $\mN \geq 2$ SCSM theory, up to two loops and at framing one these identities lead to non--trivial vanishing conditions for the correlators of the corresponding bosonic and fermionic operators (see eq.~\eqref{nontrivial}). We make the educated conjecture that these identities survive at higher orders and discuss the novel constraints that follow for three--loop, framing--one integrals, and the implications for the defect CFT defined on the Wilson contour.

As a by--product of our analysis, we provide the two--loop expression for the framing-zero and framing--one expectation values of all the fermionic BPS WLs. While the framing--one result is parameter independent, being equal to the bosonic expectation value, the framing--zero result exhibits a non--trivial dependence on the parameters that feature the couplings to matter fields.

The rest of the paper is organized as follows.
In section~\ref{secn2}, we provide the general classification of 1/2 BPS WLs in a generic $\mN=2$ SCSM theory, both on a line in Minkowski spacetime and on a circle in Euclidean space. We distinguish the cases of matter with canonical and non--canonical conformal dimensions.
In section \ref{secn2abjm} we apply the previous results to the construction of BPS WLs in $\mN=2$ orbifold ABJM theory and study the gravity duals of operators that can be obtained as orbifold quotients of 1/2 BPS WLs in ABJ(M) theory.
Section \ref{secquantum} is devoted to the perturbative calculation of the WL expectation values, the discussion of the cohomological equivalence, its stronger version and its non--trivial consequences.
Our conclusions are then collected in section \ref{SecCnD}.
In appendix~\ref{AppSpinor} we give spinor conventions in Minkowski  and Euclidean signatures.
In appendices~\ref{AppABJM} and \ref{AppN=4} we review the WLs in ABJ(M) theory and $\mN=4$ orbifold ABJ(M) theory, together with their gravity duals. As a check of our classification we also reproduce the known WLs by applying our general recipe.
Appendix~\ref{appcon} contains some details of section~\ref{secn2}.
{Finally, in appendix~\ref{appFeynman} we give the Lagrangian and Feynman rules of the general $\mN = 2$ SCSM theory that would be useful to the calculations of the integrals in appendix~\ref{appint}.}

\section{BPS WLs in $\cN =2$ CS-matter theories} \label{secn2}

For a generic $\cN =2$ SCSM quiver theory we construct the most general class of 1/2 BPS WLs featured by constant parametric couplings to matter fields.
In subsection \ref{SecLineWL},  we give full details of the classification for the case of line BPS WLs in Minkowski spacetime. We discuss cohomological equivalence between bosonic and fermionic operators and provide a toy--model example of a quiver theory to make manifest the novel features of WLs in $\cN =2$ SCSM models.
In subsection  \ref{SecCircleWL}, we introduce the circle BPS WLs in Euclidean spacetime signature. These operators are the relevant observables  in the context of localization and their non--trivial expectation value can be generically captured by  matrix model integrals.
In subsection  \ref{moreWL}, the classification is slightly generalized to the case of connections involving repeated nodes.
Finally, in subsection \ref{noncon} we briefly discuss the case of fields with non--canonical conformal dimensions.

\subsection{Line BPS WLs in Minkowski spacetime}\label{SecLineWL}

In three--dimensional Minkowski spacetime, we consider a generic quiver $\cN=2$ SCSM theory with gauge group $\prod_{a=1}^{n}U(N_a)_{k_a}$, where $k_a$ indicate the CS levels that can be non--vanishing or vanishing. The gauge sector of the theory is organized into $n$ $\cN=2$ vector multiplets, which in the Wess--Zumino gauge read (we refer to appendix~\ref{AppSpinor} for spinor conventions)
\be
\mV^{(a)} = 2\ii\bar\th\th \s^{(a)}
             + 2 \bar\th\g^\m\th A_\m^{(a)}
             + \sr{2} \ii\th^2\bar\th\bar\chi^{(a)}
             - \sr{2} \ii\bar\th^2 \th\chi^{(a)}
             + \th^2\bar\th^2D^{(a)}  \qquad a = 1 , \dots , n
\ee
whereas the matter sector is described by sets of $N_{ab}$ chiral multiplets in the bifundamental representation of arbitrary pairs of nodes $U(N_a)\times U(N_b)$
\be\label{matter}
\mZ_{(ab)}^i = Z_{(ab)}^i + \ii\bar\th\g^\m\th\p_\m Z_{(ab)}^i
                - \f14 \bar\th^2\th^2 \p^\m\p_\m  Z_{(ab)}^i
                + \sr2 \th\z_{(ab)}^i - \f{\ii}{\sr2}\th^2\bar\th\g^\m\p_\m\z_{(ab)}^i
                + \th^2 F_{(ab)}^i
\ee
with $i = 1, \dots , N_{ab}$.
In general, for a different $(ab)$ pair, the $i$ index in eq. \eqref{matter} varies in a different range. For $a=b$, $\mZ_{(aa)}^i$ describe $N_{aa}$ matter chiral multiplets in the adjoint representation of $U(N_a)$. We also allow for the presence of $N_{a0}$ matter multiplets in the fundamental representation of $U(N_a)$ and $N_{0a}$ multiplets in the antifundamental representation, denoted by $\mZ_{(a0)}^i$ and $\mZ_{(0a)}^i$ respectively.

Complex conjugated matter fields, belonging to the anti--bifundamental representation of $U(N_a)\times U(N_b)$, are defined as $[Z^{i}_{(ab)}]^\dagger = \bar Z_{i}^{(ba)}$, $[\z^{i}_{(ab)}]^\dagger = \bar\z_{i}^{(ba)}$, $[F^{i}_{(ab)}]^\dagger = \bar F_{i}^{(ba)}$, with $i=1,2,\cdots,N_{ab}$.

In superspace language the lagrangian of the theory is given by
\be
\mL = \mL_\CS + \mL_k + \mL_{sp}
\ee
where
\bea \label{lagrangian}
&& \mL_\CS = - \sum_a \f{k_a}{8\pi\ii} \int_0^1 dt \Tr\big( \bar D^\a \mV^{(a)} \ep^{t \mV^{(a)}}D_\a \ep^{-t\mV^{(a)}} \big) \big|_{\th^2\bar\th^2} \nn\\
&& \mL_k = - \sum_{a,b} \Tr\big[ \bar\mZ^{(ba)}_i \ep^{ - \mV^{(a)}} \mZ_{(ab)}^i \ep^{ \mV^{(b)}} \big]\big|_{\th^2\bar\th^2}
\eea
whereas $\mL_{sp}$ is the superpotential term that we do not write explicitly, as it is not relevant for the construction of BPS WLs and for the perturbative investigation at the order we work. Here we have defined $D_\a = \p_\a + \ii\bar\th^\b\g^\m_{\b\a}\p_\m$, $\bar D^\a = \bar\p^\a + \ii\g^\m{}^{\a\b}\th_\b\p_\m$.

For non--vanishing $k_a$ levels, writing \eqref{lagrangian} in components and extracting the equations of motion for the auxiliary fields of the vector multiplets we obtain  \footnote{Through the paper we use the convention that repeated flavor $i$ indices are summed, while summations on node indices $a,b,c\cdots$ are explicitly indicated. Repeated node indices with no explicit sum are meant to be fixed.}
\bea \label{equations}
&& \s^{(a)} = \f{2\pi}{k_a} \sum_b ( Z_{(ab)}^i \bar Z^{(ba)}_i - \bar Z^{(ab)}_i Z_{(ba)}^i ) \nn\\
&& \chi^{(a)} = - \f{4\pi}{k_a} \sum_b ( \z_{(ab)}^i \bar Z^{(ba)}_i - \bar Z^{(ab)}_i \z_{(ba)}^i )  \qquad  a = 1, \dots, n \nn\\
&& \bar \chi^{(a)} = - \f{4\pi}{k_a} \sum_b ( Z_{(ab)}^i \bar \z^{(ba)}_i - \bar \z^{(ab)}_i Z_{(ba)}^i )
\eea

General superconformal transformations of the component fields read
\bea \label{susy}
&& \d A^{(a)}_\m = \frac{1}{2} ( \bar\chi^{(a)}\g_\m\Th + \bar\Th\g_\m\chi^{(a)}), ~~ \qquad
   \d \s^{(a)} = - \frac{\ii}{2}(\bar\chi^{(a)}\Th + \bar\Th\chi^{(a)}) \nn\\
&& \d Z^{i}_{(ab)} = \ii\bar\Th\z^{i}_{(ab)}, ~~ \qquad \d\bar Z_{i}^{(ab)}=\ii\bar\z_{i}^{(ab)}\Th \nn\\
&& \d \z^{i}_{(ab)} = (-\g^\m D_\m Z^{i}_{(ab)} -\s^{(a)} Z^{i}_{(ab)}+ Z^{i}_{(ab)}\sigma^{(b)})\Th
                      - Z^{i}_{(ab)} \vth
                      + \ii F^{i}_{(ab)} \bar\Th  \nn\\
&& \d\bar \z_{i}^{(ab)} =  \bar\Th(\g^\m D_\m \bar Z_{i}^{(ab)} - \bar Z_{i}^{(ab)}\s^{(b)}+\s^{(a)}\bar Z_{i}^{(ab)})
                         - \bar\vth \bar Z_{i}^{(ab)}
                         - \ii \bar F_{i}^{(ab)}\Th
\eea
where $\Th \equiv \th+x^\m\g_\m\vth$, $\bar\Th \equiv \bar\th - \bar\vth x^\m\g_\m$ are linear combinations of $(\th, \bar{\th})$ spinors parametrizing Poincar\'e supersymmetry transformations, and $(\vth, \bar{\vth})$ ones parameterizing superconformal transformations \footnote{Here we only consider the case where matter fields in the chiral multiplets have canonical conformal dimensions. The more general case will be discussed in section~\ref{noncon}. The difference does not appear when we focus on Poincar\'e supercharges. }. The definition of covariant derivative can be found in \eqref{cov}.

\vskip 10pt
\noindent
{\bf  The 1/2 BPS WLs.} We construct WLs defined along the timelike infinite straight line $x^\m=(\t,0,0)$, which preserve half of the supersymmetries. Decomposing the spinorial charges as in \eqref{spinorspm}, without loss of generality we choose the preserved supercharges to be $Q_+, \bar{Q}_-, S_+, \bar{S}_-$, i.e. we require the operator to be invariant under
\bea \label{delta}
   \d   = \bar \th_- Q_+ +\bar Q_- \th_+ + \bar \vth_- S_+ +\bar S_- \vth_+
\eea
In the rest of the paper we will shortly identify the preserved supercharges with the corresponding $\th_+,  \bar \th_-, \vth_+, \bar \vth_-$ parameters.

The first kind of 1/2 BPS operator is the so--called bosonic WL defined as \cite{Gaiotto:2007qi}
\be\label{Wbos}
W_\bos = \mP \ep^{-\ii \int d\t L_\bos(\t)}
\ee
where the $n \times n$ connection matrix is given by
\be \label{Adef}
L_\bos = \diag( A_0^{(1)}-\s^{(1)}, A_0^{(2)}-\s^{(2)}, \cdots, A_0^{(n)}-\s^{(n)} )
\ee
It is easy to check that this operator is invariant under \eqref{delta}.

Using equations of motion \eqref{equations} for $\s^{(a)}$, $a=1,2,\cdots,n$, this generalized connection ends up including quadratic couplings to matter scalars.
However, as usually happens in three dimensions, we can look for more general operators with connections containing couplings also to fermions. We then consider the fermionic operator
\be\label{Wfer}
W_\fer = \mP \ep^{-\ii \int d\t L_\fer(\t)}
\ee
with a $n \times n$ connection   \cite{Drukker:2009hy}
\be \label{LABF}
L_\fer = L_\bos + B + F
\ee
where $L_\bos$ is given in \eqref{Adef},  whereas the $B$ and $F$ entries
\bea \label{Fdef}
&& B_{(ab)} = \sum_c ( R_{ab}^{c}{}_i{}^j Z_{(ac)}^i\bar Z^{(cb)}_j
                     + R_{ab}^{c}{}_{ij} Z_{(ac)}^i Z_{(cb)}^j \nn\\
&& \phantom{B_{(ab)} =}
                     + S_{ab}^{c}{}^i{}_j \bar Z^{(ac)}_i Z_{(cb)}^j )
                     + S_{ab}^{c}{}^{ij} \bar Z^{(ac)}_i \bar Z^{(cb)}_j ) \nn\\
&& F_{(ab)} = \bar m^{ab}_i \z_{(ab)+}^i + n^i_{ab} \bar\z^{(ab)}_{i-} \equiv [\bar M_\z]_{(ab)} + [N_{\bar\z}]_{(ab)}
\eea
contain couplings to bilinear scalars and linear fermions respectively, parametrized by to--be--determined matrices and vectors.

The requirement for $W_\fer$ to be 1/2 BPS can be traded with the search for a Grassmann odd matrix $G$ satisfying \cite{Drukker:2009hy,Lee:2010hk}
\be
\d L_\fer = \p_\t G + \ii [L_\fer, G]
\ee
Inserting decomposition \eqref{LABF} for $L_\fer$ this condition splits into a set of Grassmann even and odd constraints, respectively
\bea \label{dBdF}
&& \d B = \ii [F, G] \nn\\
&& \d F = \p_\t G + \ii [L_\bos + B , G]
\eea
From the Grassmann odd one we obtain\footnote{For the straight line it is sufficient to focus on super--Poincar\'e symmetries, since once these supercharges are preserved also the superconformal ones are automatically preserved.}
\be \label{GBG}
G = - \ii \bar M_Z \th_+ + \ii N_{\bar Z} \bar\th_- , \qquad ~~ [B , G] = 0
\ee
where we have defined
\be \label{MZNZ}
[\bar M_Z]_{(ab)} \equiv \bar m^{ab}_i Z_{(ab)}^i, ~~ \qquad [N_{\bar Z}]_{(ab)} \equiv n^i_{ab} \bar Z^{(ab)}_{i}
\ee
Using $G$ in \eqref{GBG} in the Grassmann even part of eq. \eqref{dBdF} we eventually obtain non--trivial relations among the coefficients
\bea \label{UVmmnn}
&& R_{ab}^{c}{}_i{}^j = \bar m_i^{ac} n_{cb}^j , ~~
   S_{ab}^{c}{}^i{}_j = n_{ac}^i \bar m_j^{cb} \nn\\
&& R_{ab}^{c}{}_{ij} = S_{ab}^{c}{}^{ij} =
   \bar m_i^{ac} \bar m_j^{cb} = n_{ac}^i n_{cb}^j =0
\eea
In particular, they imply the following relation
\be \label{Bdef}
B = \bar M_Z N_{\bar Z} + N_{\bar Z} \bar M_Z
\ee
It is easy to check that this expression automatically satisfies the second condition in \eqref{GBG}.

Solutions to constraints \eqref{UVmmnn} exhibit several interesting features:
\begin{itemize}
  \item Setting $a=b=c$ in the second line of \eqref{UVmmnn} we obtain $\bar m_i^{aa} = n_{aa}^i =0$ (no summation on $a = 1, \cdots , N_{aa}$). Therefore, although adjoint matter may appear in the theory, adjoint fermion fields $\z^i_{(aa)}$ or $\bar\z_i^{(aa)}$ can never appear in the connection. In other words, the diagonal blocks of the connection contain only bosonic couplings.
  \item For $a\neq b$ fixed, the $B_{(ab)} $ and $F_{(ab)} $ entries can be simultaneously non--vanishing. Therefore, in general the $L_\fer$ connection is not a superconnection, i.e. it does not give a representation of a supergroup. However, this does not contradict what has been already found for ABJM and $\cN=4$ orbifold ABJM theories where $L_\fer$ are indeed superconnections for the $U(N_1|N_2)$ supergroup. In fact, as we show below discussing a toy model, the conditions $B_{(ab)} \neq 0, F_{(ab)} \neq 0$ can occur simultaneously only when the quiver diagram contains triangles.
    \item Finally, constraints $\bar m_i^{ab} \bar m_j^{ba} = n_{ab}^i n_{ba}^j =0$ following from \eqref{UVmmnn} imply that if a positive chirality fermion appears in $F_{(ab)}$ ($\bar m_i^{ab} \neq 0$), then all fermion fields of positive chirality in $F_{(ba)}$ must be absent ($\bar m_i^{ba} =0$). Similarly, if fermion fields of both chiralities appear in $F_{(ab)}$, then $F_{(ba)}=0$.
\end{itemize}

To summarize, for a generic $\mN=2$ SCSM theory we have constructed bosonic and fermionic 1/2 BPS WLs \eqref{Wbos} and \eqref{Wfer} with connections
\bea\label{genconn}
&& L_\bos = \diag( A_0^{(1)} - \s^{(1)}, A_0^{(2)} - \s^{(2)}, \cdots, A_0^{(n)} - \s^{(n)} ) \nn \\
&& L_\fer = L_\bos + B + F, \qquad
   B = \bar M_Z N_{\bar Z} + N_{\bar Z} \bar M_Z, ~~
   F =\bar M_\z + N_{\bar\z} \nn\\
&& [\bar M_Z]_{(ab)} = \bar m^{ab}_i Z_{(ab)}^i, ~~
   [N_{\bar Z}]_{(ab)} = n^i_{ab} \bar Z^{(ab)}_{i} \nn\\
&& [\bar M_\z]_{(ab)} = \bar m^{ab}_i \z_{(ab)+}^i, ~~
   [N_{\bar\z}]_{(ab)} =  n^i_{ab} \bar\z^{(ab)}_{i-}
\eea
In particular, a generic fermionic connection has the following structure
\be \label{matrix}
L_\fer = \lt(\ba{cccc}
A_0^{(1)} - \s^{(1)} + B_{(11)} & B_{(12)} + F_{(12)}             & \cdots & B_{(1n)} + F_{(1n)}  \\
B_{(21)} + F_{(21)}             & A_0^{(2)} - \s^{(2)} + B_{(22)} & \cdots & B_{(2n)} + F_{(2n)}  \\
\vdots                          & \vdots                          & \ddots & \vdots \\
B_{(n1)} + F_{(n1)}             & B_{(n2)} + F_{(n2)}             & \cdots & A_0^{(n)} - \s^{(n)} + B_{(nn)} \\
\ea\rt)
\ee
with the caveat that some of the bosonic and fermionic couplings may be forced to be absent, as discussed above.
We note that the fundamental and antifundamental fields do not appear explicitly in the connection, but the connection may depend on them through $\s^{(a)}$.

Similarly, we can construct bosonic and fermionic 1/2 BPS WL $\td W_\bos$, $\td W_\fer$ preserving the  complementary set of supercharges $\theta_-,  \bar \theta_+,  \vartheta_-,  \bar \vartheta_+ $. The corresponding connections read
\bea
&& \td L_\bos = \diag( A_0^{(1)}+\s^{(1)}, A_0^{(2)}+\s^{(2)}, \cdots, A_0^{(n)}+\s^{(n)} ) \nn \\
&& \td L_\fer = \td L_\bos + \td B + \tilde F, \qquad
   \td B = - ( {\bar M}_Z  N_{\bar Z} +  N_{\bar Z} {\bar M}_Z ), ~~
   \tilde F =  {\bar M}_\z -  N_{\bar\z} \nn\\
&& [{\bar M}_Z]_{(ab)} = \bar m^{ab}_i Z_{(ab)}^i, ~~
   [ N_{\bar Z}]_{(ab)} = n^i_{ab} \bar Z^{(ab)}_{i} \nn\\
&& [{\bar M}_\z]_{(ab)} = \bar m^{ab}_i \z_{(ab)-}^i, ~~
   [{N}_{\bar\z}]_{(ab)} = n^i_{ab} \bar\z^{(ab)}_{i+}
\eea
where the constant parameters $\bar m_i^{ab}$, $n_{ab}^i$ satisfy the same constraints \eqref{UVmmnn}.

\vskip 10pt
\noindent
{\bf  Cohomological equivalence.} For ABJ(M) and $\mathcal{N}=4$ SCSM theories the full parametric family of fermionic BPS WLs has been shown to be (classically) cohomologically equivalent to the bosonic WL  \cite{Drukker:2009hy,Ouyang:2015qma,Cooke:2015ila,Ouyang:2015bmy,Mauri:2017whf}. This means that the difference between a given fermionic operator and the bosonic one can be written as $\mQ$(something), with $\mQ$ being a suitable linear combination of conserved supercharges shared by the two operators.

We now prove that this property holds also in the $\mathcal{N}=2$ setting: The general fermionic 1/2 BPS WL with connection \eqref{LABF} is classically   $\mathcal Q$--equivalent to the bosonic 1/2 BPS WL with connection \eqref{Adef}.

According to the analysis of \cite{Drukker:2009hy,Ouyang:2015qma}, this is the case if we manage to find $\k$, $\L$ and $\mathcal Q$ quantities that satisfy
\bea
&& \k \L^2 = B, ~~ \mathcal Q \L = F, ~~ \mathcal Q L_\bos =0 \nn\\
&& \mathcal Q F = \p_\t (\ii\k\L) + \ii [L_\bos, \ii\k\L]
\eea
Using (\ref{dBdF}), it is easy to see that in the present case a solution to the above equations is given by
\be
\k = 1, ~~ \L = \bar M_Z + N_{\bar Z}, ~~ \mathcal Q = Q_+ + \bar Q_-
\ee
This solution implies the classical identity
\be\label{eq:classicalcoho}
W_{\fer} - W_{\bos} = \mQ V
\ee
where $V$ is a known function of the gauge and matter fields of the theory.

\begin{figure}[htbp]
\centering
\includegraphics[width=0.3\textwidth]{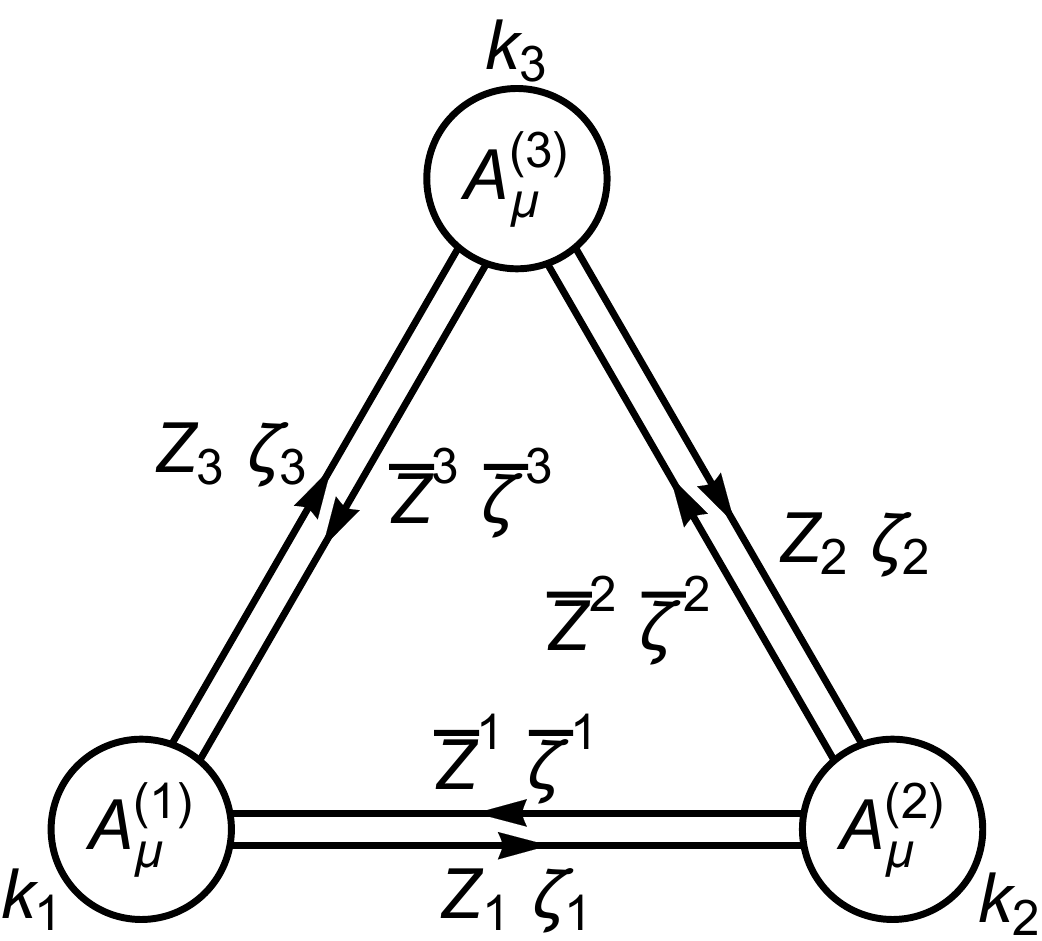}
\caption{The quiver diagram of a toy model $\mN=2$ SCSM theory.} \label{neq2toy}
\end{figure}

\vskip 10pt
\noindent{\bf  A triangular quiver toy model.}
Aimed at highlighting novel properties of the 1/2 BPS WLs that we have constructed, we consider the simple case of a $\mN=2$ SCSM theory associated to a triangle quiver diagram, as given in figure ~\ref{neq2toy}.

Specifying the general WL construction to this case, a particular solution to the BPS conditions in \eqref{UVmmnn} takes the form
\be
 L_\fer = \lt(\ba{ccc}
\mA^{(1)} ~& ~\bar m^1 \z_{1+} & 0 \\
0 & \mA^{(2)} ~& 0 ~\\
~n_3\bar\z^3_- ~&~ n_3\bar m^1\bar Z^3 Z_1 + \bar m^2 \z_{2+} ~& ~\mA^{(3)}
\ea\rt)
\ee
where
\bea
&& \mA^{(1)} = A_0^{(1)} + \f{2\pi}{k_1} ( -Z_1\bar Z^1 - Z_3 \bar Z^3 ) \nn\\
&& \mA^{(2)} = A_0^{(2)} + \f{2\pi}{k_2} ( \bar Z^1 Z_1 + \bar Z^2 Z_2 ) \nn\\
&& \mA^{(3)} = A_0^{(3)} + \f{2\pi}{k_3} ( -Z_2\bar Z^2 + \bar Z^3 Z_3 )
\eea
Notably, $L_\fer$ has a block entry which contains a sum of both bosonic and fermionic field combinations. As a consequence, the full connection ceases to be a supermatrix.

It is easy to see that a necessary condition for this feature to appear is the presence of a triangle in the quiver diagram. In fact, non--diagonal entries in \eqref{matrix} connecting adjacent nodes may contain both bilinear scalar terms $B_{(a\, a+1)}$ and linear fermion ones $F_{(a \,a+1)}$. While the fermionic entry corresponds to a fermionic arrow connecting the two nodes, the bosonic bilinear can be formed only passing by the third vertex of the triangle. Therefore, models with suitably chosen matter content allow for the existence of matrix entries in the WL connection that exhibit mixed Grassmann parity.

Operators with this structure have no counterpart  in the classification of  $\mathcal{N}=4$ and  $\mathcal{N}=6$ models,  where the underlying dynamics of the fermionic WLs seems to be  captured by a gauge supergroup (this is particularly manifest in the Higgsing derivation of the 1/2 BPS operators \cite{Lee:2010hk,Lietti:2017gtc}). It would be interesting to understand the implications of the existence of these WLs, especially in terms of a string  dual description at strong coupling.

\vskip 10pt
\noindent
{\bf  Lightlike WLs.}
We can also construct bosonic and fermionic BPS WLs along the lightlike infinite straight line $x^\mu=(\t,\t,0)$. The corresponding connections read
\bea
&&  L_\bos = \diag( A_0^{(1)}+A_1^{(1)}, A_0^{(2)}+A_1^{(2)}, \cdots, A_0^{(n)}+A_1^{(n)} ) \nn \\
&&  L_\fer =  L_\bos + B +  F, \qquad
    B =  \bar M_Z N_{\bar Z} + N_{\bar Z} \bar M_Z , ~~
    F =  {\bar M}_\z +  N_{\bar\z}  \nn\\
&& [\bar M_Z]_{(ab)} = \bar m^{ab}_i Z_{(ab)}^i, ~~
   [N_{\bar Z}]_{(ab)} = n^i_{ab} \bar Z^{(ab)}_{i} \nn\\
&& [{\bar M}_\z]_{(ab)} = \bar m^{ab}_i \z_{(ab)1}^i, ~~
   [{N}_{\bar\z}]_{(ab)} =  n^i_{ab} \bar\z^{(ab)}_{i1}
\eea
where the constant parameters $\bar m_i^{ab}$, $n_{ab}^i$ satisfy constraints \eqref{UVmmnn}.
Index $1$ in the last line indicates the first component of a spinor in the standard spinorial notation $\psi_\a$, $\a=1,2$. The lightlike bosonic WL is 3/4 BPS with preserved supercharges
\be
\th_2, ~~ \bar \th_2,  ~~ \vth_1, ~~ \bar \vth_1, ~~ \vth_2, ~~ \bar \vth_2
\ee
whereas the lightlike fermionic WL is 1/2 BPS with preserved supercharges
\be
\th_2, ~~ \bar \th_2, ~~ \vth_2, ~~ \bar \vth_2
\ee

\subsection{Circle BPS WLs in Euclidean space}\label{SecCircleWL}

The previous procedure can be easily generalized to construct 1/2 BPS WLs along the circle $x^\m=(\cos\t,\sin\t,0)$ in Euclidean space. The computational steps follow closely the ones of the Minkowskian  case, thus we report only the final result.

In a generic $\cN =2$ SCSM theory, the bosonic and fermionic WLs can be written as
\be \label{bosfereuc}
 W_\bos = \Tr\mP \ep^{-\ii\oint d\t L_\bos(\t)} , \qquad W_\fer = \Tr\mP \ep^{-\ii\oint d\t L_\fer(\t)}
\ee
with connections
\bea\label{bosferconeuc}
&& L_\bos = \diag(  A_\mu^{(1)}\dot x^\mu + \ii \s^{(1)},  A_\mu^{(2)}\dot x^\mu + \ii \s^{(2)}, \cdots,  A_\mu^{(n)}\dot x^\mu + \ii \s^{(n)} ) \nn\\
&& L_\fer = L_\bos + B + F, ~~
   B = - \ii( \bar M_Z N_{\bar Z} + N_{\bar Z} \bar M_Z ), ~~
   F = \bar M_\z - N_{\bar\z} \nn\\
&& [\bar M_Z]_{(ab)} = \bar m^{ab}_i Z_{(ab)}^i, ~~ [N_{\bar Z}]_{(ab)} = n^i_{ab} \bar Z^{(ab)}_{i} \nn\\
&& [\bar M_\z]_{(ab)} = \bar m^{ab}_i \z_{(ab)+}^i, ~~ [N_{\bar\z}]_{(ab)} =  n^i_{ab} \bar\z^{(ab)}_{i-}
\eea
Note that $\z_{(ab)+}^i = \ii u_+ \z_{(ab)}^i$, $\bar\z^{(ab)}_{i-}= \ii \bar\z^{(ab)}_{i} u_-$, with the spinorial couplings $u_\pm$ being defined in \eqref{g6}. They are non--trivial functions of the contour, whereas the scalar couplings are still contour independent.
The supercharges preserved by $W_\bos$ and $W_\fer$ are
\be \label{sc5}
\vth = -\ii \g_3 \th, ~~ \bar\vth = -\bar\th \ii \g_3
\ee
Similarly, we can construct 1/2 BPS bosonic and fermionic WLs $\td W_\bos$, $\td W_\fer$ preserving complementary supercharges, $\vth = \ii \g_3 \th, ~~ \bar\vth = \bar\th \ii \g_3$ and corresponding to
\bea
&& \td L_\bos = \diag( A_0^{(1)} - \ii \s^{(1)}, A_0^{(2)} - \ii \s^{(2)}, \cdots, A_0^{(n)} - \ii \s^{(n)} ) \nn\\
&& \td L_\fer = \td L_\bos + \td B + \td F, \qquad
   \td B = \ii ( \bar M_Z N_{\bar Z} + N_{\bar Z} \bar M_Z ), ~~
   \td F = \bar M_\z + N_{\bar\z} \nn\\
&& [\bar M_Z]_{(ab)} = \bar m^{ab}_i Z_{(ab)}^i, ~~ [N_{\bar Z}]_{(ab)} = n^i_{ab} \bar Z^{(ab)}_{i} \nn\\
&& [\bar M_\z]_{(ab)} = \bar m^{ab}_i\z_{(ab)-}^i, ~~ [N_{\bar\z}]_{(ab)} = n^i_{ab} \bar\z^{(ab)}_{i+}
\eea

In both cases it is not difficult to prove that the classical cohomological equivalence in \eqref{eq:classicalcoho} still holds.

\vskip 12pt

\subsection{More general 1/2 BPS WLs} \label{moreWL}

The previous class of 1/2 BPS WLs can be generalized to include extra operators constructed in the following way.

We arbitrarily select a subset of $n'$ nodes of the quiver diagram and label the corresponding
gauge fields as $A^{(s_{a'})},$ with $a'=1',2',\cdots,n'$ and $s_{a'} \in \{1,2,\cdots,n\}$. Note that $n'$ can be greater or smaller than $n$, and each node can be either not chosen or chosen more than once. For example, for $n=4$, we may choose $n'=3$ with $s_{a'}=1,1,3$, or $n'=5$ with $s_{a'}=1,1,3,4,4$.

Along the line $x^\m=(\t,0,0)$ in Minkowski spacetime we construct the bosonic 1/2 BPS WL with connection
\be
L_\bos = \diag( A_0^{(s_{1'})}-\s^{(s_{1'})}, A_0^{(s_{2'})}-\s^{(s_{2'})}, \cdots, A_0^{(s_{n'})}-\s^{(s_{n'})} )
\ee
It is easy to prove that this operator preserves the $\th_+, \bar\th_-, \vth_+, \bar\vth_-$ supercharges.

More generally, starting from the most general ansatz we can construct the fermionic 1/2 BPS WL with connection
\bea
&& L_\fer = L_\bos + B + F \nn\\
&& B = \bar M_Z N_{\bar Z} + N_{\bar Z} \bar M_Z, ~~
   F = \bar M_\z + N_{\bar \z} \nn\\
&& [\bar M_\z]_{(a'b')} = \bar m^{(a'b')}_i \z_{(s_{a'}s_{b'})+}^i, ~~
   [N_{\bar\z}]_{(a'b')} = n^i_{(a'b')} \bar\z^{(s_{a'}s_{b'})}_{i-} \nn\\
&& [\bar M_Z]_{(a'b')} = \bar m^{(a'b')}_i Z_{(s_{a'}s_{b'})}^i, ~~
   [N_{\bar Z}]_{(a'b')} = n^i_{(a'b')} \bar Z^{(s_{a'}s_{b'})}_{i}
\eea
Imposing the operator to preserve the $\th_+, \bar\th_-, \vth_+, \bar\vth_-$ supercharges, leads to non--trivial constraints for the constant parameters
\be
\sum_{c',s_{c'}=c} \bar m_i^{(a'c')} \bar m_j^{(c'b')} =
\sum_{c',s_{c'}=c} n^i_{(a'c')} n^j_{(c'b')} = 0
\ee
where we sum over $c'$, whereas indices $a',b',c,i,j$ are kept fixed. Again, the fundamental and antifundamental fields do not appear explicitly in the connection, but the connection may depend on them through $\s^{(s_{a'})}$.

\subsection{The case of matter fields with non-canonical dimensions }\label{noncon}

In SCSM theories with $\mN\geq 3$ supersymmetries, the matter fields always have canonical R--charges since the R--symmetry group $\rS\rO(\mN)$ is non--Abelian.
For theories with $\mN=2$ supersymmetry the situation is different, as the R--symmetry can mix with other U(1) flavor symmetries present in the theory. In this case the R--charges of matter fields have to be determined by F--maximization \cite{Jafferis:2010un} and generically they turn out to be non--canonical (we informally call these matter fields non--canonical).
In this section we briefly  discuss how to construct BPS WLs in this case.

We consider the UV theory on $\rm S^3$.
The metric on $\rm S^3$ of radius $r$ is
\begin{equation}
 ds^2=\Big( 1+\frac{|x|^2}{4 r^2} \Big)^{-2} \big[ (dx^1)^2+(dx^2)^2+(dx^3)^2 \big]
\end{equation}
where $x^\m=(x^1,x^2,x^3)$ are stereographic coordinates on $\rm S^3$ and $|x|^2=(x^1)^2+(x^2)^2+(x^3)^2$.
Given the vielbeins $E_\m^a = \big( 1+\frac{|x|^2}{4 r^2}\big)^{-1} e_\m^a$, $e^a = (dx^1,dx^2,dx^3)$, the gamma matrices are defined as
$\G_\m=E_\m^a\g_a$, $\g_\m=e_\m^a\g_a$, where $\g^\m$ are the three--dimensional gammas in eq. (\ref{gammaeuc}).

The Killing spinors $\Th$, $\bar\Th$ satisfy
\bea
&& D_\mu \Th=\G_\mu \tilde{\Th}, ~~ D_\mu\tilde{\Th}=-\frac{1}{4 r^2}\G_\mu \Th \nn\\
&& D_\mu \bar\Th= - \bar{\tilde{\Th}}\G_\mu, ~~ D_\mu\bar{\tilde{\Th}}=\frac{1}{4 r^2}\bar\Th\G_\mu
\eea
Note that $\Th$, $\bar\Th$ are independent, and are different from the ones in (\ref{susy}).
The solution can be found in \cite{Pestun:2007rz} and reads
\bea \label{ThbarTh}
&& \Th = \frac{1}{\sqrt{1+\frac{|x|^2}{4 r^2}}}(\th+{x^\m\gamma_\m} \vth), ~~
   \tilde{\Th} = \frac{1}{\sqrt{1+\frac{|x|^2}{4 r^2}}}(\vth-\frac{{x^\m\gamma_\m}}{4r^2} \th) \nn\\
&& \bar\Th = \frac{1}{\sqrt{1+\frac{|x|^2}{4 r^2}}}(\bar\th-\bar\vth{x^\m\gamma_\m}), ~~
   \bar{\tilde{\Th}} = \frac{1}{\sqrt{1+\frac{|x|^2}{4 r^2}}}(\bar\vth+\bar\th\frac{{x^\m\gamma_\m}}{4r^2})
\eea
When $r\to \infty$, $\Theta$, $\bar\Th$ go back to $\Theta$, $\bar\Th$ introduced in eq. \eqref{susy}, with $\td\Th$ going to $\vth$ and $\bar{\td\Th}$ going to $\bar\vth$. The superconformal transformations of $A^{(a)}_\mu, \sigma^{(a)}, Z^i_{(ab)}, \bar{Z}_i^{(ab)}$ can be obtained from \eqref{susy} by simply replacing $\g_\m$ with $\G_\m$ and $\Theta, \bar{\Theta}$ with (\ref{ThbarTh}), while the superconformal transformations of the fermions in the chiral multiplets are
\begin{eqnarray}\label{s3f}
\d \z^{i}_{(ab)} & =  & (-\G^\m D_\m Z^{i}_{(ab)} -\s^{(a)} Z^{i}_{(ab)}+ Z^{i}_{(ab)}\sigma^{(b)})\Th
                      - \frac{2\Delta_{(ab)}}{3} Z^{i}_{(ab)}\G^\mu D_\mu \Th
                      + \ii F^{i}_{(ab)} \bar\Th \nn\\
\d\bar \z_{i}^{(ab)} & = & \bar\Th(\G^\m D_\m \bar Z_{i}^{(ab)} - \bar Z_{i}^{(ab)}\s^{(b)}+\s^{(a)}\bar Z_{i}^{(ab)})
                         + \frac{2\Delta_{(ba)}}{3} \bar Z_{i}^{(ab)} D_\mu \bar \Th \G^\mu
                        - \ii \Th \bar F_{i}^{(ab)}
\end{eqnarray}
 where $\Delta_{(ab)}$ are the R--charges of the matter fields.

Usually, the action for the non--canonical matter is not superconformal in the UV. But one can construct actions which are invariant under supersymmetries generated by the left-invariant Killing spinors \cite{Jafferis:2010un,Hama:2010av}. In stereographic coordinates, the left-invariant Killing spinors take the form
\bea\label{left}
&& D_\mu \Th=  \frac{\ii}{2 r}\G_\mu \Th~~\Rightarrow~~\vth=\frac{\ii}{2r}\th \nn\\
&& D_\mu \bar\Th=  -\frac{\ii}{2 r}\bar\Th\G_\mu~~\Rightarrow~~\bar\vth=\frac{\ii}{2r}\bar\th
\eea
and transformations (\ref{s3f}) restricted to left-invariant Killing spinors read
\begin{align} \label{s3fbis}
   \d \z^{i}_{(ab)} =  & (-\G^\m D_\m Z^{i}_{(ab)} -\s^{(a)} Z^{i}_{(ab)}+ Z^{i}_{(ab)}\sigma^{(b)})\Th -\frac{\ii\Delta_{(ab)}}{r} Z^{i}_{(ab)} \Th
                      + \ii F^{i}_{(ab)} \bar\Th\nn\\
    \d\bar \z_{i}^{(ab)} =&\bar\Th(\G^\m D_\m \bar Z_{i}^{(ab)} - \bar Z_{i}^{(ab)}\s^{(b)}+\s^{(a)}\bar Z_{i}^{(ab)}){-\frac{\ii\Delta_{(ba)}}{r}  \bar \Th \bar Z_{i}^{(ab)}}
                         - \ii \Th \bar F_{i}^{(ab)}
\end{align}
In many cases the theory flows to an IR fixed point with superconformal symmetry. At this point the R--charges $\Delta_{(ab)}$ are determined by F--maximization \cite{Jafferis:2010un}, modulo some flat directions associated to the transformations of the R--charges which leave the partition function in the matrix model invariant \cite{Jafferis:2011zi}. This is related to the invariance of the transformartion rules in \eqref{s3fbis} under the shift
\begin{equation} \label{shift}
\Delta_{(ab)}\rightarrow\Delta_{(ab)}+\delta^{(a)}-\delta^{(b)},~~
\sigma^{(a)}\rightarrow\sigma^{(a)}-\ii \frac{\delta^{(a)}}{r}
\end{equation}

Along a general contour $x^\m(\t)$ on $\rS^3$, we construct the bosonic WL with connection
\be
L_\bos = \diag( A_\mu^{(1)}\dot x^\mu + \ii |\dot x| \s^{(1)}, A_\mu^{(2)}\dot x^\mu + \ii|\dot x| \s^{(2)}, \cdots, A_\mu^{(n)}\dot x^\mu +\ii|\dot x| \s^{(n)} )
\ee
Using the shift transformation (\ref{shift}), we can also construct fermionic WLs with non--canonical matter satisfying the condition
\begin{equation}
\Delta_{(ab)}=1/2+\delta^{(a)}-\delta^{(b)}
\end{equation}
The fermionic WL has connection
\bea \label{WLgeneral}
&& L_\fer = L_\bos + B + F +C \nn\\
&& C =  -\diag( \delta^{(1)}, \delta^{(2)}, \cdots,\delta^{(n)} )|\dot x| /r \nn\\
&& B = - \ii|\dot x| (\bar M_Z N_{\bar Z} + N_{\bar Z} \bar M_Z), ~~
   F = |\dot x| (\bar M_\z + N_{\bar\z}) \nn\\
&& [\bar M_Z]_{(ab)} = \bar m^{ab}_i  Z_{(ab)}^i, ~~ [N_{\bar Z}]_{(ab)} =   n^i_{ab}  \bar Z^{(ab)}_{i} \nn\\
&& [\bar M_\z]_{(ab)} = \bar m^{ab}_i u_+ \z_{(ab)}^i, ~~ [N_{\bar\z}]_{(ab)} =  n^i_{ab}  \bar\z^{(ab)}_{i}u_-  \nn\\
&& \bar m_i^{ac} \bar m_j^{cb} = n_{ac}^i n_{cb}^j =0
\eea
where on the $x^\m$ contour the $u_\pm(\t)$ spinors satisfy
\be \label{hahaha}
\G_\m \dot x^\m u_\pm = \pm |\dot x| u_\pm, ~~ u_+ u_- = -u_-u_+ = -\ii, ~~ u_+u_+ = u_-u_- = u_+D_\t u_- = u_-D_\t u_+ = 0
\ee
The bosonic and fermionic WLs are locally BPS with preserved supercharges
\be
\G_\m \dot x^\m \Th = - |\dot x| \Th, ~~ \bar\Th \G_\m \dot x^\m = - |\dot x| \bar\Th
\ee
To make the WLs globally BPS, we need to choose the contour to be a great circle. Here we discuss two special cases.

In the first case, we choose the great circle $x^\m=2r(\cos\t,\sin\t,0)$ in $\rm S^3$. Using (\ref{WLgeneral}) with $|\dot x|=r$, we obtain the connections of bosonic and fermionc WLs
\bea
&& L_\bos = \diag( A_\mu^{(1)}\dot x^\mu + \ii r \s^{(1)}, A_\mu^{(2)}\dot x^\mu + \ii r\s^{(2)}, \cdots, A_\mu^{(n)}\dot x^\mu +\ii r\s^{(n)} ) \nn\\
&& L_\fer = L_\bos + B + F +C \nn\\
&& C =  -\diag( \delta^{(1)}, \delta^{(2)}, \cdots,\delta^{(n)} ) \nn\\
&& B = - \ii r(\bar M_Z N_{\bar Z} + N_{\bar Z} \bar M_Z), ~~
   F = r (\bar M_\z - N_{\bar\z}) \nn\\
&& [\bar M_Z]_{(ab)} = \bar m^{ab}_i  Z_{(ab)}^i, ~~ [N_{\bar Z}]_{(ab)} =   n^i_{ab}  \bar Z^{(ab)}_{i} \nn\\
&& [\bar M_\z]_{(ab)} = \bar m^{ab}_i \z_{(ab)+}^i, ~~ [N_{\bar\z}]_{(ab)} =  n^i_{ab}  \bar\z^{(ab)}_{i-} \nn\\
&& \bar m_i^{ac} \bar m_j^{cb} = n_{ac}^i n_{cb}^j =0
\eea
with $u_\pm$ being the spinors defined in (\ref{g6}). The preserved supercharges are
\be\label{susyg}
\vth = -\frac{\ii}{2r} \g_3 \th, ~~ \bar\vth = - \frac{\ii}{2r}\bar\th\g_3
\ee
Using (\ref{left}) we can write preserved supersymmetries as $\th_2, \bar\th_1$.

As a second case, we consider the great circle obtained by the stereographic projection of the straight line $x^\m=(0,0,\tau)$ from $\rm R^3$ to $\rm S^3$, along which we define
\bea
&& u_{+\a} = \left( \ba{cc} 1 \\ 0 \ea \right), ~~
   u_{-\a} = \left( \ba{cc} 0 \\ \ii \ea \right) \nn\\
&& u_{+}^{\a} = \left( 0 , -1  \right), ~~
   u_{-}^{\a} = \left( \ii , 0  \right)
\eea
Here $|\dot x| = (1+\f{\t^2}{4r^2})^{-1}$, and (\ref{hahaha}) are satisfied. The connections of bosonic and fermionic WLs take the form
\bea
&& L_\bos = \diag( A_\mu^{(1)}\dot x^\mu + \ii |\dot x| \s^{(1)}, A_\mu^{(2)}\dot x^\mu + \ii|\dot x| \s^{(2)}, \cdots, A_\mu^{(n)}\dot x^\mu +\ii|\dot x| \s^{(n)} )\nn\\
&& L_\fer = L_\bos + B + F +C \nn\\
&& C =  -\diag( \delta^{(1)}, \delta^{(2)}, \cdots,\delta^{(n)} )|\dot x| /r \nn\\
&& B = - |\dot x| (\bar M_Z N_{\bar Z} + N_{\bar Z} \bar M_Z), ~~
   F = |\dot x| (\bar M_\z + N_{\bar\z}) \nn\\
&& [\bar M_Z]_{(ab)} = \bar m^{ab}_i  Z_{(ab)}^i, ~~ [N_{\bar Z}]_{(ab)} =   n^i_{ab}  \bar Z^{(ab)}_{i} \nn\\
&& [\bar M_\z]_{(ab)} = \bar m^{ab}_i \z_{(ab)2}^i, ~~ [N_{\bar\z}]_{(ab)} =  n^i_{ab}  \bar\z^{(ab)}_{i1} \nn\\
&& \bar m_i^{ac} \bar m_j^{cb} = n_{ac}^i n_{cb}^j =0
\eea
In the limit $r\rightarrow\inf$, the corresponding WL is along an infinite straight line and the preserved supersymmetries are half of the Poincar\'e supersymmetries, i.e. $\th_2$ and $\bar \th_1$.  Instead, conformal supersymmetries can be preserved only if there is no non--canonical matter in the theory.

\section{$\cN = 2$ orbifold ABJM theory} \label{secn2abjm}

In this section we give full details for an explicit example of $\cN = 2$ SCSM theory that admits a gravity dual description,
namely the case  of the $\cN = 2$ orbifold of ABJM theory. In this case, exploiting our knowledge of the gravity duals of the 1/2 BPS WLs in ABJM theory \cite{Drukker:2009hy,Lietti:2017gtc}, after introducing the classification we identify the gravity duals of some 1/2 BPS and non--BPS WLs in the $\mN=2$ orbifolded version.

In the rest of this section we heavily refer to appendix~\ref{AppABJM} where the construction of 1/2 BPS WLs and their gravity duals is reviewed for the ABJ(M) theory.

\subsection{1/2 BPS WLs}

The $\cN = 2$ orbifold ABJM theory can be obtained starting from $U(rN)_k\times U(rN)_{-k}$  ABJM model and performing the $\rZ_r$
quotient as in \cite{Benna:2008zy} \footnote{For ABJM two different orbifold projections can be applied, which lead to $\cN = 2$ and $\cN = 4$ quotients. Here we focus on the $\cN = 2$ orbifold, whereas the $\cN = 4$ case is reviewed in appendix~\ref{AppN=4}.}. In the usual notations of  ABJM theory, which we summarize in appendix \ref{AppABJM}, the fields are
decomposed as
\bea\label{orbifold}
&& A_\m = \diag(A_\m^{(1)},A_\m^{(2)},\cdots,A_\m^{(r)}),~~
   B_\m = \diag(B_\m^{(1)},B_\m^{(2)},\cdots,B_\m^{(r)})  \nn\\
&& \phi_{1,2} = \diag(\phi_{1,2}^{(1)},\phi_{1,2}^{(2)},\cdots,\phi_{1,2}^{(r)}) , ~~
   \psi^{1,2} = \diag(\psi^{1,2}_{(1)},\psi^{1,2}_{(2)},\cdots,\psi^{1,2}_{(r)}) \nn\\
&& \phi_3 = \lt( \ba{ccccc}
              0&&&&\phi_3^{(r)} \\
              \phi_3^{(1)}&0&&& \\
              &\phi_3^{(2)}&\ddots&& \\
              &&\ddots&0& \\
              &&&\phi_3^{(r-1)}&0 \ea \rt) , ~~
   \phi_4 = \lt( \ba{ccccc}
              0&\phi_4^{(1)}&&& \\
              &0&\phi_4^{(2)}&& \\
              &&\ddots&\ddots& \\
              &&&0&\phi_4^{(r-1)} \\
              \phi_4^{(r)}&&&&0 \ea \rt) \nn\\
&& \psi^3 = \lt( \ba{ccccc}
              0&\psi^3_{(1)}&&& \\
              &0&\psi^3_{(2)}&& \\
              &&\ddots&\ddots& \\
              &&&0&\psi^3_{(r-1)} \\
              \psi^3_{(r)}&&&&0 \ea \rt) , ~~
   \psi^4 = \lt( \ba{ccccc}
              0&&&&\psi^4_{(r)} \\
              \psi^4_{(1)}&0&&& \\
              &\psi^4_{(2)}&\ddots&& \\
              &&\ddots&0& \\
              &&&\psi^4_{(r-1)}&0 \ea \rt)
\eea
where $A_\mu$ and $B_\mu$ are the gauge connections associated to $U(rN)_k$ and $U(rN)_{-k}$ respectively, whereas in $SU(4)$
R--symmetry notations $\phi_{I= 1, \dots ,4}$ and the corresponding $\bar{\psi}_{I}$ fermions build up the matter multiplets in
the bifundamental representation of the gauge group.

\begin{figure}[htbp]
\centering
\subfigure[]{\includegraphics[width=0.6\textwidth]{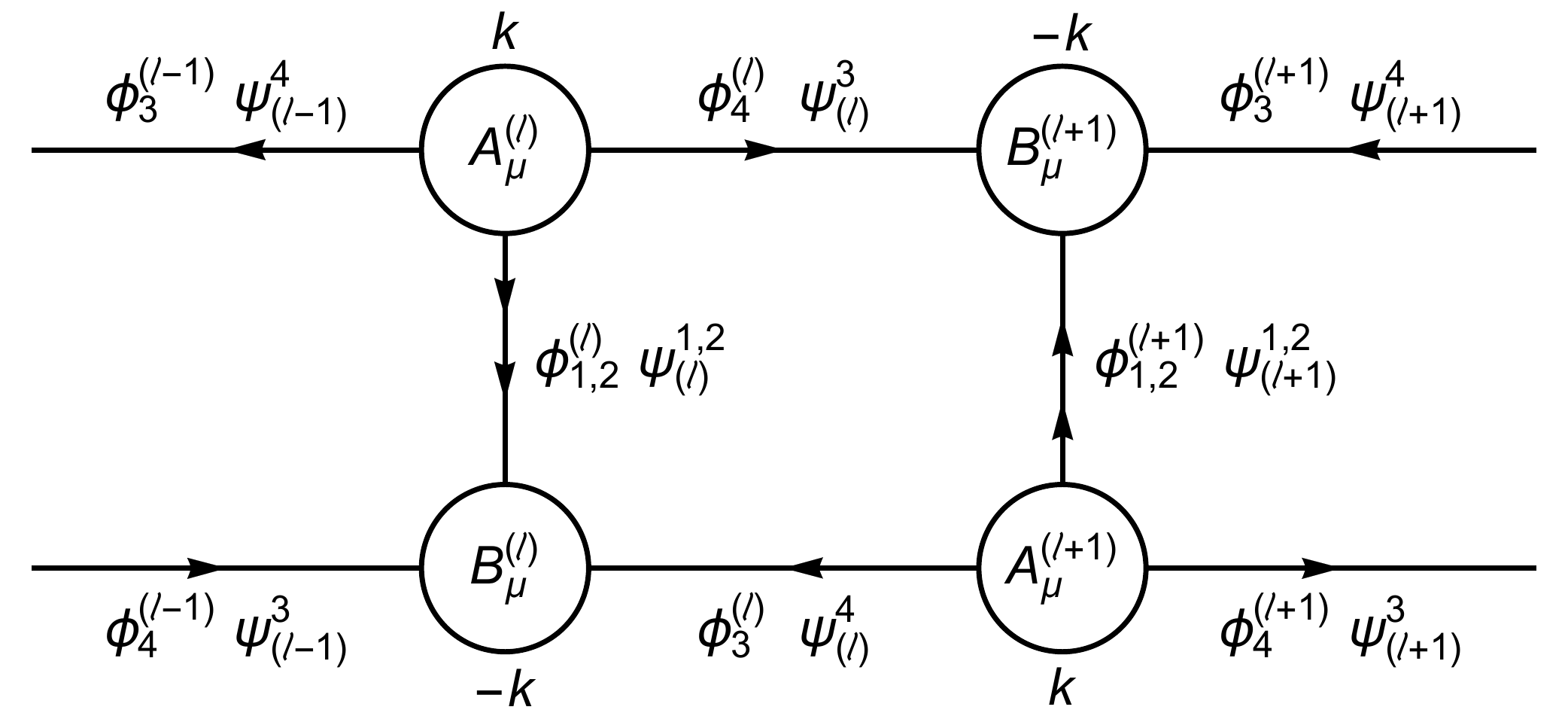} \label{n2abjm1}}
\subfigure[]{\includegraphics[width=0.6\textwidth]{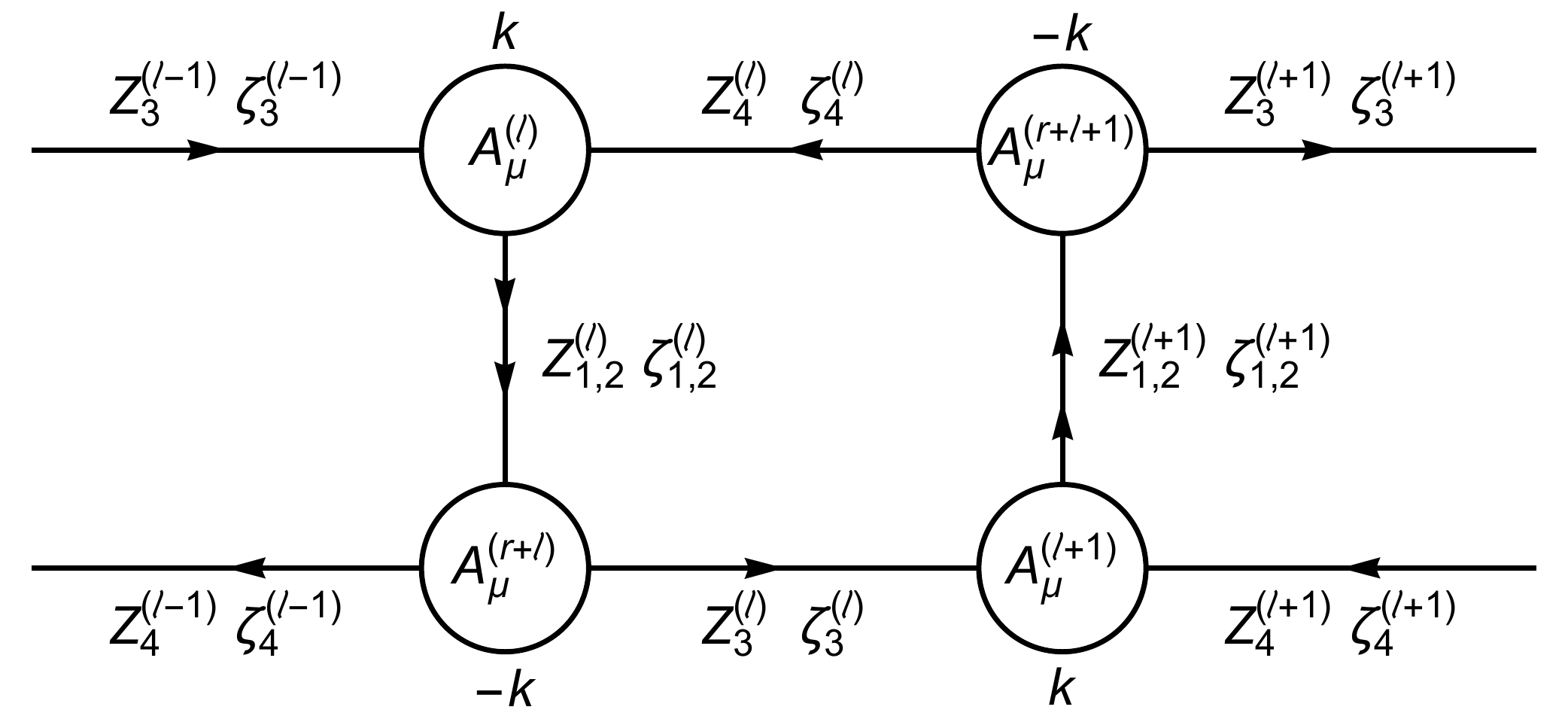} \label{n2abjm2}}
\caption{The quiver diagram of $\cN = 2$ orbifold ABJM theory in (a) ABJM notation and (b) $\cN=2$ notation. We have omitted the complex conjugates of the matter fields.} \label{abjm}
\end{figure}

Orbifold decomposition \eqref{orbifold} corresponds to choosing the unbroken supercharges as $\th^{12}=\bar\th_{34}=\th$,
$\th^{34}=\bar\th_{12}=\bar\th$, $\vth^{12}=\bar\vth_{34}=\vth$, $\vth^{34}=\bar\vth_{12}=\bar\vth$. The resulting $\cN=2$ theory
is described by the quiver diagram in figure~\ref{n2abjm1}.

In order to make contact with the classification of the previous section we have to temporarily use the alternative notation defined in the quiver
diagram of figure~\ref{n2abjm2}. In equations, the two sets of conventions are related as
\bea\label{fieldchange}
&& A_\m^{(\ell)} = A_\m^{(\ell)}, ~~
   B_\m^{(\ell)}  = A_\m^{(r+\ell)}  \nn\\
&& \phi_I^{(\ell)} = (Z_1^{(\ell)},Z_2^{(\ell)},\bar Z^3_{(\ell)},\bar Z^4_{(\ell)}) \nn\\
&& \psi^I_{(\ell)} = (-\z_2^{(\ell)},\z_1^{(\ell)},-\bar \z^4_{(\ell)},\bar \z^3_{(\ell)})
\eea
A similar change of notations for the ABJ(M) case is detailed in appendix \ref{AppABJM}.  We will refer to the notations of figure~\ref{n2abjm1} and~\ref{n2abjm2} as to  the  ABJM and  $\mN=2$ notation, respectively.

The use of $\mN=2$ notation allows to easily exploit the results of section~\ref{secn2} for writing down the connection of 1/2 BPS
fermionic WLs. They are given by the general recipe \eqref{genconn} with
\bea
&& \s^{(\ell)} = \f{2\pi}{k} ( Z_1^{(\ell)} \bar Z^1_{(\ell)}
                  + Z_2^{(\ell)} \bar Z^2_{(\ell)}
                  - \bar Z^3_{(\ell-1)} Z_3^{(\ell-1)}
                  - \bar Z^4_{(\ell)} Z_4^{(\ell)} ) \nn \\
&& \s^{(r+\ell)} = \f{2\pi}{k} ( \bar Z^1_{(\ell)} Z_1^{(\ell)}
                + \bar Z^2_{(\ell)} Z_2^{(\ell)}
                - Z_3^{(\ell)} \bar Z^3_{(\ell)}
                - Z_4^{(\ell-1)} \bar Z^4_{(\ell-1)} )
\eea
and non--vanishing blocks of $\bar M_Z$, $N_{\bar Z}$, $\bar M_\z$, $N_{\bar \z}$ matrices given by
\bea
&& \hspace{-9mm}
   [\bar M_Z]_{(\ell,r+\ell)} = \bar m^1_{(\ell)}Z_1^{(\ell)} + \bar m^2_{(\ell)}Z_2^{(\ell)}, ~~
   [\bar M_Z]_{(r+\ell,\ell+1)} = \bar m^3_{(\ell)}Z_3^{(\ell)}, ~~
   [\bar M_Z]_{(r+\ell+1,\ell)} = \bar m^4_{(\ell)}Z_4^{(\ell)} \nn\\
&& \hspace{-9mm}
   [N_{\bar Z}]_{(\ell,r+\ell+1)} = n_4^{(\ell)}\bar Z^4_{(\ell)}, ~~
   [N_{\bar Z}]_{(\ell+1,r+\ell)} = n_3^{(\ell)}\bar Z^3_{(\ell)}, ~~
   [N_{\bar Z}]_{(r+\ell,\ell)} = n_1^{(\ell)}\bar Z^1_{(\ell)} + n_2^{(\ell)}\bar Z^2_{(\ell)} \nn\\
&& \hspace{-9mm}
   [\bar M_\z]_{(\ell,r+\ell)} = \bar m^1_{(\ell)}\z_{1+}^{(\ell)} + \bar m^2_{(\ell)}\z_{2+}^{(\ell)}, ~~
   [\bar M_\z]_{(r+\ell,\ell+1)} = \bar m^3_{(\ell)}\z_{3+}^{(\ell)}, ~~
   [\bar M_\z]_{(r+\ell+1,\ell)} = \bar m^4_{(\ell)}\z_{4+}^{(\ell)} \nn\\
&& \hspace{-9mm}
   [N_{\bar \z}]_{(\ell,r+\ell+1)} = n_4^{(\ell)}\bar \z^4_{(\ell)-}, ~~
   [N_{\bar \z}]_{(\ell+1,r+\ell)} = n_3^{(\ell)}\bar \z^3_{(\ell)-}, ~~
   [N_{\bar \z}]_{(r+\ell,\ell)} = n_1^{(\ell)}\bar \z^1_{(\ell)-} + n_2^{(\ell)}\bar \z^2_{(\ell)-}
\eea
The parameters are subject to the following constraints
\be\label{constraints}
\bar m^{1,2}_{(\ell)} \bar m^{3,4}_{(\ell-1)} = \bar m^{1,2}_{(\ell)} \bar m^{3,4}_{(\ell)} =
n_{1,2}^{(\ell)}n_{3,4}^{(\ell-1)} = n_{1,2}^{(\ell)}n_{3,4}^{(\ell)} = 0
\ee

Having in mind to find out the gravity duals of these WLs we now translate the connections back to ABJM notation. Mimicking what is done in appendix \ref{AppABJM} for the ABJ(M) case (see eq. \eqref{pt1}), we first redefine the parameters as
\bea
&& \bar m^1_{(\ell)} = \sr{\f{4\pi}{k}}\bar\a_2^{(\ell)}, ~~
   \bar m^2_{(\ell)} = - \sr{\f{4\pi}{k}}\bar\a_1^{(\ell)}, ~~
   \bar m^3_{(\ell)} = - \sr{\f{4\pi}{k}} \d^4_{(\ell)}, ~~
   \bar m^4_{(\ell)} = \sr{\f{4\pi}{k}} \d^3_{(\ell)}  \nn\\
&& n_1^{(\ell)} = \sr{\f{4\pi}{k}} \b^2_{(\ell)}, ~~
   n_2^{(\ell)} = -\sr{\f{4\pi}{k}} \b^1_{(\ell)}, ~~
   n_3^{(\ell)} = \sr{\f{4\pi}{k}} \bar\g_4^{(\ell)}, ~~
   n_4^{(\ell)} = -\sr{\f{4\pi}{k}} \bar\g_3^{(\ell)}
\eea
Constraints \eqref{constraints} now read
\be
\bar\a_{1,2}^{(\ell)}\d^{3,4}_{(\ell-1)} =
\bar\a_{1,2}^{(\ell)}\d^{3,4}_{(\ell)} =
\bar\g_{3,4}^{(\ell)}\b^{1,2}_{(\ell)} =
\bar\g_{3,4}^{(\ell)}\b^{1,2}_{(\ell+1)} =0
\ee
Then, expressing the 1/2 BPS operator $W_\fer$ in terms of $\cN=2$ orbifold ABJM fields \eqref{fieldchange},  we can rewrite the matrix
connection as
\be \label{Lorb}
L_\fer = \lt(\begin{array}{cc}
      \cA & f_1 \\
      f_2 & \cB
    \end{array}\rt)
\ee
where for $r \geq 5$ the explicit expressions of the matrix blocks read
\bea
&& \cA = \lt(\begin{array}{cccccc}
           \cA^{(1)}  & 0         & h_1^{(1)}   &             &  h_3^{(r-1)} & 0 \\
           0          & \cA^{(2)} & 0           & \ddots      &              & h_3^{(r)}  \\
           h_3^{(1)}  & 0         & \cA^{(3)}   & \ddots      & h_1^{(r-3)}  &  \\
                      & \ddots    & \ddots      & \ddots      & 0            & h_1^{(r-2)}  \\
          h_1^{(r-1)} &           & h_3^{(r-3)} & 0           & \cA^{(r-1)}  & 0 \\
          0           & h_1^{(r)} &             & h_3^{(r-2)} & 0            & \cA^{(r)}
         \end{array}\rt) \nn\\
&& \cB = \lt(\begin{array}{cccccc}
           \cB^{(1)}  & 0         & h_4^{(1)}   &             &  h_2^{(r-1)} & 0 \\
           0          & \cB^{(2)} & 0           & \ddots      &              & h_2^{(r)}  \\
           h_2^{(1)}  & 0         & \cB^{(3)}   & \ddots      & h_4^{(r-3)}  &  \\
                      & \ddots    & \ddots      & \ddots      & 0            & h_4^{(r-2)}  \\
          h_4^{(r-1)} &           & h_2^{(r-3)} & 0           & \cB^{(r-1)}  & 0 \\
          0           & h_4^{(r)} &             & h_2^{(r-2)} & 0            & \cB^{(r)}
         \end{array}\rt)\\
&& f_1 = \lt(\begin{array}{ccccc}
           f_1^{(1)} & f_3^{(1)} &             & f_5^{(r)} \\
           f_5^{(1)} & f_1^{(2)} & \ddots      & \\
                     & \ddots    & \ddots      & f_3^{(r-1)} \\
           f_3^{(r)} &           & f_5^{(r-1)} & f_1^{(r)}
         \end{array}\rt) , ~~
  f_2 = \lt(\begin{array}{ccccc}
           f_2^{(1)} & f_6^{(1)} &             & f_4^{(r)} \\
           f_4^{(1)} & f_2^{(2)} & \ddots      & \\
                     & \ddots    & \ddots      & f_6^{(r-1)} \\
           f_6^{(r)} &           & f_4^{(r-1)} & f_2^{(r)}
         \end{array}\rt)\nn
\eea
with definitions
\bea \label{n2def}
&& \cA^{(\ell)} = A_0^{(\ell)} + \f{2\pi}{k} \big[  (-1+2\b^2_{(\ell)}\bar\a_2^{(\ell)})\phi_1^{(\ell)}\bar\phi^1_{(\ell)}
                                               + (-1+2\b^1_{(\ell)}\bar\a_1^{(\ell)})\phi_2^{(\ell)}\bar\phi^2_{(\ell)} \nn\\
&& \phantom{\cA^{(\ell)} =}
                                               - 2\b^1_{(\ell)}\bar\a_2^{(\ell)} \phi_1^{(\ell)}\bar\phi^2_{(\ell)}
                                               - 2\b^2_{(\ell)}\bar\a_1^{(\ell)} \phi_2^{(\ell)}\bar\phi^1_{(\ell)} \nn\\
&& \phantom{\cA^{(\ell)} =}
                                               + (1-2\d^4_{(\ell-1)}\bar\g_4^{(\ell-1)})\phi_3^{(\ell-1)}\bar\phi^3_{(\ell-1)}
                                               + (1-2\d^3_{(\ell)}\bar\g_3^{(\ell)})\phi_4^{(\ell)}\bar\phi^4_{(\ell)} \big] \nn\\
&& \cB^{(\ell)} = B_0^{(\ell)} + \f{2\pi}{k} \big[  (-1 + 2\b^2_{(\ell)}\bar\a_2^{(\ell)})\bar\phi^1_{(\ell)}\phi_1^{(\ell)}
                                             + (-1 + 2\b^1_{(\ell)}\bar\a_1^{(\ell)})\bar\phi^2_{(\ell)}\phi_2^{(\ell)} \nn\\
&& \phantom{\cB^{(\ell)} =}
                                             - 2\b^1_{(\ell)}\bar\a_2^{(\ell)}\bar\phi^2_{(\ell)}\phi_1^{(\ell)}
                                             - 2\b^2_{(\ell)}\bar\a_1^{(\ell)}\bar\phi^1_{(\ell)}\phi_2^{(\ell)} \nn\\
&& \phantom{\cB^{(\ell)} =}
                                             + (1 - 2\d^4_{(\ell)}\bar\g_4^{(\ell)})\bar\phi^3_{(\ell)}\phi_3^{(\ell)}
                                             + (1 - 2\d^3_{(\ell-1)}\bar\g_3^{(\ell-1)})\bar\phi^4_{(\ell-1)}\phi_4^{(\ell-1)}
                                             \big] \nn\\
&& h_1^{(\ell)} = \f{4\pi}{k} \d^4_{(\ell+1)}\bar\g_3^{(\ell)} \phi_4^{(\ell)} \bar\phi^3_{(\ell+1)}, ~~
   h_3^{(\ell)} = \f{4\pi}{k} \d^3_{(\ell)}\bar\g_4^{(\ell+1)} \phi_3^{(\ell+1)} \bar\phi^4_{(\ell)}  \nn\\
&& h_2^{(\ell)} = \f{4\pi}{k} \d^3_{(\ell+1)}\bar\g_4^{(\ell)} \bar\phi^4_{(\ell+1)} \phi_3^{(\ell)}, ~~
   h_4^{(\ell)} = \f{4\pi}{k} \d^4_{(\ell)}\bar\g_3^{(\ell+1)} \bar\phi^3_{(\ell)} \phi_4^{(\ell+1)} \\
&& f_1^{(\ell)} = \sr{\f{4\pi}{k}} ( \bar\a_1^{(\ell)} \psi^1_{(\ell)+}
                                   + \bar\a_2^{(\ell)} \psi^2_{(\ell)+} ), ~~
   f_3^{(\ell)} = \sr{\f{4\pi}{k}} \bar\g_3^{(\ell)} \psi^3_{(\ell)-}, ~~
   f_5^{(\ell)} = \sr{\f{4\pi}{k}} \bar\g_4^{(\ell)} \psi^4_{(\ell)-} \nn\\
&& f_2^{(\ell)} = \sr{\f{4\pi}{k}} ( \bar\psi^{(\ell)}_{1-}\b^1_{(\ell)}
                                   + \bar\psi^{(\ell)}_{2-}\b^2_{(\ell)} ), ~~
   f_4^{(\ell)} = -\sr{\f{4\pi}{k}}  \bar\psi^{(\ell)}_{3+}\d^3_{(\ell)}, ~~
   f_6^{(\ell)} = -\sr{\f{4\pi}{k}}  \bar\psi^{(\ell)}_{4+}\d^4_{(\ell)} \nn
\eea
In the case of shorter orbifold quivers ($r=2,3,4$), the connections get slightly modified by boundary effects. We report their
explicit expressions in appendix \ref{appcon}.

We note that in this case, compared with the general $\cN=2$  quiver models,  the full connection \eqref{Lorb} is still given
by a proper supermatrix.

It is interesting to observe that the subset of 1/2 BPS operators (\ref{Lorb})--(\ref{n2def}) identified by the condition
\be\label{parameters}
\bar\a_{1,2}^{(\ell)} = \bar\a_{1,2}, ~~
\b^{1,2}_{(\ell)} = \b^{1,2}, ~~
\bar\g_{3,4}^{(\ell)} = \bar\g_{3,4}, ~~
\d^{3,4}_{(\ell)} = \d^{3,4}
\ee
correspond to direct orbifold projections of the 1/6 BPS $W_{\fer}$ of the ABJ(M) theory defined in \eqref{WferABJM}.
Among them, it is possible to select the ones which arise from the orbifold projection of 1/2 BPS WLs $W_{1/2}$ in ABJ(M) defined in (\ref{W12ABJM}), after setting $\b^{1,2} = \a^{1,2}/|\a|^2, \bar{\g}_{3,4} = \d^{3,4}=0$.

Similarly, from results in section~\ref{secn2} we obtain the 1/2 BPS WLs $\td W_\fer$ that preserve supercharges complementary to
the ones of $W_\fer$, and among them we can select the WLs corresponding to the projections of the $\td W_\fer$ and $\td W_{1/2}$ operators of ABJ(M) theory defined in \eqref{tdWbosferABJM} and (\ref{W12ABJM}).

\subsection{Gravity duals}

Given the known gravity dual configurations of 1/2 BPS WLs $W_{1/2}$ and $\td W_{1/2}$ in ABJM theory, we can perform their orbifold projection and identify the (anti--)M2--brane solutions dual to  WLs in $\mN=2$ orbifold ABJM theory. In general, this operation may break SUSY. Nonetheless, as we now show some BPS configurations survive, which are dual to particular 1/2 BPS WLs constructed in the previous section.

The $\mN=2$ orbifold ABJM theory is dual to M--theory in $\goabjm$ background with metric
\be ds^2=R^2 \Big( \frac14 ds^2_{\AdS_4}+ds^2_{ {\rm S}^7/(\Z_{rk}\times\Z_r)} \Big) \ee
where the AdS$_4$ metric is given in \eqref{ads4} and the S$^7$ one in \eqref{s7}, respectively.

In this case the $\rZ_{rk}\times\rZ_r$ quotient is realized by imposing the following identification on the ${\rC}^4$ coordinates

\be
(z_1,z_2,z_3,z_4)\sim\ep^{\f{2\pi\ii}{rk}}(z_1,z_2,z_3,z_4) , ~~
(z_1,z_2,z_3,z_4)\sim(\ep^{\f{2\pi\ii}{r}}z_1,\ep^{-\f{2\pi\ii}{r}}z_2,z_3,z_4)
\ee
This is equivalent to requiring (see eq. \eqref{z1234})
\be \label{quotient}
\zeta\sim \zeta - \frac{8\pi}{rk} , ~~ \phi_1 \sim \phi_1 - \frac{4\pi}{r}
\ee

As reviewed in appendix \ref{Appduals}, the general solution to the Killing spinor equations in M-theory on $\AdS_4\times\rS^7$ background reads \cite{Drukker:2008zx,Lietti:2017gtc}
\be 
\e = u^{\f12}h(\e_1+x^{\m}\g_{\m}\e_2) - u^{-\f12} \g_3 h \e_2 \qquad \mu = 0,1,2
\ee
where $\e_1$, $\e_2$ are two constant Majorana spinors satisfying $\g^{012}\e_i=\e_i$, $i=1,2$, and $h$ is given in eq. \eqref{eq:h}.

We decompose the Killing spinors on the basis of gamma matrices eigenstates as in (\ref{sum}, \ref{dec1}) where $\th^i$ are associated to the Poincar\'e supercharges, while $\vth^i$ are the superconformal supercharges.
Imposing the orbifold projection in (\ref{quotient}) leads to the constraints
\be \mL_{\partial_\zeta}\e=\mL_{\partial_{\phi_1}}\e=0 \label{con0} \ee
where the definition of spinorial Lie derivative with respect to a Killing vector $K$ is \cite{Figueroa-OFarrill:1999klq}
\be\mL_K\epsilon=K^\mu\nabla_\mu\epsilon+\frac{1}{4}\nabla_\mu K_\nu \Gamma^{\mu\nu}\epsilon\ee
Here we convert gamma matrices with tangent space indices to the ones with curved space indices using the vielbein of $\rS^7$.
Eq.~(\ref{con0}) leads then to the constraints
\be \label{con1}
( \g_{3\natural} + \g_{58} + \g_{47} + \g_{69} ) \e_i = ( \g_{3\natural} - \g_{58} ) \e_i = 0, \qquad  i = 1,2
\ee
which on the $\eta_i$ spinors defined in \eqref{t1t2t3t4} translates into
\be
t_1 + t_2 + t_3 + t_4 = 0 , \qquad t_1 - t_2 = 0
\ee
Therefore, the only surviving eigenstates correspond to
\be
(t_1,t_2,t_3,t_4) = (++--), (--++)
\ee
and we are led to $\e_1 = \th^2 \otimes \eta_2 + \th^7 \otimes \eta_7$ and $ \e_2=\vth^2 \otimes \eta_2 + \vth^7 \otimes \eta_7$.

Redefining
\be\label{definn2abjm}
\th^2\equiv\th, ~~ \th^7\equiv\bar\th, ~~ \vth^2\equiv\vth, ~~ \vth^7\equiv\bar\vth, ~~ \eta_2\equiv\eta, ~~ \eta_7\equiv\bar\eta
\ee
we finally obtain the two Killing spinors surviving the orbifold projection
\be
\e_1 = \th \otimes \eta + \bar\th \otimes \bar\eta, \qquad \e_2 = \vth \otimes \eta + \bar\vth \otimes \bar\eta
\ee
The corresponding field theory is in fact $\mN=2$ superconformal invariant.

Comparing (\ref{definn2abjm}) with (\ref{definabjm}), it turns out  that the relation of SUSY parameters in ABJM and $\mN=2$ orbifold
ABJM theories is the following
\be
\th^{12}=\bar\th_{34}=\th, ~~ \th^{34}=\bar\th_{12}=\bar\th, ~~
\vth^{12}=\bar\vth_{34}=\vth, ~~ \vth^{34}=\bar\vth_{12}=\bar\vth
\ee

As reviewed in appendix \ref{AppABJM}, in ABJM theory the 1/2 BPS operator $W_{1/2}[\bar\a_I]$ is dual to an M2-brane wrapping the cycle in
$\rS^7/\rZ_k$ specified by $\bar\a_I$ given in eq. \eqref{unit}. Similarly, $\td W_{1/2}[\bar\a_I]$ is dual to an anti-M2-brane wrapping the same cycle \cite{Lietti:2017gtc}.
On the other hand, as described in the previous section, orbifolding $W_{1/2}[\bar\a_I]$ and $\td W_{1/2}[\bar\a_I]$ leads to WLs in $\mN=2$ orbifold ABJM theory. Therefore, performing the same orbifold projection on the corresponding gravity dual solutions, we obtain the (anti)--M2--brane configurations dual to a particular subset of WLs of the $\mN=2$ theory.

In general, when the $\bar\a_I$ parameters (\ref{aI}) appearing in the $L_{1/2}[\bar\a_I]$ connection (\ref{W12ABJM}) satisfy $\bar\a_{1}\a^1+\bar\a_{2}\a^2 \neq 0$, $\bar\a_{3}\a^3+\bar\a_{4}\a^4 \neq 0$, the resulting operators and their dual configurations are non--BPS. However, for special choices of the parameters this may happen. Precisely,
\begin{itemize}
  \item When $\bar\a_{3,4}=0$, the WL in $\mN=2$ orbifold ABJM theory and its dual M2-brane in $\goabjm$ are 1/2 BPS. The WL
      preserves supercharges $\th_+,\bar\th_-,\vth_+,\bar\vth_-$, and is just a special case of the 1/2 BPS operator $W_\fer$ of the
      previous subsection.
  \item When $\bar\a_{1,2}=0$, the WL and its dual M2-brane are 1/2 BPS. The WL preserves supercharges
      $\th_-,\bar\th_+,\vth_-,\bar\vth_+$, and is a special case of the 1/2 BPS WL $\td W_\fer$ of the previous subsection.
\end{itemize}
A similar investigation can be performed for $\td W_{1/2}[\bar\a_I]$ with $|\a|^2 \neq 0$, which leads to the same conclusions. For $\bar\a_{3,4}=0$ or $\bar\a_{1,2}=0$ the resulting WLs are BPS and preserve sets of supercharges that are complementary to the ones already listed.

\section{BPS WLs at quantum level}\label{secquantum}

We now promote WLs to quantum operators and consider the problem of evaluating their vacuum expectation values (vev's). Similarly to what happens in the ABJ(M) theory, vev's of BPS WLs along a timelike straight line introduced in section~\ref{SecLineWL} are constants that can be trivially normalized to one. Instead, non--trivial vev's can be obtained for circular BPS operators in Euclidean space defined in  section~\ref{SecCircleWL}. These can be evaluated by using localization techniques and the output turns out to be a non--trivial function of the couplings that interpolates between the weak coupling result obtained via ordinary perturbation theory and the strong coupling result possibly obtained by holographic methods.

In this context it becomes particularly important to understand whether and how the classical cohomological $\cal Q$--equivalence between $W_{\bos}$ and $W_{\fer}$ discussed in subsection \ref{SecLineWL} gets promoted at quantum level. In fact, if we manage to prove that under suitable conditions the equivalence survives quantum corrections, then using the $\cal Q$ charge to localize the functional integral we can conclude that the expectation value of the general parametric--dependent $W_{\fer}$  is identical to the one of $W_{\bos}$, making them effectively quantum equivalent and independent of the choice of the parameters.

In the case of ABJ(M) theory this problem has been extensively analysed \cite{Bianchi:2013zda,Bianchi:2013rma,Griguolo:2013sma} and it has been shown that its solution is deeply interconnected with the choice of a framing regularization for the perturbative definition of the WL. Precisely, the cohomological equivalence at quantum level {leads to} \cite{Drukker:2009hy}
\be\label{framing1equiv}
\langle W_\bos \rangle^{(\rm{ABJ(M)})}_1 = \langle W_\fer \rangle^{(\rm{ABJ(M)})}_1
\ee
where the subscript indicates that the identity holds only at framing one \cite{Kapustin:2009kz}.

With a direct perturbative computation we are now going to show that, up to two loops, the problem of studying the quantum cohomological equivalence between circular $W_{\fer}$  and $W_{\bos}$ in  generic $\mathcal{N}=2$ SCSM models  can be mapped to the parallel problem in ABJ(M) theory. Using the speculations in \cite{Bianchi:2016gpg}, we can thus conclude that the classical cohomological equivalence can be promoted at quantum level, at least at two loops, if we work at framing one.
Indeed, the two--loop result supports a stronger version of the $\mathcal Q$--equivalence, which we probe at three loops in subsection \ref{conj}.

\subsection{The perturbative analysis}
In order to test the cohomological equivalence at quantum level, it is convenient to consider the difference between the expectation values of fermionic and bosonic WLs, order by order in perturbation theory. At classical level, the cohomological equivalence
\be \label{relation1}
W_\fer - W_\bos = \mQ V
\ee
can be expanded as \cite{Drukker:2009hy}
\be \label{relation2}
\sum_{n=1}^{\infty}\Tr \mP( \ep^{-\ii\oint d\t L_\bos(\t)} W_n ) = \mQ V
\ee
where the $W_{n}$ expressions arise from the expansion of $e^{-\ii \oint (L_\fer - L_\bos)} = e^{-\ii \oint (B +F)}$ inside the WL (see eq. \eqref{LABF}). The label $n$ indicates the total number of scalar and fermion fields in the product. As follows from the explicit expressions of $B$ and $F$ in \eqref{bosferconeuc}, at a given order $n$ the $W_{n}$ function is built up by $p$ powers of $B$, which is quadratic in the scalar fields, and $q$ powers of $F$, such that $2p+q = n$ (in particular, when $n$ is odd the corresponding $W_n$ contains an odd number of spinors $F$).
As an example,  we report the first few even terms in \eqref{relation2}
\bea\label{W246}
&&   W_2 = -\ii \!\oint\! d\t B(\t)
            - \!\oint\! d\t_{1>2} F(\t_1)F(\t_2) \nn\\
&& 
   W_4 = - \!\oint\! d\t_{1>2} B(\t_1)B(\t_2)
      +\ii \!\oint\! d\t_{1>2>3} \big[ B(\t_1)F(\t_2)F(\t_3) 
                                 + F(\t_1)B(\t_2)F(\t_3) \nn\\
&& 
   \phantom{W_4 =}
                                 + F(\t_1)F(\t_2)B(\t_3) \big]
        + \!\oint\! d\t_{1>2>3>4} F(\t_1)F(\t_2)F(\t_3)F(\t_4) \nn\\
&& W_6 = \ii \!\oint\! d\t_{1>2>3} B(\t_1) B(\t_2) B(\t_3)
       + \!\oint\! d\t_{1>2>3>4} \big[ B(\t_1) B(\t_2) F(\t_3) F(\t_4) \nn\\
&& \phantom{W_6 =}
         + B(\t_1) F(\t_2) B(\t_3) F(\t_4)
         + B(\t_1) F(\t_2) F(\t_3) B(\t_4)
         + F(\t_1) B(\t_2) B(\t_3) F(\t_4) \nn\\
&& \phantom{W_6 =}
         + F(\t_1) B(\t_2) F(\t_3) B(\t_4)
         + F(\t_1) F(\t_2) B(\t_3) B(\t_4) \big] \nn\\
&& \phantom{W_6 =}
       -\ii \!\oint\! d\t_{1>2>3>4>5}
        \big[ B(\t_1) F(\t_2) F(\t_3) F(\t_4) F(\t_5)
           + F(\t_1) B(\t_2) F(\t_3) F(\t_4) F(\t_5) \nn\\
&& \phantom{W_6 =}
           + F(\t_1) F(\t_2) B(\t_3) F(\t_4) F(\t_5)
           + F(\t_1) F(\t_2) F(\t_3) B(\t_4) F(\t_5) \nn\\
&& \phantom{W_6 =}
           + F(\t_1) F(\t_2) F(\t_3) F(\t_4) B(\t_5) \big] \nn\\
&& \phantom{W_6 =}
      - \!\oint\! d\t_{1>2>3>4>5>6}
             F(\t_1) F(\t_2) F(\t_3) F(\t_4) F(\t_5) F(\t_6)
\eea
The structure of higher order terms clearly follows. Here $\oint d \t_{1>2> \dots >j}$ means integrals over the contour parameters $\t_1, \dots , \t_j$ satisfying $\t_1>\t_2> \dots > \t_j$.

It has been shown, for the first few orders in \cite{Drukker:2009hy} and proved at all orders in \cite{Ouyang:2015qma}, that at classical level the $\mQ$--equivalence in \eqref{relation1} follows from a stronger set of identities that read
\be \label{relation2bis}
\Tr \mP( \ep^{-\ii\oint d\t L_\bos(\t)} W_n ) = \mQ V_n , \qquad n = 1, 2, \dots
\ee
that is, every single term in the summation \eqref{relation2} is cohomologically trivial for a suitable choice of the $V_n$ function (see Appendix D in \cite{Ouyang:2015qma} for their explicit forms).

As already mentioned, in the ABJ(M) case when we move to quantum level and compute expectation values at framing one, no anomalies arise and relation \eqref{relation1} implies
\be\label{relation3}
\langle W_{\fer} \rangle_1 = \langle W_{\bos} \rangle_1
\ee
which in turns can be written as
\be \label{relation4}
\sum_{n=1}^{\infty} \lag \Tr \mP( \ep^{-\ii\oint d\t L_\bos(\t)} W_{2n} ) \rag_1 = 0
\ee
Here we have already neglected $W_n$ for $n$ odd, since strings of an odd number of $F$ would have vanishing expectation values.

An interesting question is whether at quantum level and framing one, identities \eqref{relation2bis} remain true separately, as this would imply the non--trivial results
\be \label{nontrivial}
\langle  \Tr \mP( \ep^{-\ii\oint d\t L_\bos(\t)} W_{2n} ) \rangle_1 = 0 , \qquad n = 1, 2, \dots
\ee
We devote the rest of this section to the two--loop evaluation of \eqref{relation4} and the discussion of identities \eqref{nontrivial}.

To this aim we assume CS levels $k_a$  to be of the same order $k$, and all the WL parameters $\bar m_i^{ab}$ and $n^i_{ab}$ to be of order $1/\sqrt{k}$. In particular, this implies that a given $W_{2m}$ term in \eqref{relation4}  starts contributing at order $O(1/k^m)$.
Therefore, in order to obtain the full result up to order $1/k^2$ we need to evaluate the terms in \eqref{relation4} involving correlators with $W_2$ and $W_4$ up to two loops and show that at framing one either their sum vanishes or they vanish separately.  As a by--product, we also compute the two--loop expectation values of bosonic and general fermionic WLs at framing zero.

We focus on euclidean circle WLs (\ref{bosfereuc}) with connections (\ref{bosferconeuc}). Moreover, for simplicity we restrict to quiver theories without adjoint or (anti-)fundamental matter. The more general case can be similarly worked out, but the results are more involved.

\begin{figure}[htbp]
  \centering
  \includegraphics[height=0.16\textwidth]{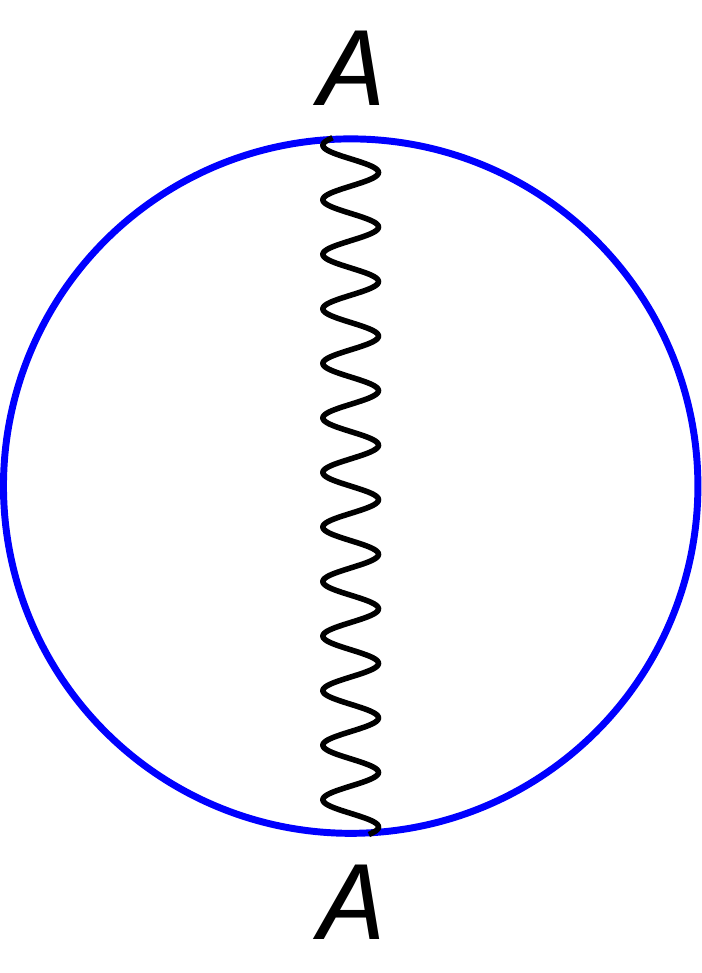}
  \caption{One-loop Feynman diagram for $\lag  W_\bos \rag_f$. Wavy lines represent gauge fields.}\label{fdAA}\label{fd1}
\end{figure}

\begin{figure}[htbp]
  \centering
  \subfigure[]{\includegraphics[height=0.16\textwidth]{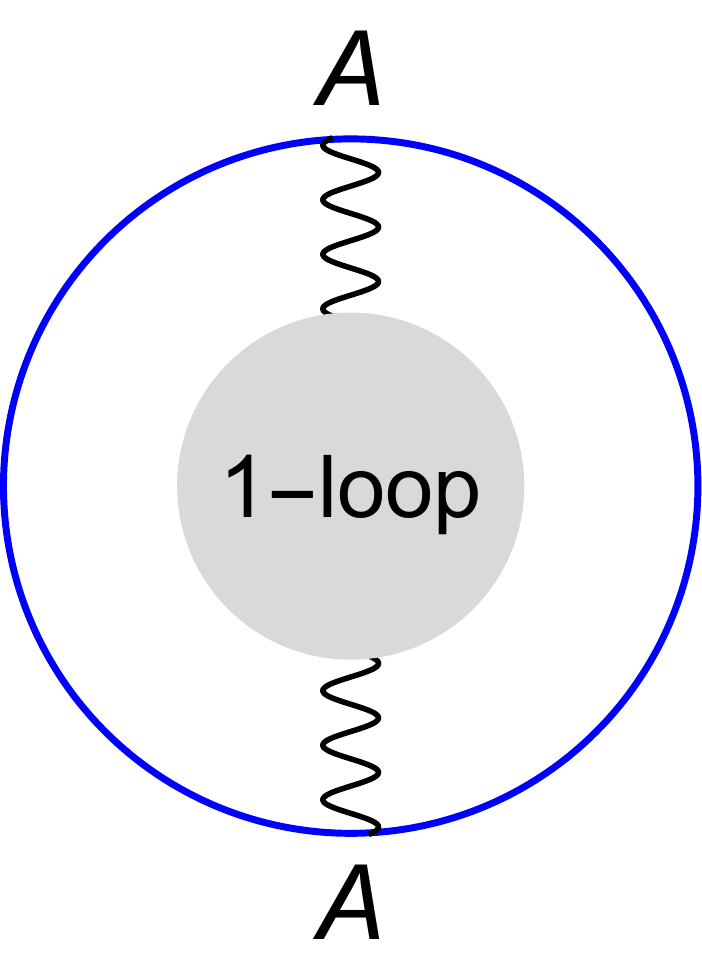}\label{fdAAp}}~~
  \subfigure[]{\includegraphics[height=0.16\textwidth]{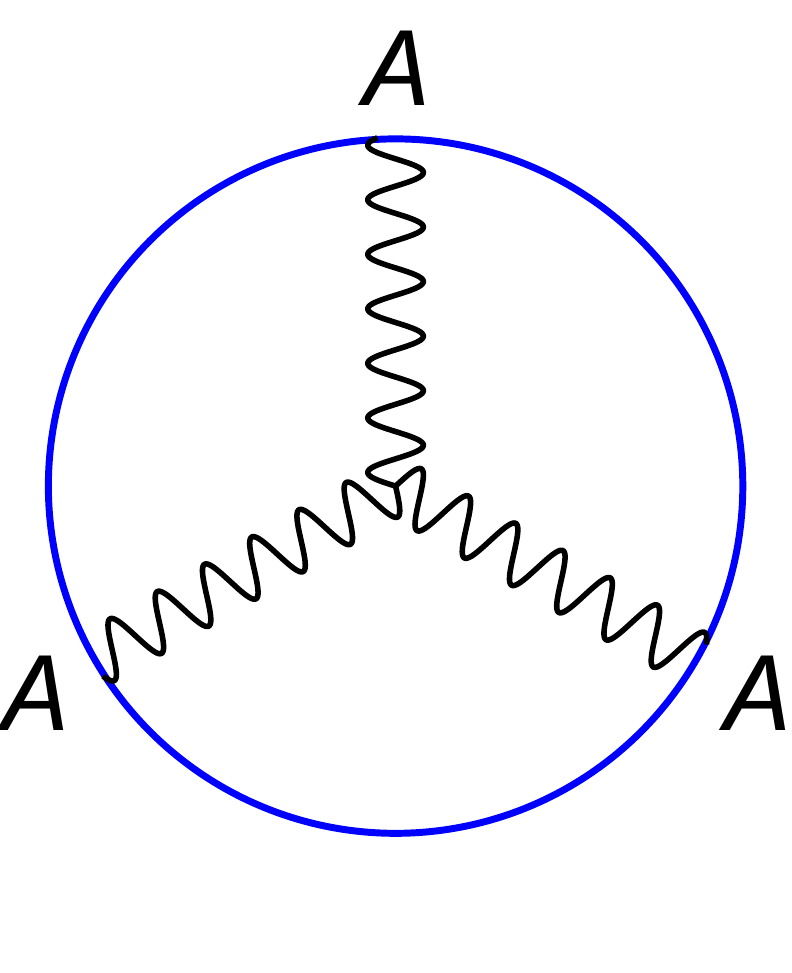}\label{fdAAA}}~~
  \subfigure[]{\includegraphics[height=0.16\textwidth]{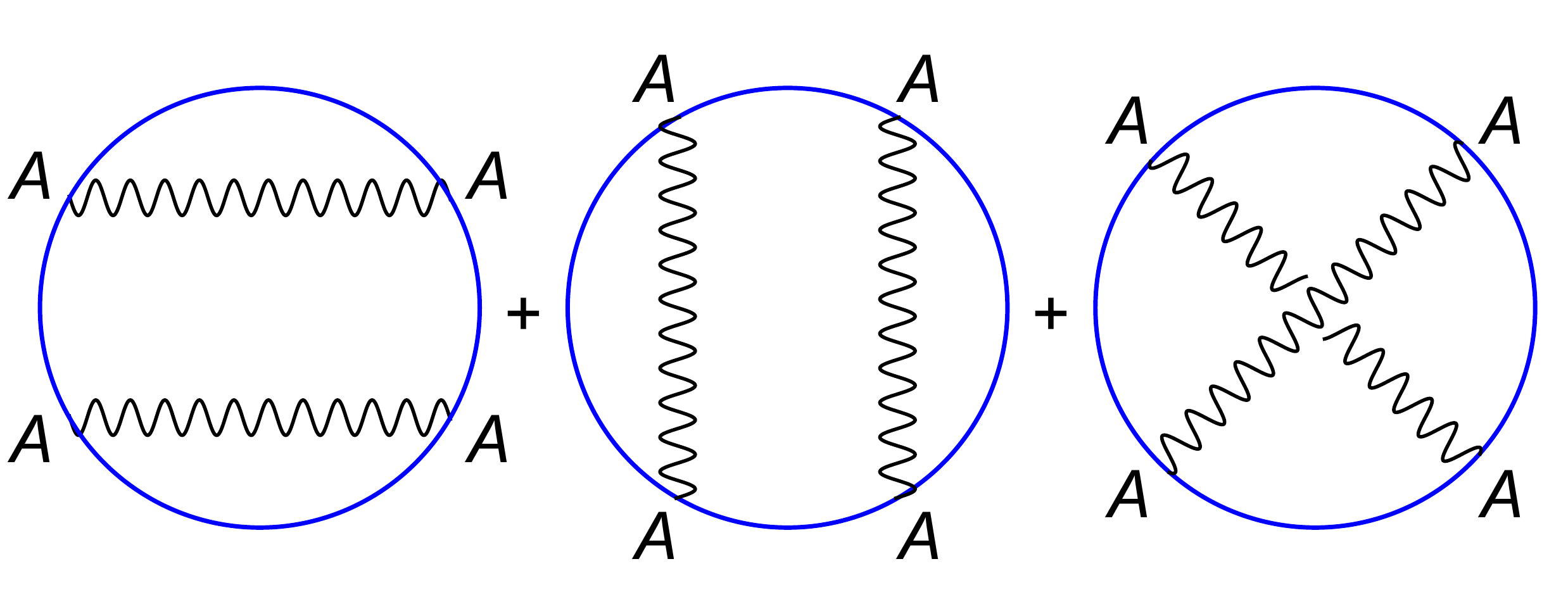}\label{fdAAAA}}~~
  \subfigure[]{\includegraphics[height=0.16\textwidth]{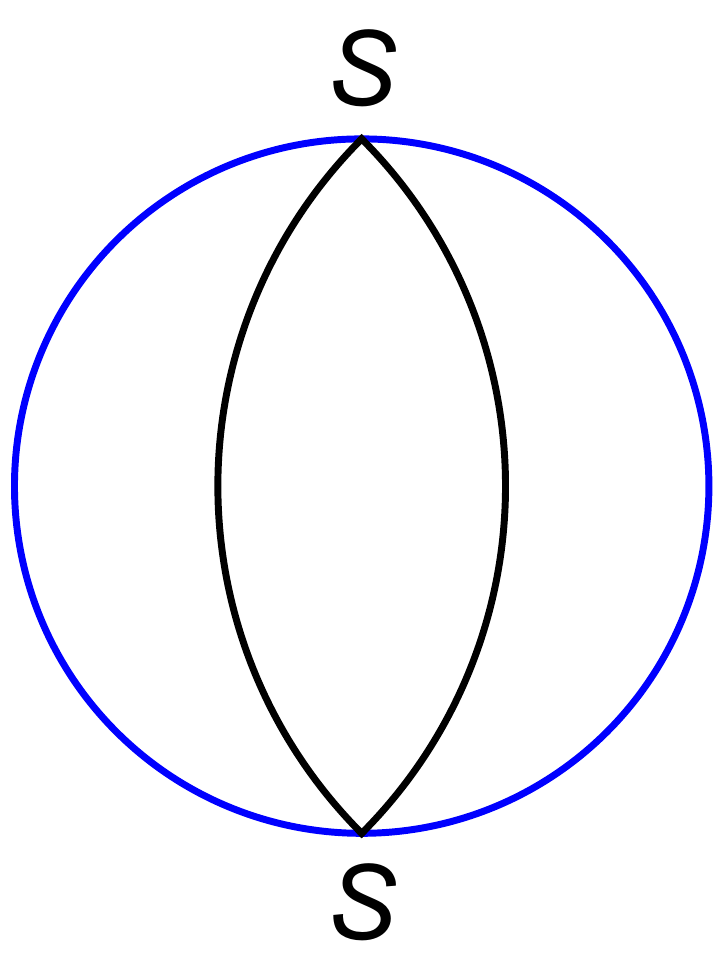}\label{fdSS}}~~
  \caption{Two-loop Feynman diagrams for $\lag  W_\bos \rag_f$. Here $S$ stays for the scalar bilinears which arise from $\s^{(a)}$, $a=1,2,\cdots,n$ in $L_\bos$ when we use equations of motion \eqref{equations}.}
\label{fd2}
\end{figure}

\begin{figure}[htbp]
  \centering
  \includegraphics[height=0.16\textwidth]{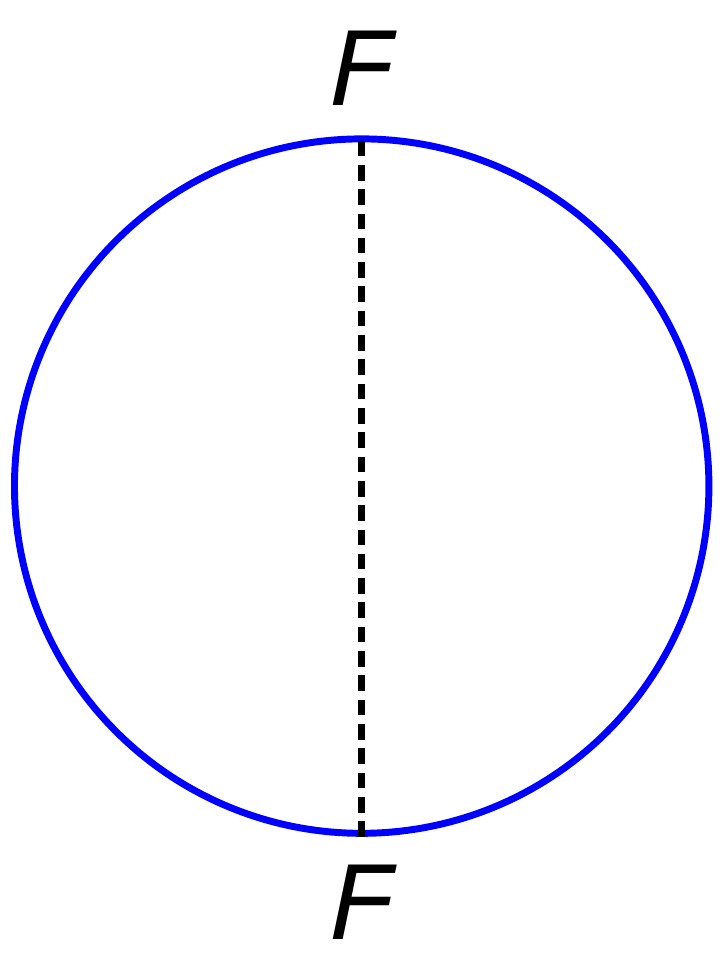}
  \caption{One-loop Feynman diagram for $\lag \Tr\cP (\ep^{-\ii\oint d \t L_\bos(\t)} W_2) \rag_f$. $F$  indicates the fermionic quantities defined in \eqref{bosferconeuc}.}\label{fdFF}\label{fd3}
\end{figure}

\begin{figure}[htbp]
  \centering
  \subfigure[]{\includegraphics[height=0.16\textwidth]{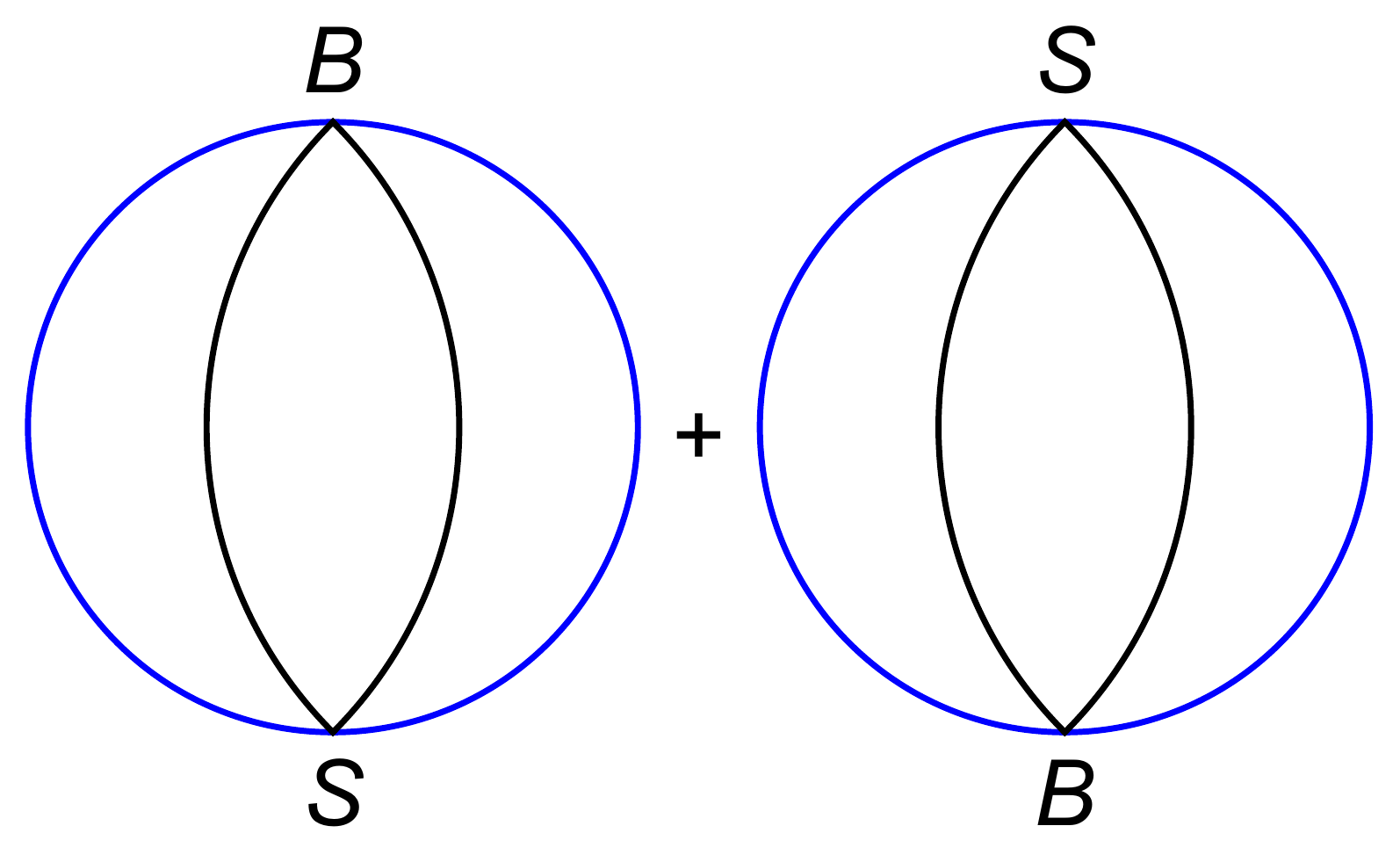}\label{fdSB}}~~
  \subfigure[]{\includegraphics[height=0.16\textwidth]{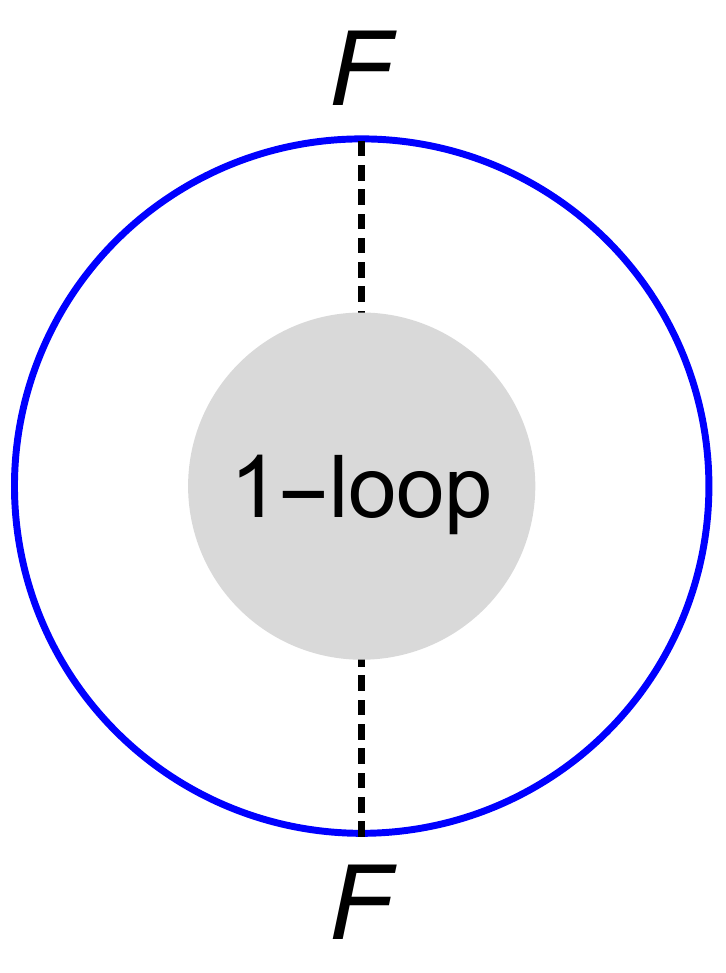}\label{fdFFp}}~~
  \subfigure[]{\includegraphics[height=0.16\textwidth]{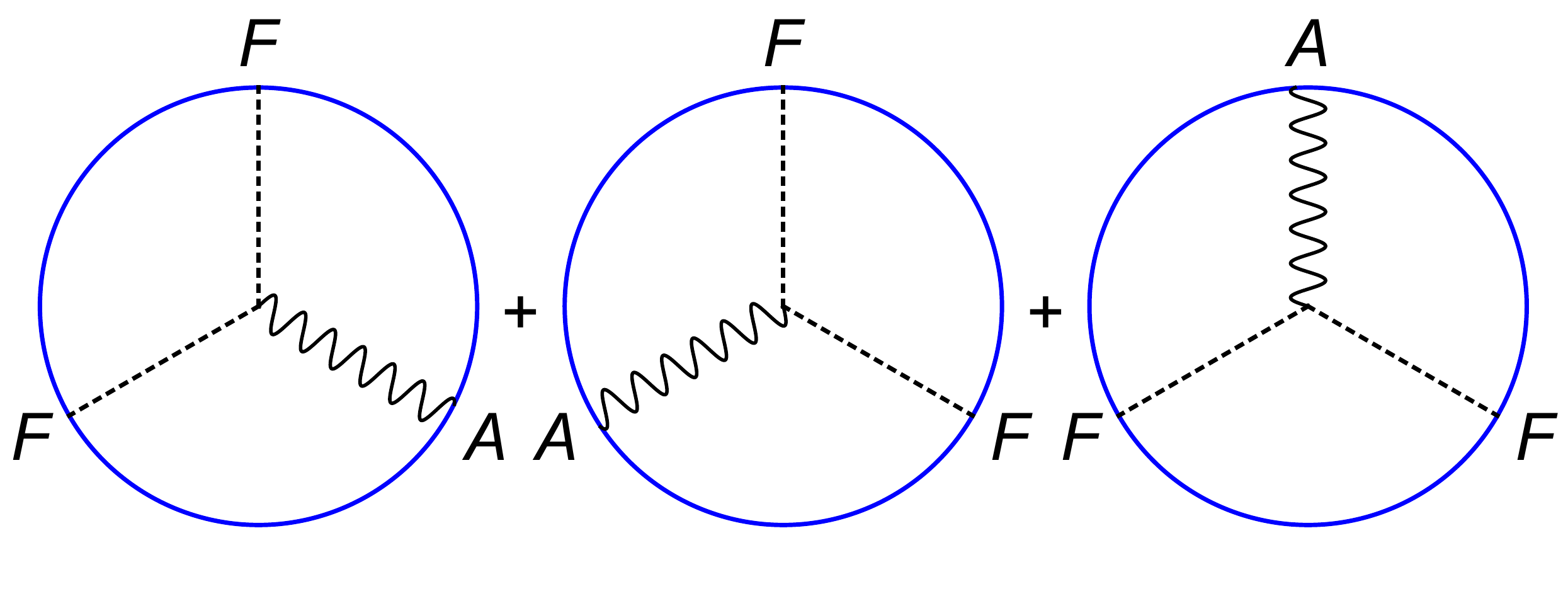}\label{fdAFF}}\\
  \subfigure[]{\includegraphics[height=0.16\textwidth]{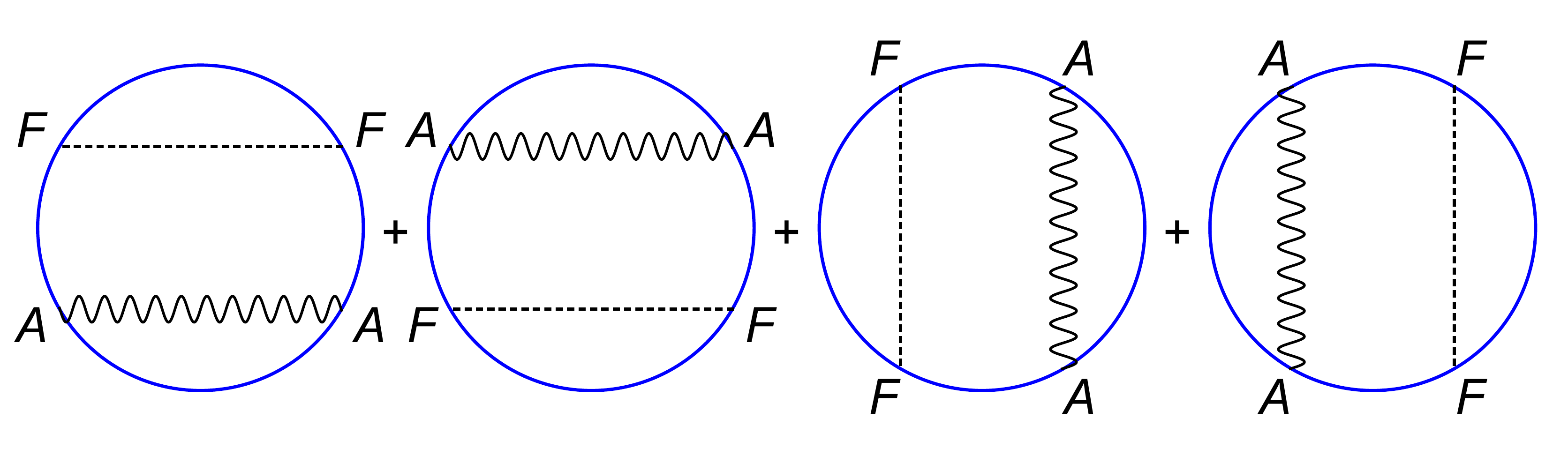}\label{fdAAFF}}
  \caption{Two-loop Feynman diagrams for $\lag \Tr \cP(\ep^{-\ii\oint d \t L_\bos(\t)} W_2) \rag_f$. Here $B$ indicates the bosonic quantities defined in \eqref{bosferconeuc}.}\label{fd4}
\end{figure}

\begin{figure}[htbp]
  \centering
  \subfigure[]{\includegraphics[height=0.16\textwidth]{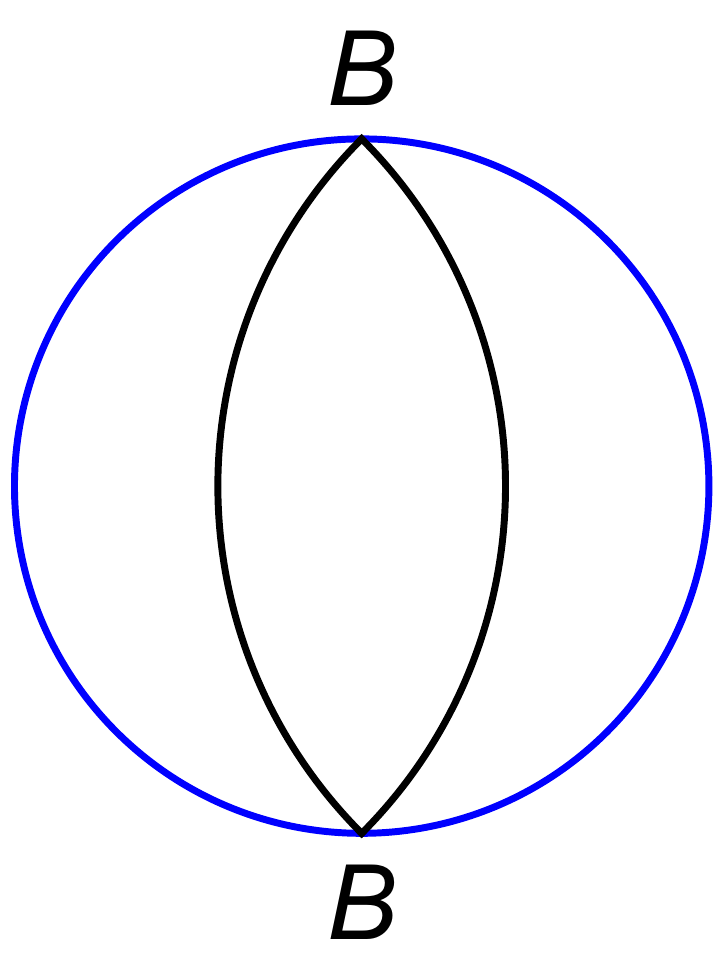}\label{fdBB}}~~
  \subfigure[]{\includegraphics[height=0.16\textwidth]{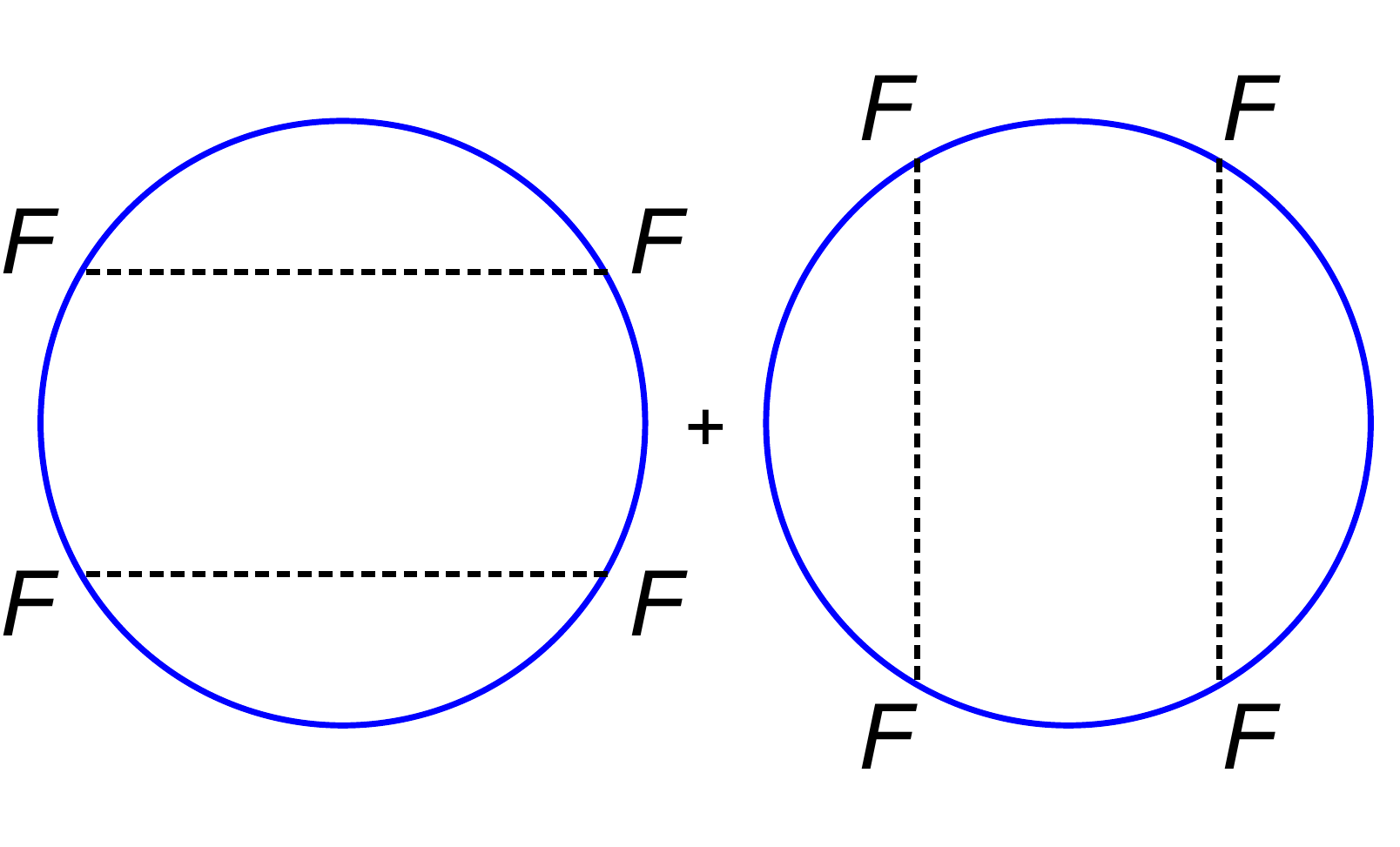}\label{fdFFFF}}\\
  \caption{Two-loop Feynman diagrams for $\lag \Tr \cP(\ep^{-\ii\oint d \t L_\bos(\t)} W_4) \rag_f$.}\label{fd5}
\end{figure}

We organize our calculation as follows.  We first compute  $W_{\bos}$ at one and two loops, with relevant  diagrams  shown in figures~\ref{fd1} and \ref{fd2}.  Then  we consider equation  \eqref{relation4} and evaluate correlators involving  $W_{2}, W_{4}$ as defined in \eqref{W246}.
We compute the correlator with a $W_2$ insertion at one and two loops (figures~\ref{fd3} and \ref{fd4}) whereas the term with  $W_4$ starts directly at two loops  (figure~\ref{fd5}).
Writing the bosonic connection as $L_\bos=A+S$ (and consequently $L_\fer=A+S+B+F$), in these figures $A$ insertions on the contour represent gauge connections, $B$ and $F$ insertions indicate quantities defined in \eqref{bosferconeuc}, whereas $S$ stays for the scalar bilinears which arise from $\s^{(a)}$, $a=1,2,\cdots,n$ in $L_\bos$ when we use equations of motion \eqref{equations}.



The interesting observation is that a full fledged computation of all these Feynman diagrams is not required, as we can heavily rely on the results already present in the literature for a WL in pure CS theory \cite{Guadagnini:1989am} and 1/2 and 1/6 WLs in ABJ(M) theory \cite{Drukker:2008zx,Chen:2008bp,Rey:2008bh,Bianchi:2013zda,Bianchi:2013rma,Griguolo:2013sma,Bianchi:2016yzj,Bianchi:2016gpg}.
In fact, at the order we are working the internal vertices involving superpotential interactions do not play any role, and we have to deal only with pure gauge vertices and minimal couplings of matter fields to gauge vectors. Therefore, the difference between diagrams in ABJ(M) theory and any $\mN=2$ SCSM model is only due to the different matter content, that is the different number of matter multiplets linking pairs of quiver nodes. It follows that at this order diagrams in the two theories differ only by the overall combinatorial factor, whereas the corresponding  integrals are the same.  Clearly, if we were to consider higher order corrections to $W_n$ correlators the specific structure of the superpotential would generally kick in and different models might display different behaviours.

A long but straightforward evaluation of the diagrams leads to the results that can be found in appendix~\ref{appint}. For each diagram we extract the color factor and the parametric dependence coming from the WL expansion, and indicate with italic capital letters the corresponding integrals that include both combinatorics and any other factors coming from the Feynman rules listed in appendix \ref{appFeynman}. We stress that some of the fermionic diagrams can give rise to more than one integral, which differ by the order of the fermions along the contour.
We present the results at framing $f$, meaning that $f$ can be either zero or one.

Summing over all the contributions in each figure and using constraints (\ref{UVmmnn}), we obtain

\noindent
for $W_\bos$
\bea\label{results1}
&& \dcircled{\ref{fdAA}} = \sum_a \f{N_a^2}{k_a} {\cal I}^{(f)}_{\rm\ref{fdAA}} \nn\\
&& \dcircled{\ref{fd2}} = \sum_{a,b} \f{(N_{ab} + N_{ba})N_a^2N_b}{k_a^2}
                                     ( \cI_{\rm\ref{fdAAp}}^{(f)} + \cI_{\rm\ref{fdSS}}^{(f)} )
                        + \sum_a \f{N_a^3}{k_a^2} ({\cal I}_{{\rm\ref{fdAAA}}}^{(f)} + {\cal I}_{\rm{\ref{fdAAAA}}}^{(f)}) \nn\\
&& \hspace{13mm}
                        + \sum_a \f{N_a}{k_a^2} ( - \cI_{{\rm\ref{fdAAA}}}^{(f)} + \cJ_{{\rm\ref{fdAAAA}}}^{(f)} )
\eea
for $W_2$
\bea   \label{results2}
&& \dcircled{\ref{fd3}} = \sum_{a,b} \bar m_i^{ab} n^i_{ba} N_a N_b {\cal I}_{\rm\ref{fdFF}}^{(f)}  \nn\\
&& \dcircled{\ref{fd4}} = \sum_{a,b} [ \bar m_i^{ab} n^i_{ba} ( {\cal I}_{\rm\ref{fdSB}}^{(f)}
                                                              + {\cal I}_{\rm\ref{fdFFp}}^{(f)}
                                                              + {\cal I}_{\rm\ref{fdAFF}}^{(f)}
                                                              + {\cal I}_{\rm\ref{fdAAFF}}^{(f)}  ) \nn \\
&& \hspace{13mm}
                                     + \bar m_i^{ba} n^i_{ab} ( - {\cal I}_{\rm\ref{fdSB}}^{(f)}
                                                                + {\cal I}_{\rm\ref{fdFFp}}^{(f)}
                                                                + {\cal J}_{\rm\ref{fdAFF}}^{(f)}
                                                                + {\cal J}_{\rm\ref{fdAAFF}}^{(f)} ) ] \f{N_a^2 N_b}{k_a}
\eea
for $W_4$
\bea \label{results3}
&& \dcircled{\ref{fd5}} = \sum_{a,b,c} \Big[ \bar m_i^{ab} n^i_{ba} \bar m_j^{ac} n^j_{ca} \Big(   {\cal I}_{\rm\ref{fdBB}}^{(f)} + {\cal I}_{\rm\ref{fdFFFF}}^{(f)}  \Big) \nn\\
&& \hspace{13mm}
                                           + \bar m_i^{ba} n^i_{ab} \bar m_j^{ca} n^j_{ac} \Big(   {\cal I}_{\rm\ref{fdBB}}^{(f)} +{\cal J}_{\rm\ref{fdFFFF}}^{(f)} \Big)
                                       \Big] N_a N_b N_c
\eea
The exact expression of the ${\cal I}, {\cal J}$ integrals can be found in appendix \ref{appint}. In particular, from their explicit expression it follows  that
\be
{\cal I}_{\rm\ref{fdSB}}^{(f)} = -\f1\pi {\cal I}_{\rm\ref{fdSS}}^{(f)} \quad \quad {\cal I}_{\rm\ref{fdBB}}^{(f)}  = \frac{1}{4\pi^2} {\cal I}_{\rm\ref{fdSS}}^{(f)}
\ee

We stress that these results are common to any $\mN \geq 2$ SCSM theory, ABJ(M) models included. Therefore, the explicit values of the integrals can be extracted from the literature on the pure Chern--Simons theory \cite{Guadagnini:1989am} and ABJ(M) theory \cite{Drukker:2008zx,Chen:2008bp,Rey:2008bh}.

At generic framing it is known that \cite{Bianchi:2016gpg}
\be\label{framingf}
{\cal I}_{\rm\ref{fdAA}}^{(f)} = {-}\pi\ii f, ~~~~
{\cal I}_{\rm\ref{fdAAp}}^{(f)} + \cI_{\rm\ref{fdSS}}^{(f)} = \f{\pi^2}{4}, ~~~~
{\cal I}_{\rm\ref{fdAAA}}^{(f)} = -\f{\pi^2}{6}, ~~~~
{\cal I}_{\rm\ref{fdAAAA}}^{(f)} = -\f{\pi^2}{2}f^2, ~~~~
{\cal J}_{\rm\ref{fdAAAA}}^{(f)} = 0
\ee
whereas the ones appearing in (\ref{results2}, \ref{results3}) are explicitly known only at framing zero \cite{Bianchi:2013zda,Bianchi:2013rma,Griguolo:2013sma,Bianchi:2016gpg}
\be\label{framing0}
{\cal I}_{\rm\ref{fdFF}}^{(0)} =
{\cal I}_{\rm\ref{fdSB}}^{(0)} =
{\cal I}_{\rm\ref{fdFFp}}^{(0)} =
{\cal I}_{\rm\ref{fdAAFF}}^{(0)} =
{\cal J}_{\rm\ref{fdAAFF}}^{(0)} =0, ~~~~
{\cal I}_{\rm\ref{fdAFF}}^{(0)}  = - {\cal J}_{\rm\ref{fdAFF}}^{(0)} = -\f{\pi}{2}, ~~~~
{\cal I}_{\rm\ref{fdFFFF}}^{(0)} = {\cal J}_{\rm\ref{fdFFFF}}^{(0)} = \f{3}{32}
\ee
However, it can be checked numerically that both ${\cal I}_{\rm\ref{fdAAp}}^{(f)}$ and $\cI_{\rm\ref{fdSS}}^{(f)}$ are framing independent. Therefore, since at framing zero $\cI_{\rm\ref{fdSS}}^{(0)}=\cI_{\rm\ref{fdSB}}^{(0)}=\cI_{\rm\ref{fdBB}}^{(0)}=0$, at generic framing we obtain
\be \label{eq139}
{\cal I}_{\rm\ref{fdAAp}}^{(f)} = \f{\pi^2}{4}, ~~ \cI_{\rm\ref{fdSS}}^{(f)}=\cI_{\rm\ref{fdSB}}^{(f)}=\cI_{\rm\ref{fdBB}}^{(f)}=0
\ee

Using redefinitions \eqref{pt1} for the parameters and choosing $\b^I = \a^I/|\a|^2$, $\bar\g_I=\d^I=0$, expressions (\ref{results2}--\ref{results3}) reduce to $W_2, W_4$ terms for the expansion of $\langle W_{1/2}  - W_\bos \rangle_f$ in ABJ(M) theory. Therefore, using the arguments and the well--based speculations of \cite{Bianchi:2016gpg} for 1/2 BPS WLs in ABJ(M) theory, we obtain that at framing one
\be  \label{eq140}
{\cal I}_{\rm\ref{fdFF}}^{(1)} =
{\cal I}_{\rm\ref{fdFFp}}^{(1)} + {\cal I}_{\rm\ref{fdAFF}}^{(1)} =
{\cal I}_{\rm\ref{fdFFp}}^{(1)} + {\cal J}_{\rm\ref{fdAFF}}^{(1)} =
{\cal I}_{\rm\ref{fdAAFF}}^{(1)} = {\cal J}_{\rm\ref{fdAAFF}}^{(1)} =
{\cal I}_{\rm\ref{fdFFFF}}^{(1)} = {\cal J}_{\rm\ref{fdFFFF}}^{(1)} = 0
\ee
Therefore, we can conclude that at framing one diagrams~\ref{fd3}--\ref{fd5} are identically vanishing
\be \label{fd345}
\dcircled{\ref{fd3}} = \dcircled{\ref{fd4}} = \dcircled{\ref{fd5}} = 0 ~~ \quad {\rm for}~f=1
\ee
and the cohomological equivalence between $W_\bos$ and $W_{\fer}$ holds for any $\cN \geq 2$ SCSM theory, i.e.
\be
\lag W_\fer \rag_1   -  \lag W_\bos \rag_1   =0 ~~ \quad {\rm up~to~ two~ loops}
\ee

Actually, we find a much stronger result. In fact, results \eqref{fd345} imply that
\be
\langle  \Tr \mP( \ep^{-\ii\oint d\t L_\bos(\t)} W_{2} ) \rangle_1 =  \langle  \Tr \mP( \ep^{-\ii\oint d\t L_\bos(\t)} W_{4} ) \rangle_1 = 0
\ee
separately. This is nothing but eq. (\ref{nontrivial}) that we expect to hold quantum mechanically as a consequence of the classical relations in (\ref{relation2bis}).

We stress that the cohomological equivalence up to two loops and framing one and its stronger version are still valid when there are adjoint and fundamental matter fields in the theory.

Specializing these findings to the ABJ(M) theory, we find that $W_\bos$ is cohomological equivalent not only to the 1/2 BPS fermionic operator \cite{Drukker:2009hy}, but also to generic 1/6 BPS fermionic WLs introduced in \cite{Ouyang:2015iza,Ouyang:2015bmy}, at least at the order we are working.  Similarly, in $\mN=4$ circular quiver SCSM theory with alternating levels, the fermionic 1/4 BPS WLs introduced in \cite{Mauri:2017whf} are also quantum mechanically cohomological equivalent to the bosonic 1/4 BPS WL up to two loops.

\vskip 10pt
Exploiting results \eqref{framingf} and summing up all the contributions in \eqref{results1} we obtain the general result for $W_\bos$ in $\mN=2$ SCSM models up to two loops and generic framing
\bea \label{bosf1}
&& \lag W_\bos \rag_f = \sum_a \Big\{ N_a {-} \f{\pi\ii f N_a^2}{k_a} + \f{\pi^2[ -(3f^2+1)N_a^3 + N_a ]}{6k_a^2}  \Big\} \nn\\
&& \phantom{\lag W_\bos \rag_f =}
                    +\sum_{a,b} \f{\pi^2(N_{ab}+N_{ba})N_a^2N_b}{4k_a^2}
                    +O\Big(\f{1}{k^3}\Big)
\eea
Similarly, we can now use results \eqref{framing0} in eq. (\ref{results2}--\ref{results3}) and obtain the two--loop expression
\bea \label{diff1}
&& \lag W_\fer - W_\bos \rag_0 =  - \sum_{a,b} ( \bar m^{ab}_i n^i_{ba} - \bar m^{ba}_i n^i_{ab} ) \f{\pi N_a^2 N_b}{2k_a} \\
&& \phantom{\lag W_\fer - W_\bos \rag_0 =}
             + \sum_{a,b,c} ( \bar m^{ab}_i n^i_{ba}\bar m^{ac}_j n^j_{ca} + \bar m^{ba}_i n^i_{ab}\bar m^{ca}_j n^j_{ac} )
                           \f{3N_a N_b N_c}{32}
            + O\Big(\f{1}{k^3}\Big) \nn
\eea
In particular, this shows that at framing zero cohomological equivalence is in general broken.

By combining \eqref{diff1} with \eqref{bosf1} evaluated at $f=0$ we also obtain the framing zero result for  $W_\fer $  up to two loops
\bea
&& \lag W_\fer \rag_0 = \sum_a \Big[ N_a  + \f{\pi^2( - N_a^3 + N_a )}{6k_a^2}  \Big] \\
&& \phantom{\lag W_\fer \rag_0 =}
                      + \sum_{a,b} \Big[ \f{\pi^2(N_{ab}+N_{ba})N_a^2N_b}{4k_a^2}
                                       - ( \bar m^{ab}_i n^i_{ba} - \bar m^{ba}_i n^i_{ab} ) \f{\pi N_a^2 N_b}{2k_a} \Big] \nn\\
&& \phantom{\lag W_\fer \rag_0 =}
                      + \sum_{a,b,c} ( \bar m^{ab}_i n^i_{ba}\bar m^{ac}_j n^j_{ca} + \bar m^{ba}_i n^i_{ab}\bar m^{ca}_j n^j_{ac} )
                           \f{3N_a N_b N_c}{32}
                      + O\Big(\f{1}{k^3}\Big) \nn
\eea
The result displays a non--trivial parametric dependence on the fermion couplings, and thus it is different for different WLs.

More interestingly, at framing one, independently of the choice of the parametric couplings we find
\bea \label{bosferm1}
&& \lag W_\bos \rag_1 = \lag W_\fer \rag_1 = \sum_a \Big\{ N_a - \f{\pi\ii  N_a^2}{k_a} + \f{\pi^2[ -4N_a^3 + N_a ]}{6k_a^2}  \Big\} \nn\\
&& \phantom{\lag W_\bos \rag_1 = \lag W_\fer \rag_1 =}
                    +\sum_{a,b} \f{\pi^2(N_{ab}+N_{ba})N_a^2N_b}{4k_a^2}
                    +O\Big(\f{1}{k^3}\Big)
\eea
In particular, both $\lag W_\bos \rag_1$ and $\lag W_\fer \rag_1$ can in principle be evaluated by using localization techniques \cite{Kapustin:2009kz}. The result, expanded at second order, should match \eqref{bosferm1}.

Specializing our results to the ABJ(M) theory and normalizing the operators properly, we find
\bea
&& \f{\lag W_\bos \rag_f}{N_1+N_2} = 1 {-} \f{\pi\ii f(N_1-N_2)}{k} + \f{\pi^2}{6k^2} \Big[ -(3f^2+1)(N_1^2+N_2^2) \nn\\
&& \phantom{\f{\lag W_\bos \rag_f}{N_1+N_2} =}
                                                                                          + (3f^2+7) N_1N_2
                                                                                          + 1 \Big]
                                    + O\Big(\f{1}{k^3}\Big) \nn\\
&& \f{\lag W_\fer - W_\bos \rag_0}{N_1+N_2} = \f{\pi^2N_1N_2}{2k^2} \Big\{  3[ (\bar\a_I\b^I)^2 + (\bar\g_I\d^I)^2 ]
                                                                          - 4( \bar\a_I\b^I + \bar\g_I\d^I ) \Big\}
                                            + O\Big(\f{1}{k^3}\Big) \nn\\
&& \f{\lag W_\fer \rag_0}{N_1+N_2} = 1 + \f{\pi^2}{6k^2}
                                 \Big\{ -N_1^2 - N_2^2 +
                                     \big\{ 9 [ (\bar\a_I\b^I)^2 + (\bar\g_I\d^I)^2 ] \nn\\
&& \phantom{\f{\lag W_\fer \rag_0}{N_1+N_2} =}
                                         -12 ( \bar\a_I\b^I + \bar\g_I\d^I )
                                         + 7
                                     \big\} N_1 N_2 + 1
                                 \Big\}
                                 + O\Big(\f{1}{k^3}\Big)
\eea

\subsection{A conjecture}\label{conj}

As already stressed, the two--loop computation of the previous section displays an important feature which could lead to far reaching consequences, if confirmed at higher orders. In fact, using in \eqref{relation4}  the shorthand $ \lag \Tr \cP(\ep^{-\ii\oint d \t L_\bos(\t)} W_{2n}) \rag_1 \equiv \lag W_{2n} \rag_1 $,  up to order $1/k^2$ we find not only $\lag W_2 \rag_1 + \lag W_4 \rag_1 =0$,  but also  $\lag W_2 \rag_1=0$ and $\lag W_4 \rag_1=0$,  separately.  At order $1/k$ this is trivially realized, since only  the $W_2$ correlator can be constructed, but at order $1/k^2$ this is a non--trivial statement.

Repeating the analysis of \cite{Bianchi:2016gpg}, this property can be understood as follows. If we temporarily consider WL operators with non--trivial  winding $m$, as a rule of thumb it can be argued that, independently of framing, a generic perturbative diagram has a polynomial dependence on $m$, whose leading power is $m^{2[r/2]}$ where $r$ is the number of contour insertions, that is the number of contour integrations. Up to two loops this property has been tested explicitly in \cite{Bianchi:2016gpg}.
Focusing on the two--loop diagrams in figure \ref{fd4} and \ref{fd5} contributing to the $W_2$ and $W_4$ correlators, at framing one the only potentially non--vanishing contributions come from diagrams \ref{fdFFp}, \ref{fdAFF} and \ref{fdFFFF}.
These diagrams produce contributions to $W_2$ and $W_4$ correlators that display different leading powers in the winding number.  Precisely, contributions to  $\lag W_2 \rag$ go as $m^2$, while the ones contributing to $\lag W_4 \rag$ go as $m^4$. Thus, at non--trivial winding, cohomological equivalence in ABJ(M) forces the two correlators to vanish separately. We note that at this order what plays a crucial role in  assigning different powers of $m$ to $W_2$ and $W_4$ correlators is the fact that scalar eye--like diagrams \ref{fdSB} and \ref{fdBB} vanish. If this were not the case, since they appear in both correlators the winding argument would get spoiled. However, the eye--like diagram is vanishing at framing zero by analytical continuation in dimensional regularization, and is shown to be framing-independent numerically.

Reinforced by this preliminary result, we can then reasonably believe that a similar pattern may survive at higher orders, so allowing to  conjecture that $\langle W_{2n} \rangle_1=0$ at any order in perturbation theory, with the definition of $W_{2n}$ in (\ref{relation2}).

The validity of these identities implies strong constraints on the (unknown) integrals at a given perturbative order. As a non--trivial example, we consider \eqref{nontrivial} at three loops. At this order we should prove that $\langle W_2 \rangle_1$, $\langle W_4 \rangle_1$ and $\langle W_6 \rangle_1$ vanish separately.

\begin{figure}[htbp]
  \centering
  \subfigure[]{\includegraphics[height=0.16\textwidth]{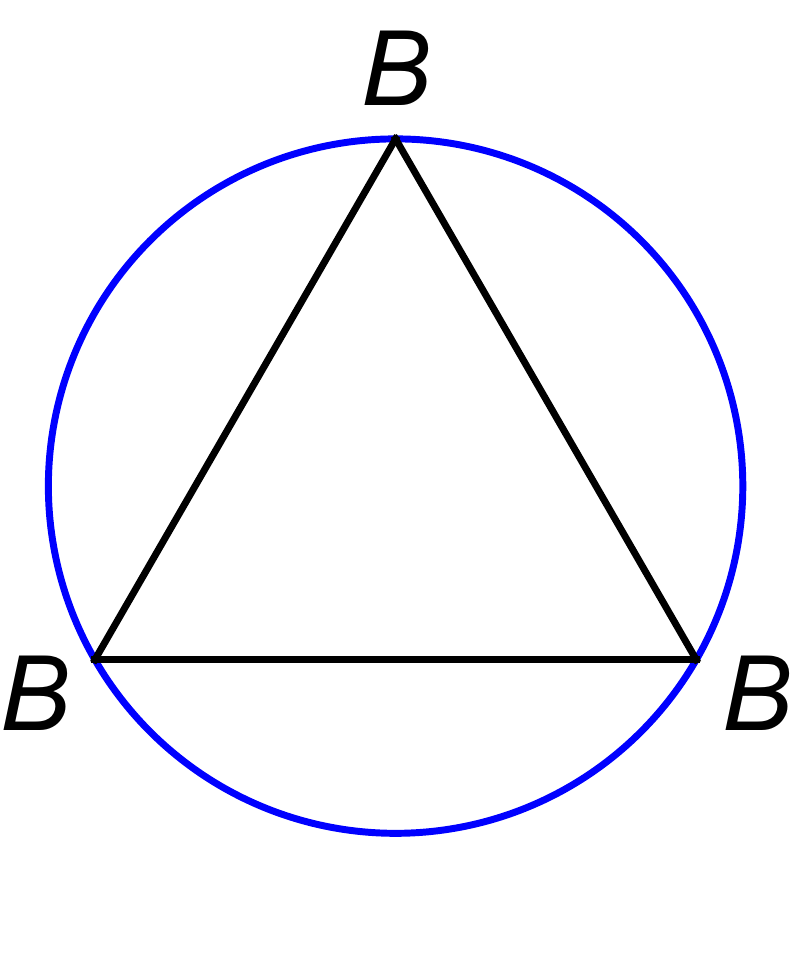}\label{fdBBB}}~~
  \subfigure[]{\includegraphics[height=0.16\textwidth]{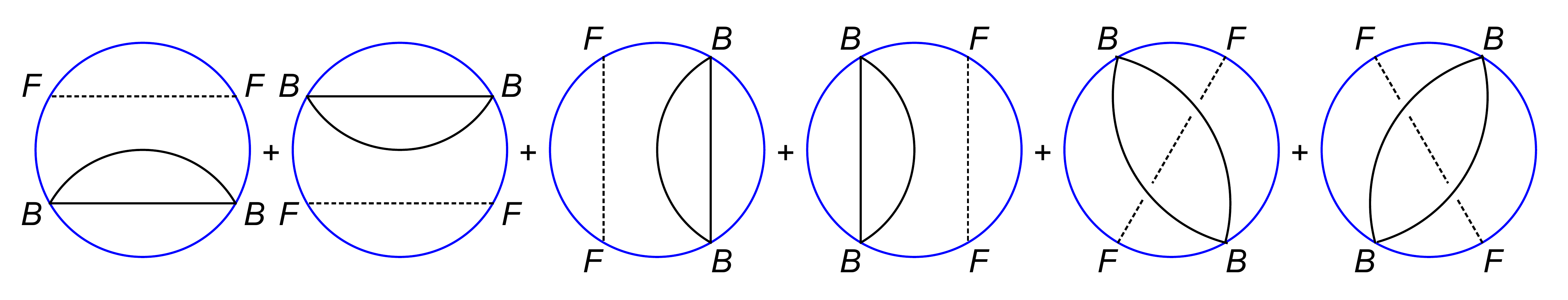}\label{fdBBFF}}\\
  \subfigure[]{\includegraphics[height=0.16\textwidth]{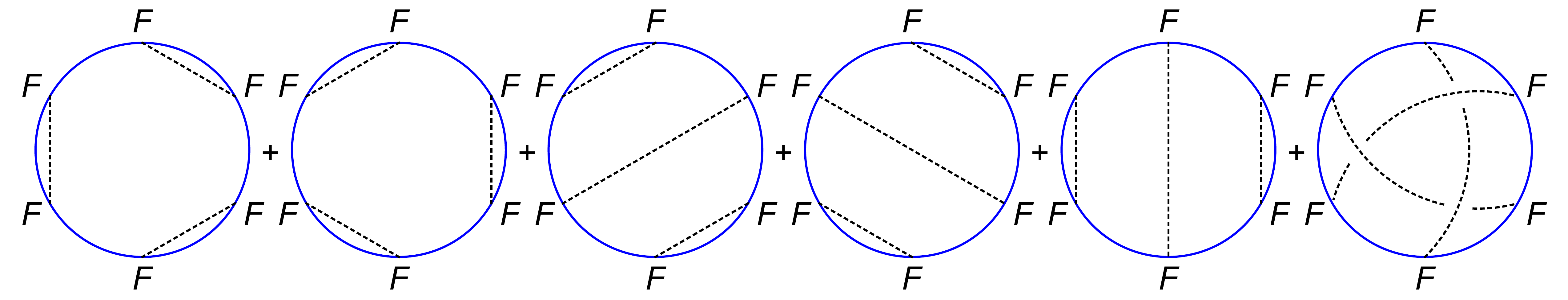}\label{fdFFFFFF}}
  \caption{Three-loop Feynman diagrams for $\lag \Tr \cP(\ep^{-\ii\oint d \t L_\bos(\t)} W_6) \rag_f$.}\label{fd6}
\end{figure}

In particular, focusing on $\langle W_6 \rangle_1$ the corresponding contributions arise from diagrams in figure \ref{fd6}. We note that these diagrams do not contain superpotential vertices, and are then common to all $\mathcal N \geq 2$ SCSM theories, ABJ(M) theory included.

{The corresponding analytic expressions read
\bea
&& \dcircled{\ref{fd6}} =
             \sum_{a,b,c,d} [ \bar m^{ab}_i n^i_{ba} \bar m^{ac}_j n^j_{ca} \bar m^{ad}_k n^k_{da}
                              ( \cI_{\rm\ref{fdBBB}}^{(f)} + \cI_{\rm\ref{fdFFFFFF}}^{(f)} ) \nn\\
&& \hspace{13mm}
                            + \bar m^{ba}_i n^i_{ab} \bar m^{ca}_j n^j_{ac} \bar m^{da}_k n^k_{ad}
                              ( \cI_{\rm\ref{fdBBB}}^{(f)} + \cJ_{\rm\ref{fdFFFFFF}}^{(f)} ) \nn\\
&& \hspace{13mm}
                            + \bar m^{ab}_i n^i_{ba} \bar m^{ac}_j n^j_{ca} \bar m^{db}_k n^k_{bd}
                              ( \cI_{\rm\ref{fdBBFF}}^{(f)} + \cK_{\rm\ref{fdFFFFFF}}^{(f)} ) ] N_a N_b N_c N_d \nn\\
&& \hspace{13mm}
           + \sum_{a,b}(\bar m^{ab}_i n^i_{ba})^3 N_a N_b
                       ( 2 \cI_{\rm\ref{fdBBB}}^{(f)} + \cJ_{\rm\ref{fdBBFF}}^{(f)}+  \cL_{\rm\ref{fdFFFFFF}}^{(f)} )
\eea
where the explicit expression of the integrals can be found in appendix~\ref{appint} (see eq. (\ref{intfd6})).
It follows that conjecture \eqref{nontrivial} for $n=3$ forces the framing--one integrals to satisfy the non--trivial identities
\be\label{framing1bis}
\cI_{\rm\ref{fdBBB}}^{(1)} + \cI_{\rm\ref{fdFFFFFF}}^{(1)} =
\cI_{\rm\ref{fdBBB}}^{(1)} + \cJ_{\rm\ref{fdFFFFFF}}^{(1)} =
\cI_{\rm\ref{fdBBFF}}^{(1)} + \cK_{\rm\ref{fdFFFFFF}}^{(1)} =
2 \cI_{\rm\ref{fdBBB}}^{(1)} + \cJ_{\rm\ref{fdBBFF}}^{(1)} +  \cL_{\rm\ref{fdFFFFFF}}^{(1)} = 0
\ee}%

Again we expect to be able to refine this set of constraints by a direct analysis of the explicit expression of the integrals.
Moreover, it would be important to check these relations by evaluating explicitly all the integrals. This is a quite hard task which goes beyond the scopes of the present paper.

We conclude with an important observation.
If rigorously proved, identities \eqref{nontrivial} would have strong implications for the defect CFT defined on the bosonic Wilson contour.
In fact, for any local or nonlocal operator $\mO$ localized on the contour of the WL, the expression
\be
\lagg \mO \ragg_f \equiv \f{\lag \Tr\mP( \ep^{-\ii\oint d \t L_\bos(\t)} \mO ) \rag_f}{\lag \Tr\mP \ep^{-\ii\oint d \t L_\bos(\t)} \rag_f}
\ee
defines correlations functions in the one-dimensional defect CFT on the circle. Interestingly, the defect CFT would depend on the framing we choose. In this language identities \eqref{nontrivial} would read
\be \label{stronger}
\lagg W_{2n} \ragg_1 = 0, ~~ n=1,2,\cdots
\ee
These equations would then impose strong constraints on the defect CFT with interesting implications on the corresponding boostrap program.

\section{Conclusions}\label{SecCnD}

In this paper we have investigated the class of 1/2 BPS WLs in $\mathcal N = 2$ SCSM theories featured by constant parametric couplings to matter scalars and fermions. Beyond the bosonic WL that contains only couplings to scalars, we have found an infinite family of fermionic WLs.
In general, the corresponding connections cannot be decomposed as double--node connections and cannot be interpreted as superconnections of a supergroup. Nonetheless, the new fermionic 1/2 BPS WLs are classically cohomologically equivalent to the bosonic 1/2 BPS WLs.

In order to exemplify the general results, we have revisited the case of ABJ(M) theory and $\mathcal N =4$ orbifold ABJ(M) in $\mathcal N =2$ language, and studied in details the $\mathcal N =2$ orbifold ABJM theory.
In $\mathcal N =4$ and $\mathcal N =2$ orbifold ABJM theories, some of the newly found BPS WLs can be obtained from the orbifold decomposition of the 1/2 BPS WLs in ABJM theory. For these operators we have identified the corresponding gravity duals by direct orbifolding the brane configurations dual to 1/2 BPS WLs in ABJM theory.
Whether gravity duals of more general BPS WLs can be identified is an important open question that requires further investigation.

We have discussed the cohomological equivalence of the fermionic and bosonic BPS WLs at quantum level by studying their expectation values up to two loops. In fact, since at this order the superpotential couplings do not enter, it happens that the arguments and well-based speculations of \cite{Bianchi:2016gpg} lead to cohomological equivalence in any $\mathcal N =2$ CS-matter models as in ABJ(M) theory.
We have further conjectured that in general the cohomological equivalence may occur in the stronger version of eq.~\eqref{nontrivial}. Since this condition would have far--reaching implications for the defect CFT defined on the bosonic WL contour, we plan to further investigate it in the future.

For the ordinary  WLs along closed curves in gauge theories, we can take the trace in any representation of the gauge group.
In four dimensional $\mN=4$ super Yang-Mills theory, the BPS WLs in higher dimensional representations have elegant holographic description in terms of D-branes
\cite{Rey:1998ik, Drukker:2005kx, Yamaguchi:2006tq, Gomis:2006sb, Gomis:2006im} or bubbling geometries \cite{Yamaguchi:2006te, Lunin:2006xr, DHoker:2007mci}.
For 1/2 BPS fermionic WLs in ABJ(M) theory, the trace can also be taken in higher dimensional representations of the supergroup $U(N_1|N_2)$ \cite{Hatsuda:2013yua}. As we already stressed, in general the connection $L_{\fer}$ constructed here is not a superconnection with respect to a supergroup. This raises the question whether, for these WLs along closed curves, we can take the trace in some analog of higher dimensional representations mentioned above. We would like to leave this interesting question for further work.

\acknowledgments

We would like to thank Jin-Peng An, Nadav Drukker, Si Li, Hong L\"u, Zhao-Long Wang for helpful discussions.
H.O.\ gratefully acknowledges the kind hospitality of Tianjin University.
J.Z.\ would like to thank Peking University and Tianjin University for hospitality, where part of the work has been done.
The work of A.M., S.P., and J.Z.\ has been supported in part by Italian Ministero dell'Istruzione, Universit\`a e Ricerca
(MIUR), and Istituto Nazionale di Fisica Nucleare (INFN) through the ``Gauge Theories, Strings, Supergravity'' (GSS) research
projects.
The work of H.O.\ and J.--B.W.\ has been supported in part by the National Natural Science Foundation of China, Grant No.\ 11575202.
The work of J.Z.\ has been supported by Fondazione Cariplo and Regione Lombardia, Grant No.\ 2015-1253.

\newpage

\appendix

\section{Spinor conventions in three dimensions}\label{AppSpinor}

In this Appendix we collect our spinor conventions, both in Minkowski and Euclidean signatures.

\subsection{Minkowski spacetime}

In three--dimensional Minkowski spacetime we follow the convention in \cite{Ouyang:2015ada,Lietti:2017gtc}, where the reader can find more details. We use coordinates $x^\m=(x^0,x^1,x^2)$ and metric $\eta_{\m\n}=\diag(-++)$. We choose gamma matrices
\be
\g^{\m~\b}_{~\a}=(\ii\s^2,\s^1,\s^3)
\ee
where $\s^{1,2,3}$ are the Pauli matrices.
They satisfy $\g^\m\g^\n=\eta^{\m\n}+\ve^{\m\n\r}\g_\r$ with $\ve^{012}=1$.
Note the spinor index $\a=1,2$.

The charge conjugate of spinors is defined as
\be
\bar\th_\a = \th^*_\a, ~~ \bar\th_\a^*=\th_\a
\ee
The spinor indices are raised and lowered as
\be
\th^\a=\ve^{\a\b}\th_\b \qquad , \qquad  \th_\a=\ve_{\a\b}\th^\b
\ee
where $\ve^{12} = - \ve_{12} = 1$.
We also define the shorthand notation
\be
\th\psi=\th^\a\psi_\a, ~~
(\g^\m\th)_\a = \g^{\m~\b}_{~\a}\th_\b, ~~
(\th\g^\m)^\a = \th^\b\g^{\m~\a}_{~\b}, ~~
\th\g^\m\psi = \th^\a\g^{\m~\b}_{~\a} \psi_\b
\ee


On the straight line $x^\m=(\t,0,0)$, we introduce the bosonic spinors
\be \label{g3}
u_{\pm\a} = \f{1}{\sqrt{2}} \left( \ba{cc} 1 \\ \mp\ii \ea \right) , ~~
u_{\pm}^{\a} = \f{1}{\sqrt{2}} \left( \mp\ii, -1  \right)
\ee
and decompose a generic spinor as
\be \label{spinorspm}
\th_\a = u_{+\a} \th_- + u_{-\a} \th_+
\ee
where $\th_\pm$ are one--component Grassmann numbers. The product of two spinors now reads
\be\label{product}
\th \psi = \ii \left( \th_+ \psi_- - \th_- \psi_+ \right)
\ee

\subsection{Euclidean space}

In Euclidean space we follow the spinor conventions of \cite{Ouyang:2015ada}.
We use coordinates $x^\m=(x^1,x^2,x^3)$ and metric $\d_{\m\n}=\diag(+++)$. We choose gamma matrices
\be \label{gammaeuc}
\g^{\m~\b}_{~\a}=(-\s^2,\s^1,\s^3)
\ee
that satisfy $\g^\m\g^\n=\d^{\m\n}+\ii\ve^{\m\n\r}\g_\r$ with $\ve^{123}=1$.

The spinor indices are raised and lowered as
\be
\th^\a=\ve^{\a\b}\th_\b, ~~ \th_\a=\ve_{\a\b}\th^\b
\ee
where $\ve^{12} = - \ve_{12} = 1$.

In Euclidean space $\bar\th$ and $\th$ are independent spinors. For a general spinor $\th$ one can define $\th^\dagger$ that satisfies
\be
\th_\a^*=\th^{\dagger\a}, ~~ \th^{\a*}=-\th^\dagger_\a, ~~
\th^{\dagger\a*}=\th_\a, ~~ \th^{\dagger*}_\a=-\th^{\a}
\ee
Formally one has $\th^{\dagger\dagger}=-\th$.

We use the shorthand notation
\be
\th\psi=\th^\a\psi_\a, ~~
(\g^\m\th)_\a = \g^{\m~\b}_{~\a}\th_\b, ~~
(\th\g^\m)^\a = \th^\b\g^{\m~\a}_{~\b}, ~~
\th\g^\m\psi = \th^\a\g^{\m~\b}_{~\a} \psi_\b
\ee

For the circle $x^\m=(\cos\t,\sin\t,0)$, we choose the $u_{\pm \a}$ spinors as
\bea \label{g6}
&& u_{+\a} = \f{1}{\sqrt{2}} \left( \ba{cc} \ep^{-\f{\ii\t}2} \\ \ep^{\f{\ii\t}2} \ea \right), ~~
   u_{-\a} = \f{\ii}{\sqrt{2}} \left( \ba{cc} -\ep^{-\f{\ii\t}2} \\ \ep^{\f{\ii\t}2} \ea \right) \nn\\
&& u_{+}^{\a} = \f{1}{\sqrt{2}} \left( \ep^{\f{\ii\t}2}, -\ep^{-\f{\ii\t}2}  \right), ~~
   u_{-}^{\a} = \f{\ii}{\sqrt{2}} \left( \ep^{\f{\ii\t}2}, \ep^{-\f{\ii\t}2}  \right)
\eea
We decompose spinors in Euclidean space formally in the same way as in Minkowski spacetime, see eq. (\ref{spinorspm}). The product of two spinors is still given in \eqref{product}.
However, the $u_\pm$ spinors are defined differently, and in particular in Euclidean space they are not constant.

\section{BPS WLs in ABJ(M) theory}\label{AppABJM}

In this appendix, we review 1/6 and 1/2 BPS WLs in ABJ(M) theory, including both line WLs in Minkowski spacetime and circle WLs in Euclidean space \cite{Drukker:2008zx,Chen:2008bp,Rey:2008bh,Drukker:2009hy,Ouyang:2015iza,Ouyang:2015bmy}. We also reproduce these known WLs using the general construction of sections~\ref{SecLineWL} and~\ref{SecCircleWL} valid for generic $\cN=2$ SCSM theories. This requires a notational translation from what we call ABJ(M) notations to $\cN=2$ notations, which we describe in details. For 1/2 BPS line WLs in Minkowski spacetime we also review the construction of (anti--)M2--brane duals in M--theory.

\subsection{1/6 BPS line WLs in Minkowski spacetime}

\begin{figure}[htbp]
\centering
\subfigure[]{\includegraphics[width=0.3\textwidth]{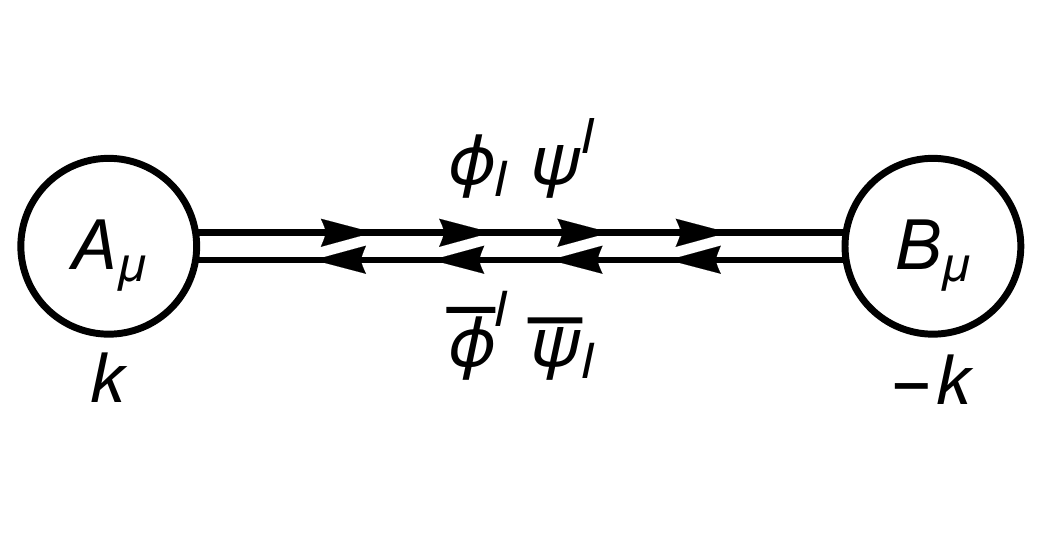} \label{abjm1}}
\subfigure[]{\includegraphics[width=0.3\textwidth]{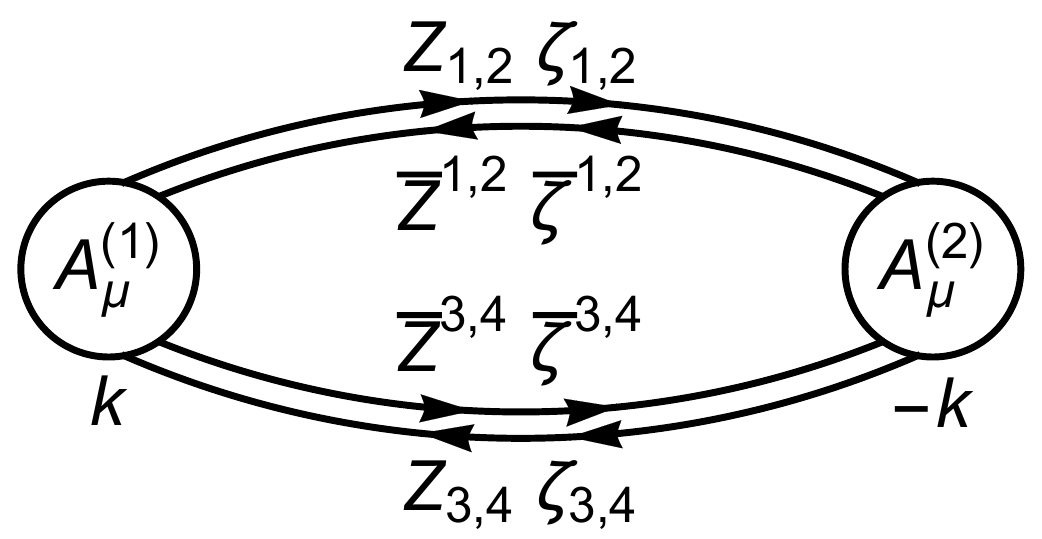} \label{abjm2}}
\caption{The quiver diagram of ABJ(M) theory in (a) ABJM notations and (b) $\cN=2$ notations.}\label{qdabjm}
\end{figure}

The $U(N_1)_k \times U(N_2)_{-k}$ ABJ(M) theory \cite{Aharony:2008ug,Hosomichi:2008jb,Aharony:2008gk} is usually written in manifest $SU(4)$ R--symmetry notations. Gauge fields $A_\m$, $B_\m$ corresponding to the two nodes of
the quiver diagram in figure~\ref{abjm1} are linked by $\phi_I$, $\psi^I$, $I=1,2,3,4$ bosonic and fermionic matter fields in the bifundamental representation of the gauge group and in the fundamental of the R--symmetry group. The SUSY parameters are $\th^{IJ}$, $\bar\th_{IJ}$, $\vth^{IJ}$, $\bar\vth_{IJ}$ with
\bea\label{susycharges}
&& \th^{IJ}=-\th^{JI}, ~~ (\th^{IJ})^*=\bar \th_{IJ}, ~~ \bar\th_{IJ}=\f{1}{2}\e_{IJKL}\th^{KL} \nn\\
&& \vth^{IJ}=-\vth^{JI}, ~~ (\vth^{IJ})^*=\bar \vth_{IJ}, ~~ \bar\vth_{IJ}=\f{1}{2}\e_{IJKL}\vth^{KL}
\eea
Here $\th^{IJ}$, $\bar\th_{IJ}$ are related to Poincar\'e supercharges, and $\vth^{IJ}$, $\bar\vth_{IJ}$ are related to superconformal charges.

In Minkowski spacetime, the bosonic 1/6 BPS WL along the line $x^\m=(\t,0,0)$ is defined as in \eqref{Wbos} with connection matrix \cite{Drukker:2008zx,Chen:2008bp,Rey:2008bh}
\bea \label{WbosABJM}
&& L_\bos = \diag\big( A_0 + \f{2\pi}{k} R^I{}_J \phi_I \bar\phi^J  ,
                    B_0 + \f{2\pi}{k} R^I{}_J \bar\phi^J \phi_I  \big) \nn\\
&& R^I{}_J = \diag(-1,-1,1,1) \label{RIJ}
\eea
Fermionic 1/6 BPS WL $W_\fer$ can also be constructed as in \eqref{Wfer}, which correspond to the superconnection \cite{Ouyang:2015iza,Ouyang:2015bmy}
\be \label{WferABJM}
L_\fer = \lt( \ba{cc} A_0 + \f{2\pi}{k} U^I_{~J} \phi_I\bar\phi^J &
         \sqrt{\f{4\pi}{k}}( \bar\a_I\psi^I_+ + \bar\g_I\psi^I_- )  \\
         \sqrt{\f{4\pi}{k}}( \bar\psi_{I-}\b^I - \bar\psi_{I+}\d^I )  &
         B_0 + \f{2\pi}{k} U^I_{~J} \bar\phi^J \phi_I  \ea \rt)
\ee
where the couplings to matter are featured by constant parameters
\bea \label{UIJaIbIgIdI}
&& U^I_{~J} = \lt( \ba{cccc}
     -1 + 2\b^2\bar\a_2 & -2\b^1\bar\a_2   &                    & \\
     -2\b^2\bar\a_1   & -1 + 2\b^1\bar\a_1 &                    & \\
                      &                  & 1 - 2 \d^4\bar\g_4 & 2\d^3\bar\g_4 \\
                      &                  &  2\d^4\bar\g_3     & 1 - 2 \d^3\bar\g_3
\ea \rt) \\
&& \bar\a_I=(\bar\a_1,\bar\a_2,0,0), ~~ \b^I=(\b^1,\b^2,0,0), ~~
   \bar\g_I=(0,0,\bar\g_3,\bar\g_4), ~~ \d^I=(0,0,\d^3,\d^4) \nn
\eea
satisfying the BPS constraints
\be \label{e36}
\bar\a_{1,2}\d^{3,4} = \bar\g_{3,4}\b^{1,2} = 0
\ee
The corresponding preserved supercharges are $\th^{12}_+,  ~\th^{34}_-,  ~\vth^{12}_+, ~\vth^{34}_- $.

Similarly, along the line $x^\m=(\t,0,0)$ we can define $\td W_\bos$, $\td W_\fer$ operators with connections
\bea \label{tdWbosferABJM}
&& \td L_\bos = \diag\big( A_0 - \f{2\pi}{k} R^I{}_J \phi_I \bar\phi^J  ,
                           B_0 - \f{2\pi}{k} R^I{}_J \bar\phi^J \phi_I  \big) \nn\\
&& \td L_\fer = \lt( \ba{cc} A_0 - \f{2\pi}{k} U^I_{~J} \phi_I\bar\phi^J  &
                             \sqrt{\f{4\pi}{k}}( \bar\a_I\psi^I_- + \bar\g_I\psi^I_+ )  \\
                             \sqrt{\f{4\pi}{k}}( - \bar\psi_{I+}\b^I + \bar\psi_{I-}\d^I )  &
                             B_0 - \f{2\pi}{k} U^I_{~J} \bar\phi^J \phi_I  \ea \rt)
\eea
where the constant parameters are the same as in (\ref{RIJ}), (\ref{UIJaIbIgIdI}) and satisfy the same constraints (\ref{e36}).
The preserved supercharges $\th^{12}_-, ~ \th^{34}_+, ~ \vth^{12}_-, ~ \vth^{34}_+$ are complementary to the ones preserved by $W_\bos, W_\fer$.

$W_\bos$ and $W_\fer$ operators are cohomologically equivalent, and that is their difference is $\cal{Q}$(something), where $\cal{Q}$ is a supercharge preserved by both WLs.
Similarly, one can prove that $\td W_\bos$, $\td W_\fer$ are cohomologically equivalent.

\vskip 12pt
In order to make contact with the general WL construction presented in the main text, we rewrite the ABJ(M) theory in $\cN=2$ superspace formalism. This is obtained by identifying $\cN= 6$ and $\cN=2$ SUSY parameters as
\be
\th^{12}=\bar\th_{34}=\th \qquad \th^{34}=\bar\th_{12}=\bar\th \qquad \vth^{12}=\bar\vth_{34}=\vth \qquad \vth^{34}=\bar\vth_{12}=\bar\vth
\ee

The ABJ(M) theory in $\cN=2$ superspace formalism has only $SU(2)$ R--symmetry invariance manifest and corresponds to the quiver diagram in figure~\ref{abjm2}. The two sets of fields are related by
\bea \label{rlt1}
A_\m = A_\m^{(1)}, \quad
B_\m = A_\m^{(2)}, \qquad
\phi_I = ( Z_1,Z_2,\bar Z^3,\bar Z^4 ), \quad
\psi^I = ( -\z_2,\z_1,-\bar\z^4,\bar\z^3 )
\eea

Using the general construction of WL operators in section~\ref{secn2} and adapting it to the ABJ(M) case, we find that the $W_\bos$ and  $W_\fer$ operators along the line $x^\m=(\t,0,0)$ and preserving supercharges $\th_+,  ~\bar{\th}_-,  ~\vth_+, ~\bar{\vth}_- $, have connections \eqref{genconn}, with the gauge auxiliary fields given by
\bea
&& \s^{(1)} = \f{2\pi}{k} ( Z_1 \bar Z^1 + Z_2 \bar Z^2 - \bar Z^3 Z_3 - \bar Z^4 Z_4 ) \nn\\
&& \s^{(2)} =   \f{2\pi}{k} (  \bar Z^1 Z_1 + \bar Z^2 Z_2 - Z_3 \bar Z^3 - Z_4 \bar Z^4 )
\eea
and matrix couplings
\bea
&& \bar M_Z = \lt( \ba{cc} & \bar m^1 Z_{1} + \bar m^2 Z_{2} \\ \bar m^3 Z_{3} + \bar m^4 Z_{4} & \ea \rt) \nn\\
&& N_{\bar Z} = \lt( \ba{cc} & n_3 \bar Z^3 + n_4 \bar Z^4 \\ n_1 \bar Z^1 + n_2 \bar Z^2 & \ea \rt) \nn\\
&& \bar M_\z = \lt( \ba{cc} & \bar m^1 \z_{1+} + \bar m^2 \z_{2+} \\ \bar m^3 \z_{3+} + \bar m^4 \z_{4+} & \ea \rt) \nn\\
&& N_{\bar \z} = \lt( \ba{cc} & n_3 \bar\z^3_- + n_4 \bar\z^4_- \\ n_1 \bar\z^1_- + n_2 \bar\z^2_- & \ea \rt)
\eea
The coupling parameters satisfy the BPS constraints
\be \label{p2}
\bar m^{1,2} \bar m^{3,4} = n_{1,2} n_{3,4} = 0
\ee
It is now easy to verify that redefining the parameters as
 \bea \label{pt1}
&& \bar m^1 = \sr{\f{4\pi}{k}}\bar\a_2, ~~
   \bar m^2 = - \sr{\f{4\pi}{k}}\bar\a_1, ~~
   \bar m^3 = - \sr{\f{4\pi}{k}} \d^4, ~~
   \bar m^4 = \sr{\f{4\pi}{k}} \d^3  \nn\\
&& n_1 = \sr{\f{4\pi}{k}} \b^2, ~~
   n_2 = -\sr{\f{4\pi}{k}} \b^1, ~~
   n_3 = \sr{\f{4\pi}{k}} \bar\g_4, ~~
   n_4 = -\sr{\f{4\pi}{k}} \bar\g_3
\eea
and using relations (\ref{rlt1}) between the two sets of conventions, we reproduce exactly the ABJ(M) WLs in (\ref{WbosABJM}), (\ref{WferABJM}).
Similarly, from the general construction in section~\ref{secn2} for BPS WLs $\td W_\bos$, $\td W_\fer$ and applying the same notational translation, we reproduce the ABJ(M) WLs with connections (\ref{tdWbosferABJM}). This is a consistency check of our general construction.

\subsection{1/2 BPS line WLs and their gravity duals}\label{Appduals}

For special values of the parameters in \eqref{UIJaIbIgIdI} the number of supercharges preserved by $W_\fer$, $\td W_\fer$ can enhance. For instance, it has been proved \cite{Drukker:2009hy} that connections
\bea \label{W12ABJM}
&& L_{1/2}[\bar\a_I] = \lt( \ba{cc} A_0 + \f{2\pi}{k} \big( \d^I_J - \f{2\a^I\bar\a_J}{|\a|^2} \big) \phi_I\bar\phi^J &
                                    \sqrt{\f{4\pi}{k}}\bar\a_I\psi^I_+ \\
                                    \sqrt{\f{4\pi}{k}} \bar\psi_{I-}\f{\a^I}{|\a|^2} &
                                    B_0 + \f{2\pi}{k} \big( \d^I_J - \f{2\a^I\bar\a_J}{|\a|^2} \big) \bar\phi^J \phi_I \ea \rt) \nn\\
&& \td L_{1/2}[\bar\a_I] = \lt( \ba{cc} A_0 - \f{2\pi}{k} \big( \d^I_J - \f{2\a^I\bar\a_J}{|\a|^2} \big) \phi_I\bar\phi^J &
                                        \sqrt{\f{4\pi}{k}}\bar\a_I\psi^I_- \\
                                      - \sqrt{\f{4\pi}{k}} \bar\psi_{I+}\f{\a^I}{|\a|^2} &
                                        B_0 - \f{2\pi}{k} \big( \d^I_J - \f{2\a^I\bar\a_J}{|\a|^2} \big) \bar\phi^J \phi_I \ea \rt)
\eea
with constant parameters
\be \label{aI}
\bar\a_I = (\bar\a_1,\bar\a_2,\bar\a_3,\bar\a_4), ~~ \a^I \equiv (\bar\a_I)^*, ~~ |\a|^2 \equiv \bar\a_I\a^I \neq 0
\ee
give rise to 1/2 BPS fermionic operators $W_{1/2}[\bar\a_I]$, $\td W_{1/2}[\bar\a_I]$ that preserve complementary supercharges
\bea
&& \bar\a_I\th^{IJ}_+, ~~ \e_{IJKL}\a^J\th^{KL}_-, ~~ \bar\a_I\vth^{IJ}_+, ~~ \e_{IJKL}\a^J\vth^{KL}_- \nn\\
&& \bar\a_I\th^{IJ}_-, ~~ \e_{IJKL}\a^J\th^{KL}_+, ~~ \bar\a_I\vth^{IJ}_-, ~~ \e_{IJKL}\a^J\vth^{KL}_+
\eea
These operators can be obtained from \eqref{WferABJM} and \eqref{tdWbosferABJM} with constant parameters \eqref{UIJaIbIgIdI} by setting $\b^{1,2} = \frac{\a^{1,2}}{|\a|^2}$, $\bar\g_{3,4} = \d^{3,4} = 0$ and performing a R--symmetry rotation $(\bar\a_1, \bar\a_2, 0,0) \to
(\bar\a_1, \bar\a_2, \bar\a_3 , \bar\a_4)$.

\vskip 10pt
For these 1/2 BPS operators gravity duals have been found \cite{Drukker:2009hy,Lietti:2017gtc}, which we now review.

The ABJM theory is dual to M--theory in $\gabjm$ background
\be
ds^2=R^2 \Big( \frac14 ds^2_{\AdS_4}+ds^2_{ {\rm S}^7/\Z_k} \Big)
\ee
where the metric of $\AdS_4$ is
\be \label{ads4}
ds^2_{\AdS_4}= u^2 (-dt^2 + dx_1^2 + dx_2^2) + \f{du^2}{u^2}
\ee
and embedding S$^7$ in ${\rC}^4$ as \cite{Drukker:2008zx}
\bea \label{z1234}
&& z_1 = \cos\frac{\b}{2}\cos\frac{\theta_1}{2} \ep^{\ii\xi_1}, ~~
   \xi_1 = -\frac14(2\phi_1+\chi+\zeta)                              \nn\\
&& z_2 = \cos\frac{\b}{2}\sin\frac{\theta_1}{2} \ep^{\ii\xi_2}, ~~
   \xi_2 = -\frac14(-2\phi_1+\chi+\zeta)                             \nn\\
&& z_3 = \sin\frac{\b}{2}\cos\frac{\theta_2}{2} \ep^{\ii\xi_3}, ~~
   \xi_3 = -\frac14(2\phi_2-\chi+\zeta)                              \nn\\
&& z_4 = \sin\frac{\b}{2}\sin\frac{\theta_2}{2} \ep^{\ii\xi_4}, ~~
   \xi_4 = -\frac14(-2\phi_2-\chi+\zeta)
\eea
we can write
\bea \label{s7}
&& ds^2_{\rm{S}^7}=\frac14 \Big[ d\b^2+\cos^2\frac{\b}{2} \big( d\theta_1^2+\sin^2\theta_1 d\varphi_1^2 \big)
                                +\sin^2\frac{\b}{2} \big( d\theta_2^2+\sin^2\theta_2 d\varphi_2^2 \big) \nn\\
&& \phantom{ds^2_{S^7}=}
+\sin^2\frac{\b}{2}\cos^2\frac{\b}{2} (d\chi+\cos\theta_1d\varphi_1-\cos\theta_2d\varphi_2)^2 \nn\\
&& \phantom{ds^2_{S^7}=}
+ \Big( \frac12d\zeta+\cos^2\frac{\b}2\cos\theta_1d\varphi_1+\sin^2\frac{\b}2\cos\theta_2d\varphi_2+\frac12\cos\b d\chi \Big)^2
\Big]
\eea
Here $\b,\th_{1,2} \in [0, \pi]$, $\xi_{1,2,3,4} \in [0,2\pi]$, so that $\phi_{1,2} \in [0, 2\pi]$, $\chi\in [0, 4\pi]$, $\zeta \in
[0, 8\pi]$. The M--theory cycle is taken along the $\z$ direction.

The orbifold projection is realized by the $\rZ_k$ identification
\be\label{Zkorb}
(z_1,z_2,z_3,z_4)\sim\ep^{\f{2\pi\ii}{k}}(z_1,z_2,z_3,z_4)
\ee
or equivalently  $\z \sim \z - \f{8\pi}{k}$.

The general solution to the Killing spinor equations in M-theory on $\AdS_4\times\rS^7$ background reads \cite{Drukker:2008zx,Lietti:2017gtc}
\be \label{ksads4s7}
\e = u^{\f12}h(\e_1+x^{\m}\g_{\m}\e_2) - u^{-\f12} \g_3 h \e_2 \qquad \quad \mu = 0,1,2
\ee
where $\e_1$, $\e_2$ are two constant Majorana spinors satisfying $\g^{012}\e_i=\e_i$, $i=1,2$, and
\be\label{eq:h}
h = \ep^{\frac{\b}{4}(\g_{34}-\g_{7\natural})}
       \ep^{\frac{\theta_1}4 (\g_{35}-\g_{8\natural})}
       \ep^{\frac{\theta_2}4(\g_{46}+\g_{79}) }
       \ep^{\frac{\xi_1}2\g_{3\natural}}
       \ep^{\frac{\xi_2}{2}\g_{58}}
       \ep^{\frac{\xi_3}2\g_{47}}
       \ep^{\frac{\xi_4}2\g_{69}}
\ee
Here $\g_A$, $A=0,\cdots,9,\natural$ are the eleven dimensional gamma matrices, and $\g_{0123456789\natural}=1$.

We decompose $\g_A$ in terms of gamma matrices $\g_{a=0,1,2}$ in $\rR^{1,2}$ and $\G_{p=3,\cdots,9,\natural}$ in $\rC^4\cong\rR^8$ as
\bea
&& \g_{a} = -\g_{a} \otimes \G, ~~ a = 0,1,2 \nn\\
&& \g_{p} = \bo \otimes \G_{p}, ~~ p = 3,\cdots,9,\natural
\eea
Correspondingly, $\e_1, \e_2$ get decomposed into direct products of Grassmann odd spinors $\th$, $\vth$ in R$^{1,2}$ and Grassmann
even spinors $\eta$ in $\rm{C}^4\cong\rm{R}^8$
\be\label{decomposition}
\e_1 \sim \th \otimes \eta \qquad \e_2 \sim \vth \otimes \eta
\ee
The $\e_1$ decomposition is related to Poincar\'e supercharges $\th$ and the $\e_2$ one is related to
superconformal charges $\vth$.

We write the $\eta$ spinors in terms of gamma matrix eigenstates (a similar procedure applies also to $\vth$)
\be\label{dec1}
\G_{3\natural} \eta = \ii t_1 \eta, ~~
\G_{58} \eta = \ii t_2 \eta, ~~
\G_{47} \eta = \ii t_3 \eta, ~~
\G_{69} \eta = \ii t_4 \eta
\ee
with $t_I = \pm$ for $I=1,2,3,4$. They satisfy $t_1t_2t_3t_4=1$ as a consequence of  the constraint $\g^{012}\e_1=\e_1$.
The $\e_i$ Killing spinors can then be expressed as a linear combination of eight eigenstates
\bea \label{t1t2t3t4}
&& (t_1,t_2,t_3,t_4) = (++++), (++--), (+-+-), (+--+), \nn\\
&& \phantom{(t_1,t_2,t_3,t_4) = }
                       (-++-), (-+-+), (--++), (----)
\eea%
where each of them corresponds to one real degree of freedom. In the order of (\ref{t1t2t3t4}), we then write
\be\label{sum}
\e_1 = \sum_{i=1}^8 \th^i \otimes \eta_i \quad , \quad \e_2 = \sum_{i=1}^8 \vth^i \otimes \eta_i
\ee
For the Killing spinor (\ref{ksads4s7}), the quotient \eqref{Zkorb} leads to the constraint
\be \mL_{\partial_\zeta}\e=0 \ee
which gives
\be
( \g_{3\natural} + \g_{58} + \g_{47} + \g_{69} ) \e_i = 0, ~~ i=1,2
\ee
which on the decomposition (\ref{dec1}) translates into
\be
t_1 + t_2 + t_3 + t_4 = 0
\ee
The surviving states in (\ref{t1t2t3t4}) are then
\bea
&& (t_1,t_2,t_3,t_4) = (++--), (+-+-), (+--+), \nn\\
&& \phantom{(t_1,t_2,t_3,t_4) = }
                       (-++-), (-+-+), (--++)
\eea
Defining
\bea \label{definabjm}
&& \th^2 = \th^{12} = -\th^{21}, ~~
   \th^3 = \th^{13} = -\th^{31}, ~~
   \th^4 = \th^{14} = -\th^{41} \nn\\
&& \th^5 = \th^{23} = -\th^{32}, ~~
   \th^6 = -\th^{24} = \th^{42}, ~~
   \th^7 = \th^{34} = -\th^{43} \nn\\
&& \vth^2 = \vth^{12} = -\vth^{21}, ~~
   \vth^3 = \vth^{13} = -\vth^{31}, ~~
   \vth^4 = \vth^{14} = -\vth^{41} \nn\\
&& \vth^5 = \vth^{23} = -\vth^{32}, ~~
   \vth^6 = -\vth^{24} = \vth^{42}, ~~
   \vth^7 = \vth^{34} = -\vth^{43} \nn\\
&& \eta_2 = \eta_{12} = -\eta_{21}, ~~
   \eta_3 = \eta_{13} = -\eta_{31}, ~~
   \eta_4 = \eta_{14} = -\eta_{41} \nn\\
&& \eta_5 = \eta_{23} = -\eta_{32}, ~~
   \eta_6 = -\eta_{24} = \eta_{42}, ~~
   \eta_7 = \eta_{34} = -\eta_{43}
\eea
we obtain
\be
\e_1 = \f12 \th^{I} \otimes \eta_{I} \quad , \quad
\e_2 = \f12 \vth^{I} \otimes \eta_{I}
\ee
where $I$ runs from 2 to 7.

The 1/2 BPS operator $W_{1/2}[\bar{\a}_I]$ along the line $x^\m=(\t,0,0)$ and corresponding to the superconnection $L_{1/2}[\bar{\a}_I]$ in \eqref{W12ABJM} is dual to an M2--brane embedded as
$t=\sigma^0, x_1=x_2=0, u=\sigma^1, \zeta=\sigma^2$ and localized at a point specified by the complex vector \cite{Lietti:2017gtc}
\bea\label{unit}
&& \f{\a^I}{|\a|} = \Big(\cos\frac{\b}{2}\cos\frac{\theta_1}{2} \ep^{-\frac{\ii}{4}(2\phi_1+\chi+\zeta)  },
        \cos\frac{\b}{2}\sin\frac{\theta_1}{2} \ep^{-\frac{\ii}{4}(-2\phi_1+\chi+\zeta) }, \nn \\
&& \phantom{\f{\a^I}{|\a|} = }
        \sin\frac{\b}{2}\cos\frac{\theta_2}{2} \ep^{-\frac{\ii}{4}(2\phi_2-\chi+\zeta) },
        \sin\frac{\b}{2}\sin\frac{\theta_2}{2} \ep^{-\frac{\ii}{4}(-2\phi_2-\chi+\zeta) } \Big)
\eea
The 1/2 BPS WL $\td W_{1/2}[\bar{\a}_I]$ is dual to an anti--M2--brane at the same position that is specified by $\bar{\a}_I$.

The gravity duals of the 1/2 BPS WLs in ABJ(M) theory are helpful for identifying the gravity duals of some BPS WLs in $\mN=4$ orbifold ABJM theory \cite{Lietti:2017gtc,Mauri:2017whf}, as we will review in the next appendix, and for identifying the gravity duals of some WLs in $\mN=2$ orbifold ABJM theory as described in section~\ref{secn2abjm}.

\subsection{Circle WLs in Euclidean space}

In the Euclidean version of the ABJ(M) theory we can define BPS WLs along the circle
\be
x^\m=( \cos\t, \sin\t, 0 )
\ee
In ABJ(M) notations, 1/6 BPS operators $W_\bos$, $W_\fer$, $\td W_\bos$, $\td W_\fer$ correspond to superconnections
\bea
&& L_\bos = \diag\big( A_\m \dot x^\m - \f{2\pi\ii}{k} R^I{}_J \phi_I \bar\phi^J  ,
                       B_\m \dot x^\m - \f{2\pi\ii}{k} R^I{}_J \bar\phi^J \phi_I   \big) \nn\\
&& L_\fer = \lt( \ba{cc} A_\m \dot x^\m - \f{2\pi\ii}{k} U^I_{~J} \phi_I\bar\phi^J  &
                         \sqrt{\f{4\pi}{k}}( \bar\a_I \psi^I_+ + \bar\g_I \psi^I_- )  \\
                         \sqrt{\f{4\pi}{k}} (  - \bar\psi_{I-} \b^I + \bar\psi_{I+} \d^I )  &
                         B_\m \dot x^\m - \f{2\pi\ii}{k} U^I_{~J} \bar\phi^J \phi_I  \ea \rt) \nn\\
&& \td L_\bos = \diag\big( A_\m \dot x^\m + \f{2\pi\ii}{k} R^I{}_J \phi_I \bar\phi^J  ,
                           B_\m \dot x^\m + \f{2\pi\ii}{k} R^I{}_J \bar\phi^J \phi_I  \big) \nn\\
&& \td L_\fer = \lt( \ba{cc} A_\m \dot x^\m + \f{2\pi\ii}{k} U^I_{~J} \phi_I\bar\phi^J  &
                             \sqrt{\f{4\pi}{k}}( \bar\a_I \psi^I_- + \bar\g_I \psi^I_+ )  \\
                             \sqrt{\f{4\pi}{k}} ( \bar\psi_{I+} \b^I - \bar\psi_{I-} \d^I )  &
                             B_\m \dot x^\m + \f{2\pi\ii}{k} U^I_{~J} \bar\phi^J \phi_I  \ea \rt)
\eea
with the same constant parameters $R^I{}_J,U^I{}_J,\bar\a_I,\b^I,\bar\g_I,\d^I$ in (\ref{RIJ}), (\ref{UIJaIbIgIdI}).
$W_\bos$, $W_\fer$ operators preserve supercharges
\be
\vth^{12} = - \ii \g_3 \th^{12}, ~~ \vth^{34} = \ii \g_3 \th^{34}
\ee
whereas $\td W_\bos$, $\td W_\fer$ preserve the complementary set
\be
\vth^{12} = \ii \g_3 \th^{12}, ~~ \vth^{34} = - \ii \g_3 \th^{34}
\ee
$W_\bos$, $\td W_\bos$ are related by a R-symmetry rotation $I=1,2 \lra I = 3,4$, whereas the $W_\fer$, $\td W_\fer$ operators are related by a R-symmetry rotation $I=1,2 \lra I = 3,4$ plus a parameter redefinition $\bar\a_I \lra \bar\g_I$, $\b^I \lra \d^I$.

1/2 BPS operators $W_{1/2}[\bar\a_I]$, $\td W_{1/2}[\bar\a_I]$ can also be defined in Euclidean signature. They correspond to connections
\bea \label{EucW12ABJM}
&& \hspace{-5mm}
   L_{1/2}[\bar\a_I] = \lt( \ba{cc}
   A_\m \dot x^\m - \f{2\pi\ii}{k} \big( \d^I_J - \f{2\a^I\bar\a_J}{|\a|^2} \big) \phi_I\bar\phi^J  &
   \sqrt{\f{4\pi}{k}}\bar\a_I \psi^I_+  \\
  -\sqrt{\f{4\pi}{k}} \bar\psi_{I-} \f{\a^I}{|\a|^2}  &
   B_\m \dot x^\m - \f{2\pi\ii}{k} \big( \d^I_J - \f{2\a^I\bar\a_J}{|\a|^2} \big) \bar\phi^J \phi_I \ea \rt) \nn\\
&& \hspace{-5mm}
   \td L_{1/2}[\bar\a_I] = \lt( \ba{cc}
   A_\m \dot x^\m + \f{2\pi\ii}{k} \big( \d^I_J - \f{2\a^I\bar\a_J}{|\a|^2} \big) \phi_I\bar\phi^J  &
   \sqrt{\f{4\pi}{k}}\bar\a_I \psi^I_-  \\
   \sqrt{\f{4\pi}{k}} \bar\psi_{I+} \f{\a^I}{|\a|^2}   &
   B_\m \dot x^\m + \f{2\pi\ii}{k} \big( \d^I_J - \f{2\a^I\bar\a_J}{|\a|^2} \big) \bar\phi^J \phi_I  \ea \rt)
\eea
The $W_{1/2}[\bar\a_I]$ operator preserves supercharges
\be
\bar\a_I\vth^{IJ} = - \ii \bar\a_I \g_3 \th^{IJ}, ~~ \e_{IJKL}\a^J\vth^{KL} = \ii \g_3 \e_{IJKL}\a^J\th^{KL}
\ee
while $\td W_{1/2}[\bar\a_I]$ preserves complementary supercharges
\be
\bar\a_I\vth^{IJ} = \ii \bar\a_I \g_3 \th^{IJ}, ~~ \e_{IJKL}\a^J\vth^{KL} = - \ii \g_3 \e_{IJKL}\a^J\th^{KL}
\ee

\section{BPS WLs in $\cN = 4$ orbifold ABJM theory}\label{AppN=4}

As a further check of our general construction we now apply it to the case of  $\cN = 4$ orbifold ABJM theory and show that, under a suitable change of notations, it reproduces the known 1/4 BPS WLs found in \cite{Mauri:2017whf}.

\begin{figure}[htbp]
\centering
\subfigure[]{\includegraphics[width=0.6\textwidth]{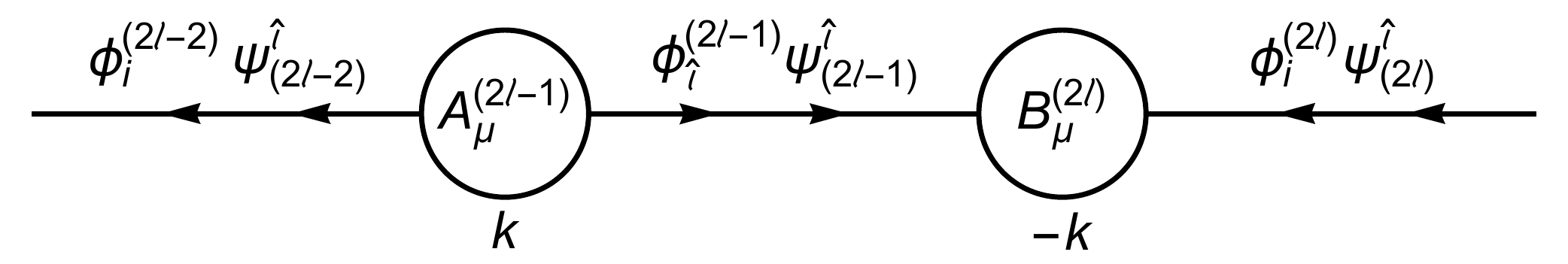} \label{n4abjm1}}
\subfigure[]{\includegraphics[width=0.6\textwidth]{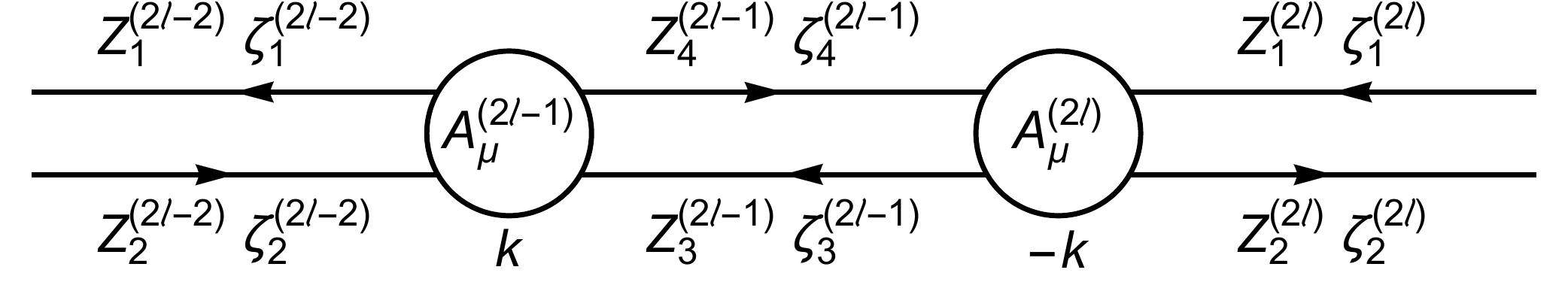} \label{n4abjm2}}
\caption{The quiver diagram of $\cN = 4$ orbifold ABJM theory in (a) ABJM notations and (b) $\cN=2$ notations.}
\end{figure}

General circular quiver $\cN = 4$ SCSM theories were constructed in \cite{Gaiotto:2008sd,Hosomichi:2008jd}, while the special case of $\cN = 4$ orbifold ABJM theory was introduced in \cite{Benna:2008zy}. For gauge group $[U(N)_k\times U(N)_{-k}]^r$ it can be obtained by applying a $\rZ_r$ quotient to the $U(rN)_k\times U(rN)_{-k}$ ABJM theory. The $SU(4)$ R-symmetry is broken to $SU(2)\times SU(2)$, and we decompose the R--symmetry index as
\be
I = 1,2,4,3 \to i=1,2, ~ \hi=\ho,\hw
\ee
We can write the theory in ABJM notations as in figure~\ref{n4abjm1},  or in $\cN=2$ notations, as in figure~\ref{n4abjm2}, under the supercharge identifications $\th^{1\ho}=\bar\th_{2\hw}=\th$, $\th^{2\hw}=\bar\th_{1\ho}=\bar\th$, $\vth^{1\ho}=\bar\vth_{2\hw}=\vth$. The two notations are related by
\bea \label{rlt2}
&& A_\m^{(2\ell-1)} = A_\m^{(2\ell-1)}, ~~
   B_\m^{(2\ell)} = A_\m^{(2\ell)} \nn\\
&& \phi_i^{(2\ell)} = (Z_1^{(2\ell)},\bar Z^2_{(2\ell)}), ~~
   \phi_\hi^{(2\ell-1)} = (Z_4^{(2\ell-1)},\bar Z^3_{(2\ell-1)}) \nn\\
&& \psi^\hi_{(2\ell)} = (\z_1^{(2\ell)},\bar \z^2_{(2\ell)}), ~~
   \psi^i_{(2\ell-1)} = (-\z_4^{(2\ell-1)},-\bar \z^3_{(2\ell-1)})
\eea

From the results in section~\ref{secn2} we obtain the 1/4 BPS WL $W_\bos$, $W_\fer$ in $\cN=2$ notations with connections
\bea
&& L_\bos = \diag( A^{(1)}_\m - \s^{(1)} , A^{(2)} - \s^{(2)}, \cdots,  A^{(2r)} - \s^{(2r)}) \nn\\
&& L_\fer = L_\bos + B + F, ~~ 
   B = \bar M_Z N_{\bar Z} + N_{\bar Z} \bar M_Z, ~~
   F = \bar M_\z + N_{\bar \z}
\eea
with
\bea
&& \s^{(2\ell-1)} = \f{2\pi}{k} ( Z_1^{(2\ell-2)} \bar Z^1_{(2\ell-2)}
                  + Z_4^{(2\ell-1)} \bar Z^4_{(2\ell-1)}
                  - \bar Z^2_{(2\ell-2)} Z_2^{(2\ell-2)}
                  - \bar Z^3_{(2\ell-1)} Z_3^{(2\ell-1)} ) \nn\\
&& \s^{(2\ell)} = \f{2\pi}{k} ( \bar Z^1_{(2\ell)} Z_1^{(2\ell)}
                + \bar Z^4_{(2\ell-1)} Z_4^{(2\ell-1)}
                - Z_2^{(2\ell)} \bar Z^2_{(2\ell)}
                - Z_3^{(2\ell-1)} \bar Z^3_{(2\ell-1)} )
\eea
The nonvanishing blocks of the matrices $\bar M_Z$, $N_{\bar Z}$, $\bar M_\z$, $N_{\bar \z}$ are
\bea
&& [\bar M_Z]_{(2\ell-1,2\ell)} = \bar m^4_{(2\ell-1)} Z_4^{(2\ell-1)}, ~~
   [\bar M_Z]_{(2\ell,2\ell+1)} = \bar m^2_{(2\ell)} Z_2^{(2\ell)} \nn\\
&& [\bar M_Z]_{(2\ell,2\ell-1)} = \bar m^3_{(2\ell-1)} Z_3^{(2\ell-1)}, ~~
   [\bar M_Z]_{(2\ell+1,2\ell)} = \bar m^1_{(2\ell)} Z_1^{(2\ell)} \nn\\
&& [N_{\bar Z}]_{(2\ell-1,2\ell)} = n_3^{(2\ell-1)} \bar Z^3_{(2\ell-1)}, ~~
   [N_{\bar Z}]_{(2\ell,2\ell+1)} = n_1^{(2\ell)} \bar Z^1_{(2\ell)} \nn\\
&& [N_{\bar Z}]_{(2\ell,2\ell-1)} = n_4^{(2\ell-1)} \bar Z^4_{(2\ell-1)}, ~~
   [N_{\bar Z}]_{(2\ell+1,2\ell)} = n_2^{(2\ell)} \bar Z^2_{(2\ell)} \nn\\
&& [\bar M_\z]_{(2\ell-1,2\ell)} = \bar m^4_{(2\ell-1)} \z_{4+}^{(2\ell-1)}, ~~
   [\bar M_\z]_{(2\ell,2\ell+1)} = \bar m^2_{(2\ell)} \z_{2+}^{(2\ell)} \nn\\
&& [\bar M_\z]_{(2\ell,2\ell-1)} = \bar m^3_{(2\ell-1)} \z_{3+}^{(2\ell-1)}, ~~
   [\bar M_\z]_{(2\ell+1,2\ell)} = \bar m^1_{(2\ell)} \z_{1+}^{(2\ell)} \nn\\
&& [N_{\bar \z}]_{(2\ell-1,2\ell)} = n_3^{(2\ell-1)} \bar \z^3_{(2\ell-1)-}, ~~
   [N_{\bar \z}]_{(2\ell,2\ell+1)} = n_1^{(2\ell)} \bar \z^1_{(2\ell)-} \nn\\
&& [N_{\bar \z}]_{(2\ell,2\ell-1)} = n_4^{(2\ell-1)} \bar \z^4_{(2\ell-1)-}, ~~
   [N_{\bar \z}]_{(2\ell+1,2\ell)} = n_2^{(2\ell)} \bar \z^2_{(2\ell)-}
\eea
with constraints on the parameters
\bea
&& \bar m^4_{(2\ell-1)} \bar m^3_{(2\ell-1)} =
   \bar m^4_{(2\ell-1)} \bar m^2_{(2\ell-2)} =
   \bar m^4_{(2\ell-1)} \bar m^2_{(2\ell)} = 0 \nn\\
&& \bar m^1_{(2\ell)} \bar m^2_{(2\ell)} =
   \bar m^1_{(2\ell)} \bar m^3_{(2\ell-1)} =
   \bar m^1_{(2\ell)} \bar m^3_{(2\ell+1)} = 0 \nn\\
&& n_3^{(2\ell-1)} n_4^{(2\ell-1)} =
   n_3^{(2\ell-1)} n_1^{(2\ell-2)} =
   n_3^{(2\ell-1)} n_1^{(2\ell)} = 0 \nn\\
&& n_2^{(2\ell)} n_1^{(2\ell)} =
   n_2^{(2\ell)} n_4^{(2\ell-1)} =
   n_2^{(2\ell)} n_4^{(2\ell+1)} = 0
\eea

By redefining the parameters as
\bea
&& \bar m^1_{(2\ell)} = \sr{\f{4\pi}{k}} \bar\a^{(2\ell)}_\ho , ~~
   \bar m^2_{(2\ell)} = - \sr{\f{4\pi}{k}} \d_{(2\ell)}^\hw \nn\\
&& \bar m^3_{(2\ell-1)} = \sr{\f{4\pi}{k}} \d_{(2\ell-1)}^2 , ~~
   \bar m^4_{(2\ell-1)} = - \sr{\f{4\pi}{k}} \bar\a^{(2\ell-1)}_1 \nn\\
&& n_1^{(2\ell)} = \sr{\f{4\pi}{k}} \b^\ho_{(2\ell)}, ~~
   n_2^{(2\ell)} = \sr{\f{4\pi}{k}} \bar\g_\hw^{(2\ell)} \nn\\
&& n_3^{(2\ell-1)} = -\sr{\f{4\pi}{k}} \bar\g_2^{(2\ell-1)} , ~~
   n_4^{(2\ell-1)} = -\sr{\f{4\pi}{k}} \b^1_{(2\ell-1)}
\eea
and taking into account relations (\ref{rlt2}), we can write the 1/4 BPS WL along the line $x^\m=(\t,0,0)$ in ABJM notations.

The connection of the bosonic 1/4 BPS WL $W_\bos$ reads
\be
L_\bos = \diag( \cA^{(1)}, \cB^{(2)}, \cdots, \cA^{(2r-1)}, \cB^{(2r)} )
\ee
whereas, for $r \geq 3$ the connection of the fermionic 1/4 BPS WL $W_\fer$ is
\be
L_\fer = \lt(\ba{cccccccc}
\cA^{(1)}    & f_1^{(1)}  & h_1^{(1)} &           &              &              & h_2^{(2r-1)} & f_2^{(2r)} \\
f_2^{(1)}    & \cB^{(2)}  & f_1^{(2)} & h_1^{(2)} &              &              &              & h_2^{(2r)} \\
h_2^{(1)}    & f_2^{(2)}  & \cA^{(3)} & f_1^{(3)} & \ddots       &              &              & \\
             & h_2^{(2)}  & f_2^{(3)} & \cB^{(4)} & \ddots       & \ddots       &              & \\
             &            & \ddots    & \ddots    & \ddots       & \ddots       & h_1^{(2r-3)} & \\
             &            &           & \ddots    & \ddots       & \ddots       & f_1^{(2r-2)} & h_1^{(2r-2)} \\
h_1^{(2r-1)} &            &           &           & h_2^{(2r-3)} & f_2^{(2r-2)} & \cA^{(2r-1)} & f_1^{(2r-1)} \\
f_1^{(2r)}   & h_1^{(2r)} &           &           &              & h_2^{(2r-2)} & f_2^{(2r-1)} & \cB^{(2r)}
\ea\rt)
\ee
and for $r=2$ it is
\be
L_\fer = \lt(\ba{cccc}
\cA^{(1)}           & f_1^{(1)}           & h_1^{(1)}+h_2^{(3)} & f_2^{(4)}           \\
f_2^{(1)}           & \cB^{(2)}           & f_1^{(2)}           & h_1^{(2)}+h_2^{(4)} \\
h_1^{(3)}+h_2^{(1)} & f_2^{(2)}           & \cA^{(3)}           & f_1^{(3)} \\
f_1^{(4)}           & h_1^{(4)}+h_2^{(2)} & f_2^{(3)}           & \cB^{(4)} \\
\ea\rt)
\ee
Here we have defined
\bea
&& \cA^{(2\ell-1)} =  A_0^{(2\ell-1)}
   - \f{2\pi}{k} \Big( U_{(2\ell-1)}{}^i_{~j} \phi_i^{(2\ell-2)}\bar\phi^j_{(2\ell-2)}
                      +U_{(2\ell-1)}{}^\hi_{~\hj} \phi_\hi^{(2\ell-1)}\bar\phi^\hj_{(2\ell-1)}
                 \Big) \nn\\
&& \cB^{(2\ell)} =  B_0^{(2\ell)}
   - \f{2\pi}{k} \Big( U_{(2\ell)}{}^i_{~j} \bar\phi^j_{(2\ell)}\phi_i^{(2\ell)}
                      +U_{(2\ell)}{}^\hi_{~\hj}\bar\phi^\hj_{(2\ell-1)}\phi_\hi^{(2\ell-1)}
                 \Big) \nn\\
&& f_1^{(2\ell-1)} = \sqrt{\f{4\pi}{k}} \Big( \bar\a_1^{(2\ell-1)}\psi^1_{(2\ell-1)+}
                                            + \bar\g_2^{(2\ell-1)}\psi^2_{(2\ell-1)-} \Big)\nn\\
&& f_1^{(2\ell)} = \sqrt{\f{4\pi}{k}} \Big( \bar \psi_{\ho-}^{(2\ell)} \b^\ho_{(2\ell)}
                                          - \bar \psi_{\hw+}^{(2\ell)} \d^\hw_{(2\ell)} \Big) \nn\\
&& f_2^{(2\ell-1)} = \sqrt{\f{4\pi}{k}} \Big( \bar \psi_{1-}^{(2\ell-1)} \b^1_{(2\ell-1)}
                                            - \bar \psi_{2+}^{(2\ell-1)} \d^2_{(2\ell-1)} \Big)\nn\\
&& f_2^{(2\ell)} = \sqrt{\f{4\pi}{k}} \Big( \bar\a_\ho^{(2\ell)} \psi^\ho_{(2\ell)+}
                                          + \bar\g_\hw^{(2\ell)} \psi^\hw_{(2\ell)-} \Big) \nn\\
&& h_1^{(2\ell-1)} = -\f{2\pi}{k} U_{(2\ell-1)}{}^\hi_{~j} \phi_\hi^{(2\ell-1)}\bar\phi^j_{(2\ell)}, ~~
   h_1^{(2\ell)} = -\f{2\pi}{k} U_{(2\ell)}{}^\hi_{~j} \bar\phi^j_{(2\ell)}\phi_\hi^{(2\ell+1)}\nn\\
&& h_2^{(2\ell-1)} = -\f{2\pi}{k} U_{(2\ell-1)}{}^i_{~\hj} \phi_i^{(2\ell)}\bar\phi^\hj_{(2\ell-1)}, ~~
   h_2^{(2\ell)} = -\f{2\pi}{k} U_{(2\ell)}{}^i_{~\hj} \bar\phi^\hj_{(2\ell+1)}\phi_i^{(2\ell)}
\eea
with constant parameters
\bea
&& U_{(2\ell-1)}{}^i_{~j}=  {\rm diag} \big(
     1-2\bar\a_\ho^{(2\ell-2)}\b^\ho_{(2\ell-2)},
     -1+ 2\bar\g_\hw^{(2\ell-2)}\d^\hw_{(2\ell-2)} \big)  \nn \\
&& U_{(2\ell-1)}{}^\hi_{~\hj} = {\rm diag} \big(
     1-2\bar\a_1^{(2\ell-1)}\b^1_{(2\ell-1)},
     -1+ 2\bar\g_2^{(2\ell-1)}\d^2_{(2\ell-1)} \big) \nn \\
&& U_{(2\ell)}{}^i_{~j} =  {\rm diag} \big(
     1-2\bar\a_\ho^{(2\ell)}\b^\ho_{(2\ell)} ,
     -1 + 2\bar\g_\hw^{(2\ell)}\d^\hw_{(2\ell)}
     \big) \nn\\
&& U_{(2\ell)}{}^\hi_{~\hj} =  {\rm diag} \big(
     1-2\bar\a_1^{(2\ell-1)}\b^1_{(2\ell-1)} ,
     -1+ 2\bar\g_2^{(2\ell-1)}\d^2_{(2\ell-1)}
     \big)  \nn\\
&& U_{(2\ell-1)}{}^\hi_{~j} =  {\rm diag} \big(
     2\bar\a_1^{(2\ell-1)}\b^\ho_{(2\ell)} ,
     -2\bar\g_2^{(2\ell-1)}\d^\hw_{(2\ell)}
     \big)  \nn\\
&& U_{(2\ell-1)}{}^i_{~\hj} =  {\rm diag} \big(
     2\bar\a_\ho^{(2\ell)}\b^1_{(2\ell-1)} ,
     -2\bar\g_\hw^{(2\ell)}\d^2_{(2\ell-1)}
     \big)  \nn\\
&& U_{(2\ell)}{}^\hi_{~j} =  {\rm diag} \big(
     2\bar\a_1^{(2\ell+1)}\b^\ho_{(2\ell)} ,
     -2\bar\g_2^{(2\ell+1)}\d^\hw_{(2\ell)}
     \big)  \nn\\
&& U_{(2\ell)}{}^i_{~\hj} = {\rm diag} \big(
     2\bar\a_\ho^{(2\ell)}\b^1_{(2\ell+1)} ,
     -2\bar\g_\hw^{(2\ell)}\d^2_{(2\ell+1)}
     \big)
\eea
The parameters are subject to the constraints
\bea
&& \bar\a_1^{(2\ell-1)} \d^2_{(2\ell-1)} =
   \bar\a_1^{(2\ell-1)} \d^\hw_{(2\ell-2)} =
   \bar\a_1^{(2\ell-1)} \d^\hw_{(2\ell)} = 0 \nn\\
&& \bar\a_\ho^{(2\ell)} \d^2_{(2\ell)} =
   \bar\a_\ho^{(2\ell)} \d^\hw_{(2\ell-1)} =
   \bar\a_\ho^{(2\ell)} \d^\hw_{(2\ell+1)} = 0 \nn\\
&& \bar\g_2^{(2\ell-1)} \b^1_{(2\ell-1)} =
   \bar\g_2^{(2\ell-1)} \b^\ho_{(2\ell-2)} =
   \bar\g_2^{(2\ell-1)} \b^\ho_{(2\ell)} = 0 \nn\\
&& \bar\g_\hw^{(2\ell)} \b^1_{(2\ell)} =
   \bar\g_\hw^{(2\ell)} \b^\ho_{(2\ell-1)} =
   \bar\g_\hw^{(2\ell)} \b^\ho_{(2\ell+1)} = 0
\eea
We have exactly reproduced the fermionic 1/4 BPS WL in \cite{Mauri:2017whf} preserving supercharges
\be
\th^{1\ho}_+, ~~ \th^{2\hw}_-, ~~ \vth^{1\ho}_+, ~~ \vth^{2\hw}_-
\ee

Similarly, from the 1/2 BPS WL $\td W_\fer$ in section~\ref{secn2}, we can construct a 1/4 BPS WL in $\mN=4$ orbifold ABJM theory that preserves supercharges $\th^{1\ho}_-$, $\th^{2\hw}_+$, $\vth^{1\ho}_-$, $\vth^{2\hw}_+$.

\vskip 10pt
In general, we do not know how to construct the gravity duals of BPS WLs in $\mN=4$ orbifold ABJM, directly. However, for WLs that can be obtained by an orbifold quotient of the 1/2 BPS operators of the ABJM theory, we can exploit their known gravity duals \cite{Lietti:2017gtc,Mauri:2017whf} and obtain the corresponding ones in $\cN=4$ orbifold ABJM theory by taking their orbifold quotient.

The $\cN=4$ orbifold ABJM theory is dual to M--theory in $\goabjm$ spacetime \cite{Benna:2008zy,Imamura:2008nn,Terashima:2008ba}
\be
ds^2=R^2 \Big( \frac14 ds^2_{\AdS_4}+ds^2_{ {\rm S}^7/(\Z_{rk} \times \Z_r)} \Big)
\ee
where the metric of ${\AdS_4}$ is given in (\ref{ads4}), the metric of $\rS^7$ in (\ref{s7}), and the $\Z_{rk} \times \Z_r$ quotient is generated by
\be
(z_1,z_2,z_3,z_4)\sim\ep^{\f{2\pi\ii}{rk}}(z_1,z_2,z_3,z_4) , ~~
(z_1,z_2,z_3,z_4)\sim(\ep^{\f{2\pi\ii}{r}}z_1,\ep^{\f{2\pi\ii}{r}}z_2,z_3,z_4)
\ee
which is equivalent to
\be
\zeta\sim \zeta - \frac{8\pi}{rk} , ~~ \chi\sim \chi - \frac{4\pi}{r} , ~~ \zeta\sim\zeta - \frac{4\pi}{r}
\ee
Performing the quotient of the 1/2 BPS operators $W_{1/2}[\bar\a_I]$ and $\td W_{1/2}[\bar\a_I]$ in ABJM theory corresponding to connections \eqref{W12ABJM}, we obtain 1/2 or 1/4 BPS WL in $\mN=4$ ABJM theory, depending on the value of $\bar\a_I$. The operator coming from $W_{1/2}[\bar\a_I]$ is dual to an M2--brane that wraps a cycle in the internal space specified by $\bar\a_I$, whereas the operator from $\td W_{1/2}[\bar\a_I]$ is dual to an anti--M2--brane that wraps the same interval cycle.

\section{Connections for 1/2 BPS WL in $\mN=2$ orbifold ABJM}\label{appcon}

In this appendix we collect the connections for 1/2 BPS WL in $\mN=2$ orbifold ABJM when $r=4,3,2$.
With definitions (\ref{n2def}), we have for $r=4$
\bea
&& \cA = \lt(\begin{array}{cccc}
           \cA^{(1)}             & 0                     & h_1^{(1)} + h_3^{(3)} & 0 \\
           0                     & \cA^{(2)}             & 0                     & h_1^{(2)} + h_3^{(4)} \\
           h_1^{(3)} + h_3^{(1)} & 0                     & \cA^{(3)}             & 0 \\
           0                     & h_1^{(4)} + h_3^{(2)} & 0                     & \cA^{(4)}
         \end{array}\rt) \nn\\
&& \cB = \lt(\begin{array}{cccc}
           \cB^{(1)}             & 0                     & h_4^{(1)} + h_2^{(3)} & 0 \\
           0                     & \cB^{(2)}             & 0                     & h_4^{(2)} + h_2^{(4)} \\
           h_4^{(3)} + h_2^{(1)} & 0                     & \cB^{(3)}             & 0 \\
           0                     & h_4^{(4)} + h_2^{(2)} & 0                     & \cB^{(4)}
         \end{array}\rt) \nn\\
&& f_1 = \lt(\begin{array}{cccc}
           f_1^{(1)} & f_3^{(1)} & 0         & f_5^{(4)} \\
           f_5^{(1)} & f_1^{(2)} & f_3^{(2)} & 0    \\
           0         & f_5^{(2)} & f_1^{(3)} & f_3^{(3)} \\
           f_3^{(4)} & 0         & f_5^{(3)} & f_1^{(4)}
         \end{array}\rt), ~~
   f_2 = \lt(\begin{array}{cccc}
           f_2^{(1)} & f_6^{(1)} & 0         & f_4^{(4)} \\
           f_4^{(1)} & f_2^{(2)} & f_6^{(2)} & 0   \\
           0         & f_4^{(2)} & f_2^{(3)} & f_6^{(3)} \\
           f_6^{(4)} & 0         & f_4^{(3)} & f_2^{(4)}
         \end{array}\rt)
\eea
for $r =3$
\bea
&& \cA = \lt(\begin{array}{ccc}
           \cA^{(1)} & h_3^{(2)} & h_1^{(1)} \\
           h_1^{(2)} & \cA^{(2)} & h_3^{(3)} \\
           h_3^{(1)} & h_1^{(3)} & \cA^{(3)}
         \end{array}\rt) , ~~
   \cB = \lt(\begin{array}{ccc}
           \cB^{(1)} & h_2^{(2)} & h_4^{(1)} \\
           h_4^{(2)} & \cB^{(2)} & h_2^{(3)} \\
           h_2^{(1)} & h_4^{(3)} & \cB^{(3)}
         \end{array}\rt) \nn\\
&& f_1 = \lt(\begin{array}{ccc}
           f_1^{(1)} & f_3^{(1)} & f_5^{(3)} \\
           f_5^{(1)} & f_1^{(2)} & f_3^{(2)} \\
           f_3^{(3)} & f_5^{(2)} & f_1^{(3)}
           \end{array}\rt), ~~
   f_2 = \lt(\begin{array}{ccc}
           f_2^{(1)} & f_6^{(1)} & f_4^{(3)} \\
           f_4^{(1)} & f_2^{(2)} & f_6^{(2)} \\
           f_6^{(3)} & f_4^{(2)} & f_2^{(3)}
         \end{array}\rt)
\eea
and for $r = 2$
\bea
&& \cA = \lt( \ba{cc} \cA^{(1)} + h_1^{(1)} + h_3^{(1)} & 0 \\
                  0 & \cA^{(2)} + h_1^{(2)} + h_3^{(2)} \ea\rt) \nn\\
&& \cB = \lt( \ba{cc} \cB^{(1)} + h_2^{(1)} + h_4^{(1)} & 0 \\
                  0 & \cB^{(2)} + h_2^{(2)} + h_4^{(2)} \ea\rt) \nn\\
&& f_1 = \lt(\begin{array}{cc}
           f_1^{(1)}             & f_3^{(1)} + f_5^{(2)} \\
           f_3^{(2)} + f_5^{(1)} & f_1^{(2)}
           \end{array}\rt) \nn\\
&& f_2 = \lt(\begin{array}{cc}
           f_2^{(1)}             & f_6^{(1)} + f_4^{(2)} \\
           f_6^{(2)} + f_4^{(1)} & f_2^{(2)}
         \end{array}\rt)
\eea

\section{{Lagrangian and Feynman rules}}  \label{appFeynman}

\begin{figure}[htbp]
\centering
\subfigure{\includegraphics[height=0.04\textwidth]{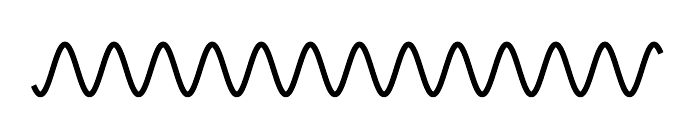}} ~~
\subfigure{\includegraphics[height=0.04\textwidth]{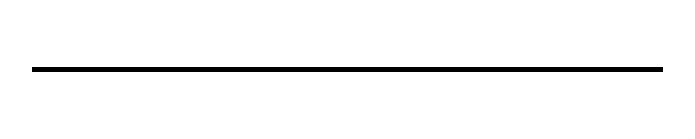}} ~~
\subfigure{\includegraphics[height=0.04\textwidth]{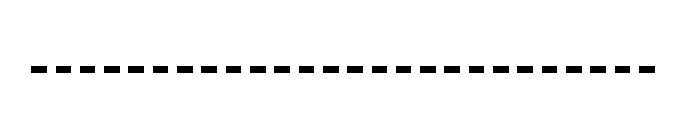}} \\
\subfigure{\includegraphics[height=0.1\textwidth]{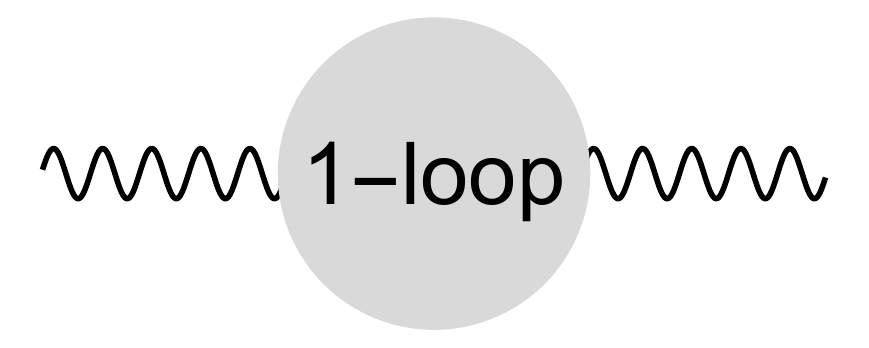}} ~~
\subfigure{\includegraphics[height=0.1\textwidth]{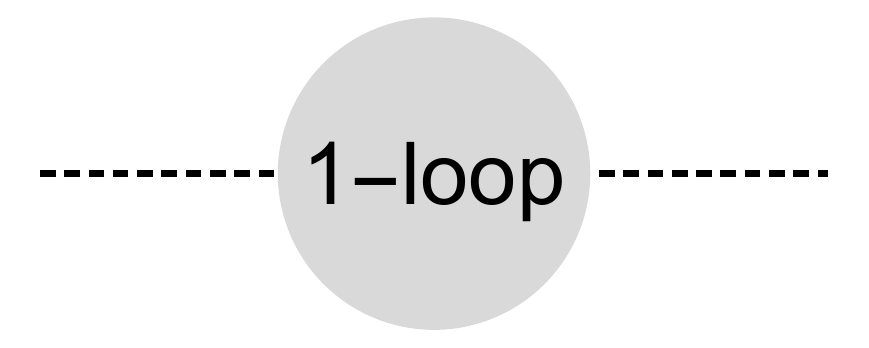}}
\caption{{Above: The tree propagators of gauge, scalar and fermionic fields. Below: The one--loop propagators of gauge and fermionic fields, respectively.}}\label{pg}
\end{figure}

In Minkowski spacetime, from the superspace lagrangian (\ref{lagrangian}) we obtain the relevant terms in components
\bea
&& \cL_\CS = \sum_a \f{k_a}{4\pi} \ve^{\m\n\r} \Tr\Big( A_\m^{(a)} \p_\n A_\r^{(a)}
                                                      + \f{2\ii}{3} A_\m^{(a)} A_\n^{(a)} A_\r^{(a)} \Big) \nn\\
&& \cL_k = \sum_{a,b}\Tr ( - D_\m \bar Z_i^{(ba)} D^\m Z^i_{(ab)} + \ii \bar \z_i^{(ba)} \g^\m D_\m \z^i_{(ab)}  )
\eea
By standard Wick rotation, the lagrangian in Euclidean space is given by
\bea\label{Lcomponents}
&& \cL_\CS = - \sum_a \f{ \ii k_a}{4\pi} \ve^{\m\n\r} \Tr\Big( A_\m^{(a)} \p_\n A_\r^{(a)}
                                                      + \f{2\ii}{3} A_\m^{(a)} A_\n^{(a)} A_\r^{(a)} \Big) \nn\\
&& \cL_k = \sum_{a,b}\Tr ( D_\m \bar Z_i^{(ba)} D^\m Z^i_{(ab)} - \ii \bar \z_i^{(ba)} \g^\m D_\m \z^i_{(ab)}  )
\eea
Here the definitions of covariant derivatives are
\bea \label{cov} D_\m Z^i_{(ab)}&=&\p_\m Z^i_{(ab)}+\ii A^{(a)}_\m Z^i_{(ab)}-\ii Z^i_{(ab)} A^{(b)}_\m,\\
D_\m \z^i_{(ab)}&=&\p_\m \z^i_{(ab)}+\ii A^{(a)}_\m \z^i_{(ab)}-\ii \z^i_{(ab)} A^{(b)}_\m \eea
{We work in Landau gauge for vector fields. Tree and one--loop propagators are drawn in figure~\ref{pg}.
In dimensional regularization, $d=3-2\e$, their explicit expressions at tree level are
\bea
&& \lag A_\m^{(a)}{}_p{}^q(x) A_\n^{(b)}{}_r{}^s(y) \rag^{(0)} = \d^{ab}\d_p^s\d_r^q
                                                                   \f{\ii}{k_a}
                                                                   \f{\G(\f32-\e)}{\pi^{\f12-\e}}
                                                                   \f{\ve_{\m\n\r}(x-y)^\r}{|x-y|^{3-2\e}} \nn\\
&& \lag Z_{(ab)}^i{}_p{}^q(x) \bar Z^{(cd)}_j{}_r{}^s(y) \rag^{(0)} = \d_a^d\d_b^c\d^i_j\d_p^s \d_r^q
                                                                      \f{\G(\f12-\e)}{4\pi^{\f32-\e}}
                                                                      \f{1}{|x-y|^{1-2\e}} \nn\\
&& \lag \z_{(ab)}^i{}_p{}^q{}_\a(x) \bar \z^{(cd)}_j{}_r{}^s{}^\b(y) \rag^{(0)} = \ii \d_a^d\d_b^c\d^i_j\d_p^s \d_r^q
                                                                                       \f{\G(\f32-\e)}{2\pi^{\f32-\e}}
                                                                                       \f{\g_\m{}_\a{}^\b(x-y)^\m}{|x-y|^{3-2\e}}
\eea
whereas their one--loop corrections read
\bea
&& \hspace{-3mm}
   \lag A_\m^{(a)}{}_p{}^q(x) A_\n^{(a)}{}_r{}^s(y) \rag^{(1)}
   = \d_p^s\d_r^q \sum_b\f{(N_{ab}+N_{ba}) N_b}{k_a}
     \f{\G^2(\f12-\e)}{4\pi^{1-2\e}} 
     \Big( \f{\d_{\m\n}}{|x-y|^{2-4\e}} - \f{\p_\m\p_\n|x-y|^{4\e}}{4\e(1+2\e)} \Big) \nn\\
&& \hspace{-3mm}
   \lag \z_{(ab)}^i{}_p{}^q{}_\a(x) \bar \z^{(cd)}_j{}_r{}^s{}^\b(y) \rag^{(1)}
   = -\ii \d_a^d\d_b^c\d^i_j\d_p^s \d_r^q \d_a^\b
     \Big( \f{N_a}{k_a} + \f{N_b}{k_b} \Big)
     \f{\G^2(\f12-\e)}{8\pi^{2-2\e}}
     \f{1}{|x-y|^{2-4\e}}
\eea}
Here the latic indices $p, q, r, s$ are color indices.

\begin{figure}[htbp]
\centering
\subfigure{\includegraphics[height=0.16\textwidth]{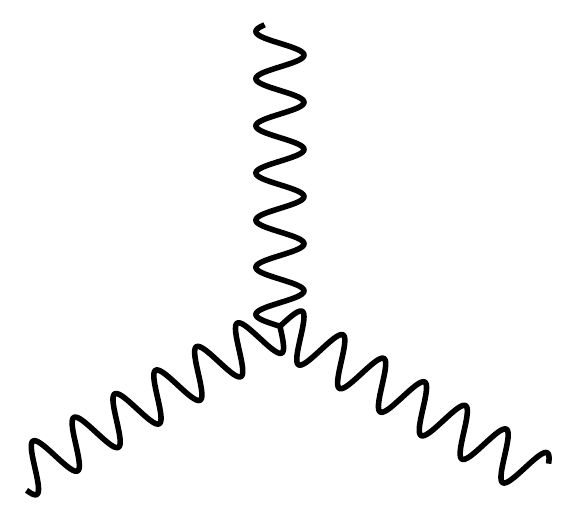}} ~~
\subfigure{\includegraphics[height=0.16\textwidth]{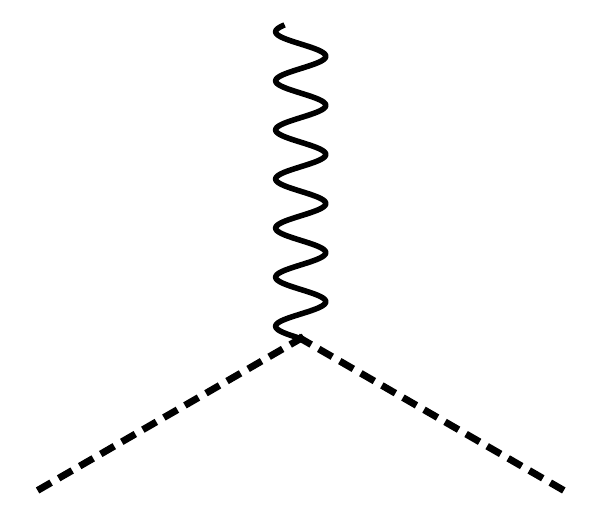}}
\caption{{The pure gauge vertex, and the mixed gauge--fermion vertex coming from the minimal coupling lagrangian.}}\label{vt}
\end{figure}

From the lagrangians in \eqref{Lcomponents} the cubic vertices of figure \ref{vt} are given by
\bea
&& - \sum_a \f{k_a}{6\pi} \!\int\! d^3x \ve^{\m\n\r} A_\m^{(a)}{}_p{}^q(x) A_\n^{(a)}{}_q{}^r(x) A_\r^{(a)}{}_r{}^p(x) \nn\\
&& - \sum_{a,b} \!\int\! d^3x \bar\z^{(ba)}_i{}_p{}^q(x) \g^\m [ A_\m^{(a)}{}_q{}^r(x) \z^i_{(ab)}{}_r{}^p(x)
                                                                 - \z^i_{(ab)}{}_q{}^r(x) A_\m^{(b)}{}_r{}^p(x) ]
\eea

\section{Details on the perturbative computation} \label{appint}

{In this appendix we give the explicit expressions of the intergrals corresponding to diagrams in figures \ref{fd1}--\ref{fd6}.
The intergrals are defined as
\be
\dcircled{\ref{fdAA}} = \sum_a \f{N_a^2}{k_a} \cI_{\rm\ref{fdAA}}^{(f)}
\ee
\bea
&& \dcircled{\ref{fdAAp}} =  \sum_{a,b} \f{(N_{ab} + N_{ba})N_a^2N_b}{k_a^2} \cI_{\rm\ref{fdAAp}}^{(f)}, ~~
   \dcircled{\ref{fdAAA}} = \sum_a \f{N_a^3 - N_a}{k_a^2} \cI_{\rm\ref{fdAAA}}^{(f)} \nn\\
&& \dcircled{\ref{fdAAAA}} = \sum_a \f{N_a^3}{k_a^2} \cI_{\rm\ref{fdAAAA}}^{(f)}
                           + \sum_a \f{N_a}{k_a^2} \cJ_{\rm\ref{fdAAAA}}^{(f)}, ~~
   \dcircled{\ref{fdSS}} =  \sum_{a,b} \f{(N_{ab} + N_{ba})N_a^2N_b}{k_a^2} \cI_{\rm\ref{fdSS}}^{(f)}
\eea
\be
\dcircled{\ref{fdFF}} = \sum_{a,b} \bar m_i^{ab} n^i_{ba} N_a N_b \cI_{\rm\ref{fdFF}}^{(f)}
\ee
\bea
&& \dcircled{\ref{fdSB}} = \sum_{a,b} ( \bar m_i^{ab} n^i_{ba} - \bar m_i^{ba} n^i_{ab} )
                                      \f{N_a^2 N_b}{k_a} \cI_{\rm\ref{fdSB}}^{(f)} \nn\\
&& \dcircled{\ref{fdFFp}} = \sum_{a,b} ( \bar m_i^{ab} n^i_{ba} + \bar m_i^{ba} n^i_{ab} )
                                       \f{N_a^2 N_b}{k_a} \cI_{\rm\ref{fdFFp}}^{(f)} \nn\\
&& \dcircled{\ref{fdAFF}} = \sum_{a,b} ( \bar m_i^{ab} n^i_{ba} \cI_{\rm\ref{fdAFF}}^{(f)}
                                       + \bar m_i^{ba} n^i_{ab} \cJ_{\rm\ref{fdAFF}}^{(f)} ) \f{N_a^2 N_b}{k_a} \nn\\
&& \dcircled{\ref{fdAAFF}} =  \sum_{a,b} ( \bar m_i^{ab} n^i_{ba} \cI_{\rm\ref{fdAAFF}}^{(f)}
                                         + \bar m_i^{ba} n^i_{ab} \cJ_{\rm\ref{fdAAFF}}^{(f)} ) \f{N_a^2 N_b}{k_a}
\eea
~
\bea
&& \dcircled{\ref{fdBB}} = \sum_{a,b,c} ( \bar m_i^{ab} n^i_{ba} \bar m_j^{ac} n^j_{ca}
                                        + \bar m_i^{ba} n^i_{ab} \bar m_j^{ca} n^j_{ac} ) N_a N_b N_c \cI_{\rm\ref{fdBB}}^{(f)} \nn\\
&& \dcircled{\ref{fdFFFF}} = \sum_{a,b,c} ( \bar m_i^{ab} n^i_{ba} \bar m_j^{ac} n^j_{ca} \cI_{\rm\ref{fdFFFF}}^{(f)}
                                          + \bar m_i^{ba} n^i_{ab} \bar m_j^{ca} n^j_{ac} \cJ_{\rm\ref{fdFFFF}}^{(f)} ) N_a N_b N_c
\eea
~
\bea
&& \dcircled{\ref{fdBBB}} = \Big[ \sum_{a,b}2(\bar m^{ab}_i n^i_{ba})^3 N_a N_b
           + \sum_{a,b,c,d} ( \bar m^{ab}_i n^i_{ba} \bar m^{ac}_j n^j_{ca} \bar m^{ad}_k n^k_{da} \nn\\
&& \hspace{13mm}
                            + \bar m^{ba}_i n^i_{ab} \bar m^{ca}_j n^j_{ac} \bar m^{da}_k n^k_{ad} ) N_a N_b N_c N_d  \Big]
                            \cI_{\rm\ref{fdBBB}}^{(f)} \nn\\
&& \dcircled{\ref{fdBBFF}} = + \sum_{a,b,c,d} \bar m^{ab}_i n^i_{ba} \bar m^{ac}_j n^j_{ca} \bar m^{db}_k n^k_{bd}
                                              N_a N_b N_c N_d \cI_{\rm\ref{fdBBFF}}^{(f)}
                             + \sum_{a,b}(\bar m^{ab}_i n^i_{ba})^3 N_a N_b \cJ_{\rm\ref{fdBBFF}}^{(f)} \nn\\
&& \dcircled{\ref{fdFFFFFF}} =
             \sum_{a,b,c,d} [ \bar m^{ab}_i n^i_{ba} \bar m^{ac}_j n^j_{ca} \bar m^{ad}_k n^k_{da} \cI_{\rm\ref{fdFFFFFF}}^{(f)}
                            + \bar m^{ba}_i n^i_{ab} \bar m^{ca}_j n^j_{ac} \bar m^{da}_k n^k_{ad} \cJ_{\rm\ref{fdFFFFFF}}^{(f)} \nn\\
&& \hspace{13mm}
                            + \bar m^{ab}_i n^i_{ba} \bar m^{ac}_j n^j_{ca} \bar m^{db}_k n^k_{bd} \cK_{\rm\ref{fdFFFFFF}}^{(f)} ] N_a N_b N_c N_d
           + \sum_{a,b}(\bar m^{ab}_i n^i_{ba})^3 N_a N_b \cL_{\rm\ref{fdFFFFFF}}^{(f)}
\eea}

{Using the Feynman rules in appendix~\ref{appFeynman}, the integral from figure~\ref{fd1} is given by
\be
\cI_{\rm\ref{fdAA}}^{(f)} = - \ii \f{\G(\f32-\e)}{\pi^{\f12-\e}}
                              \!\oint\! d \t_{1>2} \f{\ve_{\m\n\r} \dot x_1^\m \dot x_2^\n x_{12}^\r}{|x_{12}|^{3-2\e}}
\ee
where we have defined $x_i \equiv x(\t_i)$, $x_{ij} \equiv x_i - x_j$ and $\oint d \t_{1>2}$ means a double contour integral over $\t_1 > \t_2$.

With similar notations, from figure~\ref{fd2} we obtain
\bea
&& \cI_{\rm\ref{fdAAp}}^{(f)} = - \f{\G^2(\f12-\e)}{4\pi^{1-2\e}}
                                  \!\oint\! d \t_{1>2} \f{\dot x_1 \cdot \dot x_2}{|x_{12}|^{2-4\e}}  \nn\\
&& \cI_{\rm\ref{fdAAA}}^{(f)} = - \f{\G^3(\f32-\e)}{2\pi^{\f52-3\e}}
                                  \!\oint\! d \t_{1>2>3} \!\!\int\!\! d^3x
                                  \f{\dot x_1^\m\dot x_2^\n\dot x_3^\r
                                     \ve^{\a\b\g}\ve_{\m\a\s}\ve_{\n\b\l}\ve_{\r\g\eta}
                                     (x-x_1)^\s(x-x_2)^\l(x-x_3)^\eta}
                                    {|x-x_1|^{3-2\e}|x-x_2|^{3-2\e}|x-x_3|^{3-2\e}} \nn\\
&& \cI_{\rm\ref{fdAAAA}}^{(f)} = - \f{\G^2(\f32-\e)}{\pi^{1-2\e}}
                                   \!\oint\! d \t_{1>2>3>4}
                                   \Big( \f{\ve_{\m\n\l} \dot x_1^\m \dot x_2^\n \dot x_{12}^\l
                                            \ve_{\r\s\eta} \dot x_3^\r \dot x_4^\s \dot x_{34}^\eta}
                                           {|x_{12}|^{3-2\e}|x_{34}|^{3-2\e}}
                                        + (1432) \Big)    \nn\\
&& \cJ_{\rm\ref{fdAAAA}}^{(f)} = - \f{\G^2(\f32-\e)}{\pi^{1-2\e}}
                                   \!\oint\! d \t_{1>2>3>4}
                                   \f{\ve_{\m\n\l} \dot x_1^\m \dot x_3^\n \dot x_{13}^\l
                                      \ve_{\r\s\eta} \dot x_2^\r \dot x_4^\s \dot x_{24}^\eta}
                                     {|x_{13}|^{3-2\e}|x_{24}|^{3-2\e}} \nn\\
&& \cI_{\rm\ref{fdSS}}^{(f)} = \f{\G^2(\f12-\e)}{4\pi^{1-2\e}}
                               \!\oint\! d \t_{1>2} \f{|\dot x_1| |\dot x_2|}{|x_{12}|^{2-4\e}}
\eea
with the symbol $(1423)$ in $\cI_{\rm\ref{fdAAAA}}^{(f)}$ indicating the term obtained from the first one by permuting $\t_{1,2,3,4} \to \t_{1,4,2,3}$.

Similarly, from figures~\ref{fd3}--\ref{fd6} we obtain
\be
\cI_{\rm\ref{fdFF}}^{(f)} = - \ii \f{\G(\f32-\e)}{2\pi^{\f32-\e}}
                              \!\oint\! d\t_{1>2}
                              \Big( \f{|\dot x_1||\dot x_2|u_+(\t_1)\g_\m u_-(\t_2)x_{12}^\m}{|x_{12}|^{3-2\e}} - (21) \Big)
\ee
\bea
&& \cI_{\rm\ref{fdSB}}^{(f)} = - \f{\G^2(\f12-\e)}{4\pi^{2-2\e}}
                                 \!\oint\! d \t_{1>2} \f{|\dot x_1| |\dot x_2|}{|x_{12}|^{2-4\e}} \nn\\
&& \cI_{\rm\ref{fdFFp}}^{(f)} = \ii \f{\G^2(\f12-\e)}{8\pi^{2-2\e}}
                                \!\oint\! d\t_{1>2}
                                \Big( \f{|\dot x_1||\dot x_2|u_+(\t_1)u_-(\t_2)}{|x_{12}|^{2-4\e}} - (21)  \Big) \nn\\
&& \cI_{\rm\ref{fdAFF}}^{(f)} = - \f{\G^3(\f32-\e)}{4\pi^{\f72-3\e}}
                                  \!\oint\! d \t_{1>2>3} \!\!\int\!\! d^3x
                                  \Big[
                                    \f{\dot x_1^\m \ve_{\m\n\r}(x-x_1)^\r}
                                      {|x-x_1|^{3-2\e}} \nn\\
&& \phantom{\cI_{\rm\ref{fdAFF}}^{(f)} =}
                                  \times
                                  \f{|\dot x_2| |\dot x_3| u_+(\t_2)\g_\s\g^\n\g_\l u_-(\t_3) (x-x_2)^\s(x-x_3)^\l}
                                    {|x-x_2|^{3-2\e}|x-x_3|^{3-2\e}}
                                  - (231) + (312)
                                  \Big] \nn\\
&& \cJ_{\rm\ref{fdAFF}}^{(f)} = - \f{\G^3(\f32-\e)}{4\pi^{\f72-3\e}}
                                  \!\oint\! d \t_{1>2>3} \!\!\int\!\! d^3x
                                  \Big[
                                    \f{\dot x_1^\m \ve_{\m\n\r}(x-x_1)^\r}
                                      {|x-x_1|^{3-2\e}} \nn\\
&& \phantom{\cJ_{\rm\ref{fdAFF}}^{(f)} =}
                                  \times
                                  \f{|\dot x_3| |\dot x_2| u_+(\t_3)\g_\s\g^\n\g_\l u_-(\t_2) (x-x_3)^\s(x-x_2)^\l}
                                    {|x-x_3|^{3-2\e}|x-x_2|^{3-2\e}}
                                  - (231) + (312)
                                  \Big] \nn\\
&& \cI_{\rm\ref{fdAAFF}}^{(f)} = - \f{\G^2(\f32-\e)}{2\pi^{2-2\e}}
                                   \!\oint\! d \t_{1>2>3>4}
                                   \Big(
                                     \f{\dot x_1^\m \dot x_2^\n |\dot x_3| |\dot x_4|
                                        \ve_{\m\n\r} x_{12}^\r
                                        u_+(\t_3)\g_\s u_-(\t_4)x_{34}^\s}
                                       {|x_{12}|^{3-2\e}|x_{34}|^{3-2\e}} \nn\\
&& \phantom{\cI_{\rm\ref{fdAAFF}}^{(f)} =}
                                   + (3412) + (4123) - (2341)
                                   \Big) \nn\\
&& \cJ_{\rm\ref{fdAAFF}}^{(f)} = - \f{\G^2(\f32-\e)}{2\pi^{2-2\e}}
                                   \!\oint\! d \t_{1>2>3>4}
                                   \Big(
                                     \f{\dot x_1^\m \dot x_2^\n |\dot x_4| |\dot x_3|
                                        \ve_{\m\n\r} x_{12}^\r
                                        u_+(\t_4)\g_\s u_-(\t_3)x_{43}^\s}
                                       {|x_{12}|^{3-2\e}|x_{43}|^{3-2\e}} \nn\\
&& \phantom{\cJ_{\rm\ref{fdAAFF}}^{(f)} =}
                                   + (3412) + (4123) - (2341)
                                   \Big)
\eea
\bea
&& \cI_{\rm\ref{fdBB}}^{(f)} =  \f{\G^2(\f12-\e)}{16\pi^{3-2\e}}
                                \!\oint\! d \t_{1>2} \f{|\dot x_1| |\dot x_2|}{|x_{12}|^{2-4\e}} \nn\\
&& \cI_{\rm\ref{fdFFFF}}^{(f)} = - \f{\G^2(\f32-\e)}{4\pi^{3-2\e}}
                                   \!\oint\! d\t_{1>2>3>4}
                                   \Big(
                                     \f{|\dot x_1||\dot x_2|
                                        u_+(\t_1)\g_\m u_-(\t_2)x_{12}^\m}
                                       {|x_{12}|^{3-2\e}} \nn\\
&& \phantom{\cI_{\rm\ref{fdFFFF}}^{(f)} =}
                                     \times
                                     \f{|\dot x_3||\dot x_4|
                                        u_+(\t_3)\g_\n u_-(\t_4)x_{34}^\n}
                                       {|x_{34}|^{3-2\e}}
                                     - (2341)
                                    \Big) \nn\\
&& \cJ_{\rm\ref{fdFFFF}}^{(f)} = - \f{\G^2(\f32-\e)}{4\pi^{3-2\e}}
                                   \!\oint\! d\t_{1>2>3>4}
                                   \Big(
                                     \f{|\dot x_2||\dot x_1|
                                        u_+(\t_2)\g_\m u_-(\t_1)x_{12}^\m}
                                       {|x_{21}|^{3-2\e}} \nn\\
&& \phantom{\cJ_{\rm\ref{fdFFFF}}^{(f)} =}
                                     \times
                                     \f{|\dot x_4||\dot x_3|
                                        u_+(\t_4)\g_\n u_-(\t_3)x_{43}^\n}
                                       {|x_{43}|^{3-2\e}}
                                     - (2341)
                                    \Big)
\eea
\bea \label{intfd6}
&& \cI_{\rm\ref{fdBBB}}^{(f)} = - \f{\G^3(\f12-\e)}{64\pi^{\f92-3\e}}
                                  \!\oint\! d \t_{1>2>3}
                                  \f{|\dot x_1| |\dot x_2| |\dot x_3|}
                                    {|x_{12}|^{1-2\e}|x_{13}|^{1-2\e}|x_{23}|^{1-2\e}} \nn\\
&& \cI_{\rm\ref{fdBBFF}}^{(f)} = - \ii \f{\G^2(\f12-\e)\G(\f32-\e)}{32\pi^{\f92-3\e}}
                                   \!\oint\! d \t_{1>2>3>4} \Big(
                                   \f{|\dot x_1| |\dot x_2| |\dot x_3||\dot x_4|u_+(\t_3)\g_\m u_-(\t_4)x_{34}^\m}
                                     {|x_{12}|^{2-4\e}|x_{34}|^{3-2\e}} \nn\\
&& \phantom{\cI_{\rm\ref{fdBBFF}}^{(f)} =}
                                   +(3412)-(4123)-(2341)-(2143)-(4321)-(1432)+(3214)  \Big) \nn\\
&& \cJ_{\rm\ref{fdBBFF}}^{(f)} = - \ii \f{\G^2(\f12-\e)\G(\f32-\e)}{32\pi^{\f92-3\e}}
                                   \!\oint\! d \t_{1>2>3>4} \Big(
                                   \f{|\dot x_1| |\dot x_3| |\dot x_2||\dot x_4|u_+(\t_2)\g_\m u_-(\t_4)x_{24}^\m}
                                     {|x_{13}|^{2-4\e}|x_{24}|^{3-2\e}} \nn\\
&& \phantom{\cJ_{\rm\ref{fdBBFF}}^{(f)} =}
                                   -(2341)-(1432)+(2143) \Big) \nn\\
&& \cI_{\rm\ref{fdFFFFFF}}^{(f)} = \ii \f{\G^3(\f32-\e)}{8\pi^{\f92-3\e}}
                                   \!\oint\! d \t_{1>2>3>4>5>6} \Big(
                                   \f{|\dot x_1||\dot x_2|u_+(\t_1)\g_\m u_-(\t_2)x_{12}^\m}{|x_{12}|^{3-2\e}} \nn\\
&& \phantom{\cI_{\rm\ref{fdFFFFFF}}^{(f)} =}\times
                                   \f{|\dot x_3||\dot x_4|u_+(\t_3)\g_\m u_-(\t_4)x_{34}^\m}{|x_{34}|^{3-2\e}}
                                   \f{|\dot x_5||\dot x_6|u_+(\t_5)\g_\m u_-(\t_6)x_{56}^\m}{|x_{56}|^{3-2\e}}
                                   -(234561)
                                   \Big) \nn\\
&& \cJ_{\rm\ref{fdFFFFFF}}^{(f)} =- \ii \f{\G^3(\f32-\e)}{8\pi^{\f92-3\e}}
                                   \!\oint\! d \t_{1>2>3>4>5>6} \Big(
                                   \f{|\dot x_2||\dot x_1|u_+(\t_2)\g_\m u_-(\t_1)x_{21}^\m}{|x_{21}|^{3-2\e}} \nn\\
&& \phantom{\cJ_{\rm\ref{fdFFFFFF}}^{(f)} =}\times
                                   \f{|\dot x_4||\dot x_3|u_+(\t_4)\g_\m u_-(\t_3)x_{43}^\m}{|x_{43}|^{3-2\e}}
                                   \f{|\dot x_6||\dot x_5|u_+(\t_6)\g_\m u_-(\t_5)x_{65}^\m}{|x_{65}|^{3-2\e}}
                                   -(234561)
                                   \Big) \nn\\
&& \cK_{\rm\ref{fdFFFFFF}}^{(f)} =- \ii \f{\G^3(\f32-\e)}{8\pi^{\f92-3\e}}
                                   \!\oint\! d \t_{1>2>3>4>5>6} \Big(
                                   \f{|\dot x_1||\dot x_2|u_+(\t_1)\g_\m u_-(\t_2)x_{12}^\m}{|x_{12}|^{3-2\e}} \nn\\
&& \phantom{\cK_{\rm\ref{fdFFFFFF}}^{(f)} =}\times
                                   \f{|\dot x_3||\dot x_6|u_+(\t_3)\g_\m u_-(\t_6)x_{36}^\m}{|x_{36}|^{3-2\e}}
                                   \f{|\dot x_5||\dot x_4|u_+(\t_5)\g_\m u_-(\t_4)x_{54}^\m}{|x_{54}|^{3-2\e}} \nn\\
&& \phantom{\cK_{\rm\ref{fdFFFFFF}}^{(f)} =}
                                   +(561234)-(612345)-(216543)-(654321)+(165432)
                                   \Big) \nn\\
&& \cL_{\rm\ref{fdFFFFFF}}^{(f)} = \ii \f{\G^3(\f32-\e)}{8\pi^{\f92-3\e}}
                                   \!\oint\! d \t_{1>2>3>4>5>6} \Big(
                                   \f{|\dot x_1||\dot x_4|u_+(\t_1)\g_\m u_-(\t_4)x_{14}^\m}{|x_{14}|^{3-2\e}} \\
&& \phantom{\cL_{\rm\ref{fdFFFFFF}}^{(f)} =}\times
                                   \f{|\dot x_5||\dot x_2|u_+(\t_5)\g_\m u_-(\t_2)x_{52}^\m}{|x_{52}|^{3-2\e}}
                                   \f{|\dot x_3||\dot x_6|u_+(\t_3)\g_\m u_-(\t_6)x_{36}^\m}{|x_{36}|^{3-2\e}}
                                   -(456123)
                                   \Big) \nn
\eea}

\providecommand{\href}[2]{#2}\begingroup\raggedright\endgroup




\end{document}